\documentclass[aps,pra,reprint,amsfonts,amsmath,amssymb,longbibliography]{revtex4-2}
\usepackage{xcolor}

\usepackage{hyperref}
\usepackage{graphicx}
\usepackage{bm}
\usepackage{newtxtext}
\usepackage[varg]{newtxmath}
\usepackage[normalem]{ulem}
\usepackage{lipsum}
\usepackage{slashed}
\usepackage{mathrsfs}
\usepackage{blkarray}
\usepackage{mleftright}
\usepackage{bbold}
\mleftright

\hypersetup{colorlinks=true,linkcolor=blue,citecolor=blue,urlcolor=blue}

\begin{document}
\title{Macroscopic Schr\"{o}dinger-cat states of nonequilibrium electrons induced by cat-state\\ optical driving and projective measurements on the light field}
\author{Shohei Imai}
\affiliation{Department of Physics, University of Tokyo, Hongo, Tokyo 113-0033, Japan}
\date{\today}

\begin{abstract}
We show that projective measurements on quantum light can induce macroscopic cat states in many-electron systems driven by such light.
Here we investigate the quantum dynamics of $N$ independent two-level electrons interacting with Schr\"{o}dinger-cat or -kitten states of light.
Without measurement, a macroscopic cat state of the electrons appears only in an ultrashort time window.
In contrast, we demonstrate that photon-number parity or quadrature projective measurements can restore a macroscopic cat state in nonequilibrium electrons, even in the thermodynamic limit.
These dynamics are captured by an external-field approximation, in which the electronic system evolves into a Rabi-oscillation cat state.
Our results highlight the need for precise quantum measurement techniques for light to control macroscopic quantum states of matter driven by quantum light.
\end{abstract}
\maketitle

\section{Introduction} \label{sec:introduction}
Light provides a versatile tool for controlling material properties through its various degrees of freedom.
One striking example is circular polarization, which enables control over magnetic properties and electron orbital motion (topology) by breaking time-reversal symmetry~\cite{Oka2009, McIver2020}.
This naturally raises the question: can quantum states of light directly generate or manipulate quantum states of matter?

Among the various manifestations of quantumness, we focus on quantum superpositions between classically distinguishable states, i.e., superpositions of coherent states (and their electronic analogs).
Recent progress combining attosecond science with quantum-measurement techniques has enabled the generation of large-amplitude Schr\"{o}dinger-cat states of light~\cite{Lewenstein2021, Rivera-Dean2022, Lamprou2025} (see also theoretical developments in Refs.~\cite{Stammer2022a, Stammer2022, Anonymous2025a} and review articles~\cite{Stammer2023, Helversen2023, Lewenstein2024, Cruz-Rodriguez2024a}).
Notably, nonlinear optical phenomena induced by such cat-state light have already been observed experimentally~\cite{Lamprou2025}.

Despite these advances, most previous studies have been limited to few-electron systems driven by small-amplitude cat-state light~\cite{Vidiella-Barranco1992, Gerry1993, Moya-Cessa1995, Joshi1995, Bocanegra-Garay2024, Liu2006, Mohamed2019a, Mohamed2021d, Abdel-Khalek2021, Movahedi2023, Imai2025b, Kuang1997, Horoshko2000, Tomilin2016, Tomilin2017, Tomilin2017a, Bertassoli2024, Ling1997, Rundle2021}.
Studies in Refs.~\cite{Vidiella-Barranco1992, Gerry1993, Moya-Cessa1995, Joshi1995, Bocanegra-Garay2024} have examined the population dynamics of a single two-level system interacting with cat-state light.
Other works~\cite{Liu2006, Mohamed2019a, Mohamed2021d, Abdel-Khalek2021, Movahedi2023, Imai2025b} have explored various forms of quantum correlations generated between two two-level systems under such interactions.
Additional investigations have addressed quantum correlations in Josephson junctions~\cite{Kuang1997}, resonance fluorescence from excited atoms~\cite{Horoshko2000, Tomilin2016, Tomilin2017, Tomilin2017a, Bertassoli2024}, interactions with effective bosons in the large-spin limit~\cite{Ling1997}, and informationally complete phase-space representations~\cite{Rundle2021}.

In Ref.~\cite{Imai2025b}, the authors studied the reduced electronic density matrix of a two-qubit system driven by the even cat state $|\alpha_0\rangle + |{-}\alpha_0\rangle$ of light, composed of the out-of-phase coherent states $|{\pm}\alpha_0\rangle$.
They demonstrated that, as the quantum interference term $\langle {-}\alpha_0|\alpha_0\rangle$ decreases with increasing photon number, the ability of the light to induce entanglement in the electronic system diminishes (see also Ref.~\cite{Movahedi2023}).
This indicates that large-amplitude cat-state optical driving alone results in a classical statistical mixture of classically driven electronic states, rather than a quantum superposition~\cite{Zurek2003}.

\begin{figure}[b]
\centering
\includegraphics[width=0.7\columnwidth]{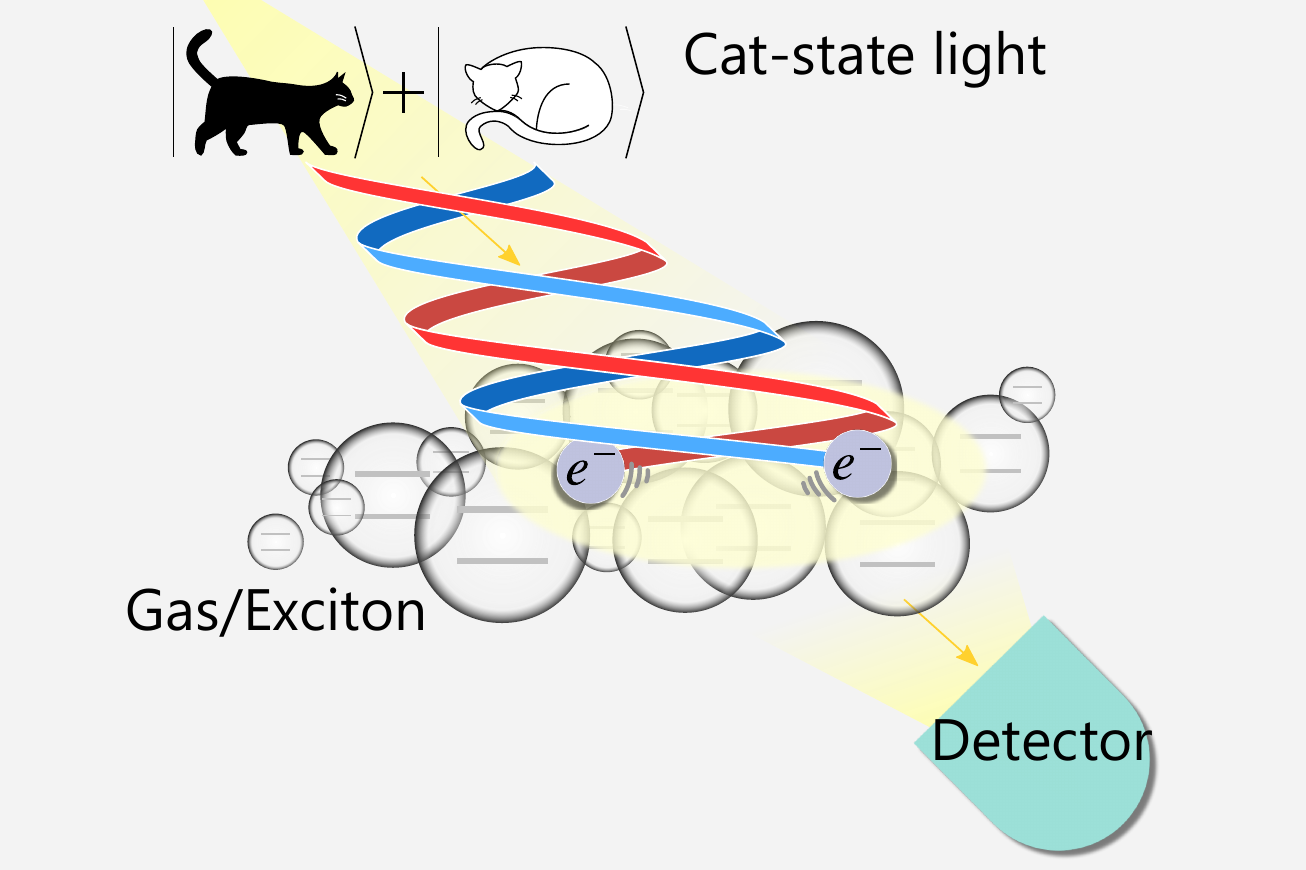}
\caption{
Schematic of measurement-assisted generation of macroscopic cat states of electrons driven by cat-state light.
A double helix represents a quantum superposition of two out-of-phase laser fields (black-cat and white-cat states).
This light drives an ensemble of $N$ independent two-level electrons (gray bubbles).
Projective measurement of the light (green region) yields a macroscopic cat state in the electronic system, while no measurement leaves a classical mixture.
}
\label{fig:cat_irradiation}
\end{figure}
Here, we demonstrate that macroscopic quantum states can be induced in many-electron systems by irradiating them with large-amplitude cat-state light and subsequently performing projective measurements on the light field (Fig.~\ref{fig:cat_irradiation}).
We show that photon-number-parity measurements or homodyne detection can project the electronic system into Rabi-oscillation cat states in an ensemble of $N$ independent two-level electrons (qubits).
The underlying mechanism can be understood through the approximate electron--photon state vector
\begin{equation}
|\Psi\rangle_{\mathrm{ep}} \sim |\psi_{+}\rangle_{\mathrm{e}} |\alpha_0\rangle_{\mathrm{p}} + |\psi_{-}\rangle_{\mathrm{e}} |{-}\alpha_0\rangle_{\mathrm{p}}, \label{eq:approx total state}
\end{equation}
derived for even cat-state optical driving, where $|\psi_{\pm}\rangle_{\mathrm{e}}$ are the electronic states driven by the classical light $|{\pm}\alpha_0\rangle_{\mathrm{p}}$.
Here, the subscripts $\mathrm{e}$, $\mathrm{p}$, and $\mathrm{ep}$ refer to the Hilbert spaces of the electrons, photons, and the total system, respectively (omitted when unambiguous).

We identify two distinct dynamical regimes: the few-photon regime and the large-photon regime, depending on the photon-to-electron ratio.
In the few-photon regime, all photons can be absorbed resonantly by the matter, and the total state becomes
\begin{equation}
|\Psi\rangle_{\mathrm{ep}} \sim (|\psi_{+}\rangle + |\psi_{-}\rangle) |0\rangle. \label{eq:approx total state:small amplitude}
\end{equation}
This is understood as a quantum-state transfer (or swapping) process, in which both energy and quantum coherence are fully transferred to the material.
Such processes have been discussed for cat states~\cite{Ling1997, Rundle2021}, and for squeezed vacuum states~\cite{Kuzmich1997, Hald1999, Hald2001, Ma2011c}.
In this work, we further provide a quantitative analysis of their real-time dynamics.

In the large-photon regime, absorption negligibly depletes the coherent-state amplitude.
Tracing out the photons then yields the reduced electronic density matrix
\begin{align}
\hat{\rho}_{\mathrm{e}} 
&\sim 
|\psi_{+}\rangle \langle \psi_{+}|
+|\psi_{-}\rangle \langle \psi_{-}| \nonumber \\
&+\langle {-}\alpha_0|\alpha_0\rangle (|\psi_{+}\rangle \langle \psi_{-}|
+ |\psi_{-}\rangle \langle \psi_{+}|). \label{eq:reduced electron density matrix}
\end{align}
The quantum interference terms $|\psi_{\pm}\rangle \langle \psi_{\mp}|$, weighted by $\langle {-}\alpha_0|\alpha_0\rangle = \exp(-2|\alpha_0|^2)$, decay exponentially with photon number.
Discarding the photons thus leaves the electronic system in a classical statistical mixture.

We show that the lost quantum coherence can be restored by suitable projective measurements, such as photon-number measurements.
Projecting onto a Fock state $|n\rangle$ ($n \in \mathbb{N}$) yields a postselected total state
\begin{equation}
|n\rangle \langle n| \Psi\rangle_{\mathrm{ep}} \sim (|\psi_{+}\rangle +(-1)^n|\psi_{-}\rangle) \otimes |n\rangle, \label{eq:n-projected state}
\end{equation}
with $\langle n|{\pm} \alpha_0\rangle = (\pm1)^n \mathrm{e}^{-|\alpha_0|^2/2} \alpha_0^{n}/\sqrt{n!}$.
When $|\psi_{\pm}\rangle_{\mathrm{e}}$ differ macroscopically in an additive observable, the postselected state is a macroscopic quantum superposition or a genuinely multipartite entangled state~\cite{Frowis2018, Shimizu2002a, Frowis2012, Toth2012, Hyllus2012}.

This conditional preparation of macroscopic quantum states in matter contrasts with previous approaches, which have typically relied on nonlinear interactions between particles~\cite{Agarwal1997} or on quantum nondemolition (QND) interactions via Raman processes followed by projective measurements~\cite{Gerry1997, Massar2003, Genes2006, Filip2008, Lemr2009, Nielsen2009, Christensen2013, McConnell2013, McConnell2015, Huang2015}.
Instead, by combining standard electromagnetic interactions with macroscopic cat-state light, our approach can be applied not only to atomic ensembles but also to more complex electronic systems, including electrons in solids or molecular gases.
This capability marks a fundamentally new paradigm for quantum material control, made possible by recent advances in the generation of macroscopic quantum light.

The remainder of this paper is organized as follows: Section~\ref{sec:XFA} formulates an external-field approximation for quantum light, yielding Eq.~\eqref{eq:approx total state}, which provides both an effective theory and an intuitive picture of many-body electronic dynamics under macroscopic quantum light.
Section~\ref{sec:model and method} describes the $N$-qubit Rabi (Dicke) and Tavis--Cummings models~\cite{Rabi1937, Dicke1954, Tavis1968}, and introduces the quantum Fisher information (QFI)~\cite{Helstrom1969, Holevo2011, Braunstein1994} as a measure of macroscopic quantumness.
Section~\ref{sec:trace_out} examines electronic dynamics under cat-state light without postselection, systematically discussing their dependence on the cat-state amplitude.
Section~\ref{sec:parity} analyzes electronic dynamics postselected on the photon-number parity, demonstrates projections onto cat states, such as the Greenberger--Horne--Zeilinger (GHZ) state~\cite{Greenberger1989}, and compares the results with an effective Rabi-oscillation-cat description.
Section~\ref{sec:quadrature} investigates electronic dynamics induced by kitten-state light combined with quadrature measurements, quantifying the role of measurement resolution.
Finally, Sec.~\ref{sec:discussion} discusses the influence of the rotating-wave approximation (Sec.~\ref{sec:bRWA}), experimental feasibility (Sec.~\ref{sec:experiment}), and general measurement frameworks (Sec.~\ref{sec:POVM}).
Section~\ref{sec:summary} concludes.

\section{External-field approximation for quantum light} \label{sec:XFA}
We derive a general effective theory in which a photon field, initially prepared in a quantum state, is treated as an external field.
This treatment, hereafter referred to as the external-field approximation (XFA), is justified when the electron--photon coupling is sufficiently weak so that backaction from the electrons to the light can be neglected.
Within the path-integral formulation, this action--backaction separation between the electronic and photonic subsystems can be implemented iteratively via a Born-series expansion.
This framework yields an approximate expression for the total-system wave function.
The present effective theory extends the density-matrix-based formulation proposed in Ref.~\cite{Imai2025b}.

We begin by outlining the setup.
Multiphoton light can induce ultrafast electronic dynamics.
Here, we focus on such coherent dynamics and analyze the Schr\"{o}dinger equation for the coupled electron--photon system, neglecting dissipation.
The total Hamiltonian is
\begin{equation}
\hat{\mathcal{H}} = \hat{\mathcal{H}}_{\mathrm{e}}[\hat{a}] + \omega \hat{a}^{\dagger} \hat{a}, \label{eq:XFA hamiltonian}
\end{equation}
where $\hat{\mathcal{H}}_{\mathrm{e}}$ includes both the Hamiltonian of the electronic system and the electron--photon interaction.
The photon field is described by the bosonic annihilation operator $\hat{a}$ and modeled as a harmonic oscillator with frequency $\omega$.
For the path-integral representation, we assume that $\hat{\mathcal{H}}_{\mathrm{e}}$ is normal-ordered with respect to the bosonic operators.
The initial state of photons is a nonequilibrium pure state $|\chi(0)\rangle_{\mathrm{p}}$, and our goal is to determine the total-system wave function $|\Psi(t)\rangle_{\mathrm{ep}}$ at time~$t$.
The Dirac constant $\hbar$ and the electric charge $q$ are set to unity.

We express the initial photonic state $|\chi(0)\rangle_{\mathrm{p}}$ in the overcomplete basis of coherent states $\{|\alpha\rangle \mid \hat{a}|\alpha\rangle = \alpha|\alpha\rangle,\ \alpha \in \mathbb{C} \}$ as
\begin{equation}
|\chi(0)\rangle_{\mathrm{p}} = \int \mathrm{d}^2 \alpha\, f(\alpha) |\alpha\rangle. \label{eq:photon state:coherent state rep}
\end{equation}
Due to the overcompleteness of this basis, the expansion coefficient $f(\alpha)$ is not unique~\cite{footnote_Fock_Bargmann}.
We revisit this point in Sec.~\ref{sec:overcomplete} and the discussion around Eq.~\eqref{eq:thermodynamics limit}.

Under this representation, we consider the time evolution of the total-system wave function $|\Psi(t)\rangle_{\mathrm{ep}}$.
Let the initial electronic state be $|\psi(0)\rangle_{\mathrm{e}}$, then
\begin{align}
|\Psi(t)\rangle_{\mathrm{ep}} =&\, \hat{\mathrm{T}}\, \mathrm{e}^{-\mathrm{i} \int_0^t \mathrm{d}\tau \{\hat{\mathcal{H}}_{\mathrm{e}}[\hat{a}]+\omega \hat{n}\}}\, 
|\psi(0)\rangle_{\mathrm{e}}|\chi(0)\rangle_{\mathrm{p}} \nonumber \\
=&\int \mathrm{d}^2 \alpha f(\alpha) \, \hat{\mathrm{T}}\, \mathrm{e}^{-\mathrm{i} \int_0^t \mathrm{d}\tau \{\hat{\mathcal{H}}_{\mathrm{e}}[\hat{a}] + \omega \hat{n}\}}\, |\psi(0)\rangle_{\mathrm{e}}  |\alpha \rangle, \label{eq:total state:time order}
\end{align}
where $\hat{\mathrm{T}}$ denotes time ordering and $\hat{n}=\hat{a}^{\dagger}\hat{a}$ for short.

We now focus on the time-ordered exponential and apply the path-integral formalism only to the photonic degrees of freedom.
This yields
\begin{align}
&\hat{\mathrm{T}}\, \mathrm{e}^{-\mathrm{i} \int_0^t \mathrm{d}\tau \{ \hat{\mathcal{H}}_{\mathrm{e}}[\hat{a}] + \omega \hat{n}\}}\, |\psi(0)\rangle_{\mathrm{e}} |\alpha \rangle \nonumber \\
=& \int_{\alpha_{\mathrm{traj}}(0)=\alpha} \mathcal{D}\alpha_{\mathrm{traj}}
\, \mathrm{e}^{\mathrm{i} \hat{S}(t)} |\psi(0)\rangle_{\mathrm{e}} |\alpha_{\mathrm{traj}}(t) \rangle, \label{eq:path integral}
\end{align}
with the action
\begin{equation}
\hat{S}(t) = \int_0^t \mathrm{d}\tau \left\{ \overline{\alpha_{\mathrm{traj}}}(\tau) \mathrm{i} \partial_\tau \alpha_{\mathrm{traj}}(\tau) - \omega |\alpha_{\mathrm{traj}}|^2 - \hat{\mathcal{H}}_{\mathrm{e}}[\alpha_{\mathrm{traj}}] \right\}. \label{eq:action}
\end{equation}
Here $\alpha_{\mathrm{traj}}(\tau)$ represents a classical trajectory starting from the initial coherent-state center $\alpha$.

We derive the approximate state vector $|\Psi_{\mathrm{XFA}}(t)\rangle_{\mathrm{ep}}$ under the condition that the electron--photon coupling is sufficiently weak, such that electronic backaction on the photon field can be neglected.
First, we consider the free evolution of the classical trajectory of the photon field $\alpha_{\mathrm{traj}}(\tau)$ under the action $\hat{S}(t)$ [Eq.~\eqref{eq:action}], corresponding to the zeroth-order Born approximation.
The equation of motion formally reads
\begin{equation}
\mathrm{i} \partial_{\tau} \alpha_{\mathrm{traj}}(\tau) - \omega \alpha_{\mathrm{traj}}(\tau) = \frac{\updelta \hat{\mathcal{H}}_{\mathrm{e}}}{\updelta \alpha_{\mathrm{traj}}} \approx 0, \label{eq:alpha_q:EoM}
\end{equation}
with the initial condition $\alpha_{\mathrm{traj}}(0) = \alpha$.
This yields the oscillatory solution
\begin{equation}
\alpha_{\mathrm{traj}}(\tau) \approx \alpha(\tau) \equiv \alpha \, \mathrm{e}^{-\mathrm{i}\omega \tau}. \label{eq:alpha(t)}
\end{equation}

We next consider the electronic dynamics in the freely-evolving photon field $\alpha(\tau)$.
Substituting Eq.~\eqref{eq:alpha(t)} into the action~\eqref{eq:action} and the path integral~\eqref{eq:path integral} yields the equation of motion for electrons:
\begin{equation}
\mathrm{i} \partial_t |\psi_{\alpha}(t)\rangle_{\mathrm{e}} = \hat{\mathcal{H}}_{\mathrm{e}}[\alpha(t)] |\psi_{\alpha}(t)\rangle_{\mathrm{e}}, \label{eq:electron state with alpha}
\end{equation}
where the operator $\hat{a}$ is replaced by the $c$-number $\alpha(t)$ [Eq.~\eqref{eq:alpha(t)}].
Inserting this solution into Eq.~\eqref{eq:total state:time order} gives the wave function of the total system under the XFA,
\begin{equation}
|\Psi_{\mathrm{XFA}}(t)\rangle_{\mathrm{ep}} = \int \mathrm{d}^2\alpha f(\alpha) |\psi_{\alpha}(t)\rangle_{\mathrm{e}} |\alpha(t)\rangle_{\mathrm{p}}, \label{eq:total state:XFA}
\end{equation}
which constitutes the central theoretical result of this paper.
Since breaking the action--backaction balance in the electron--photon system leads to a violation of exact norm conservation, we renormalize the XFA state at each time step.

The XFA state~\eqref{eq:total state:XFA} is valid when the energy transferred from the electrons to the photons (i.e., backaction) is negligible compared with the intrinsic energy scale of the photon field.
A practical criterion for this condition is that the initial photon number $\langle \hat{n} \rangle$ exceeds the number of excited electrons $N_{\mathrm{exc}}$,
\begin{equation}
\langle \hat{n} \rangle \gg N_{\mathrm{exc}}. \label{eq:photon num >> excited electron num}
\end{equation}
From the perspective of the Born-series expansion, the corresponding time window for the validity of the XFA can be expressed as $\gamma t \ll 1$, where $\gamma$ is the electron--photon coupling constant determined by the optical mode function and the electric dipole moment.

The XFA formalism~\eqref{eq:total state:XFA} reduces the standard semiclassical theory for laser-driven matter when the photonic state is a coherent state $|\alpha_0\rangle$, corresponding to $f(\alpha) = \delta^2(\alpha - \alpha_0)$ with $\delta^2(z)$ being the two-dimensional delta function.
In this case, Maxwell equations are not solved simultaneously; instead, externally prescribed fields (e.g., via the Peierls phase) drive the matter Schr\"{o}dinger equation as Eq.~\eqref{eq:electron state with alpha}.

Owing to the overcompleteness of the coherent‑state basis, the XFA total-system wave function~\eqref{eq:total state:XFA} depends on the representation of the initial photonic state in Eq.~\eqref{eq:photon state:coherent state rep}.
This dependence, however, is expected to vanish in the thermodynamic limit:
\begin{equation}
\gamma \sqrt{\langle \hat{n} \rangle}=\mathrm{const},\ \ \gamma\to 0. \label{eq:thermodynamics limit}
\end{equation}
where the intensive quantity $\gamma \sqrt{\langle \hat{n} \rangle}$, namely the field strength, is kept constant while $\gamma \to 0$ and $\langle \hat{n} \rangle \to \infty$.
Since an electronic subsystem couples only to the intensive field $\gamma\alpha$, it becomes insensitive to microscopic fluctuations $\delta \alpha \sim \mathcal{O}(1)$ [equivalently, $\delta (\gamma \alpha) = \gamma \delta \alpha \sim \mathcal{O}(\gamma)$], which are associated with the nonorthogonality of nearby coherent states; this nonorthogonality underlies the nonuniqueness of the coherent-state expansion.
By contrast, macroscopic quantumness characterized by $\delta\alpha\sim\mathcal{O}(1/\gamma)$---such as macroscopic superpositions or macroscopic squeezing---remains relevant, and its impact on the electronic dynamics is captured robustly by the XFA state.
Section~\ref{sec:overcomplete} presents numerical calculations demonstrating this convergence.

Building on the above general framework, we now turn to a concrete example: the irradiation with the even cat-state light:
\begin{equation}
|\chi(0)\rangle_{\mathrm{p}} = \frac{1}{\sqrt{\mathcal{N}}} \left( |\alpha_0\rangle + |{-}\alpha_0\rangle \right), \label{eq:even cat state}
\end{equation}
where $\mathcal{N}$ is a normalization constant, here given by $\mathcal{N}=2[1+\exp(-2|\alpha_0|^2)]$.
A natural choice for $f(\alpha)$ in Eq.~\eqref{eq:photon state:coherent state rep} is
\begin{equation}
f_{\mathrm{ecat}}(\alpha) = \frac{1}{\sqrt{\mathcal{N}}} [\delta^2(\alpha- \alpha_0) + \delta^2(\alpha+\alpha_0)]. \label{eq:even cat state:coherent state rep}
\end{equation}
Substituting this into Eq.~\eqref{eq:total state:XFA} gives
\begin{align}
&|\Psi_{\mathrm{XFA}}(t)\rangle_{\mathrm{ep}} \nonumber \\
=&  \frac{1}{\sqrt{\mathcal{N}}}\ \bigl( |\psi_{\alpha_0}(t)\rangle_{\mathrm{e}} |\alpha_0(t)\rangle 
+ |\psi_{-\alpha_0}(t)\rangle_{\mathrm{e}} |{-}\alpha_0(t)\rangle \bigr), \label{eq:derive:superposition total state}
\end{align}
with $\mathcal{N} = 2( 1 + \mathrm{Re}[\langle \psi_{\alpha_0}|\psi_{-\alpha_0} \rangle \langle \alpha_0|{-}\alpha_0 \rangle ])$, which depends on the electronic state $|\psi_{\pm\alpha_0}\rangle_{\mathrm{e}}$.
Thus, the total system is approximately a quantum superposition of product states, each formed by a classically driven electronic state $|\psi_{\pm\alpha_0}(t)\rangle_{\mathrm{e}}$ and its corresponding classical-light component $|{\pm} \alpha_0\rangle$.

Several remarks are in order.
From a quantum-information perspective, Eq.~\eqref{eq:derive:superposition total state} looks familiar with cat‑state decompositions often encountered in Schr\"{o}dinger-cat physics~\cite{Zurek2003}.
However, the XFA total-system wave function is not unique owing to the overcompleteness of the coherent-state basis, which allows for alternative but inequivalent representations.
This dependence disappears only in the thermodynamic limit discussed above (see Sec.~\ref{sec:overcomplete}).
Finally, applying the XFA at the state-vector level ensures the positivity of the resulting density matrix.
This stands in contrast with certain density-matrix formalisms, such as those using the Sudarshan--Glauber $P$-function~\cite{Imai2025b} or the generalized $P$-function~\cite{Gorlach2022b}.

\section{Model and Method} \label{sec:model and method}
\subsection{Multiple two-level systems coupled to a single bosonic mode} \label{sec:model}
We first introduce the $N$-qubit Rabi (Dicke) model~\cite{Rabi1937, Dicke1954}, given by
\begin{align}
&\hat{\mathcal{H}}_{\mathrm{RD}} = \hat{\mathcal{H}}_{\mathrm{e,RD}} + \omega \hat{a}^{\dagger}\hat{a}, \label{eq:Rabi and Dicke hamiltonian} \\
&\hat{\mathcal{H}}_{\mathrm{e,RD}} = \sum_{j=1}^{N} \varDelta \hat{S}^z_j-\hat{E}\cdot \hat{P}, \label{eq:Rabi and Dicke coupling hamiltonian} \\
&\hat{E} = \mathrm{i}\gamma\omega (\hat{a}-\hat{a}^{\dagger}),\ \ \hat{P}=\sum_{j} \mu_{\mathrm{e}} \hat{S}^x_j. \label{eq:electric field and polarization}
\end{align}
Here, the first term in $\hat{\mathcal{H}}_{\mathrm{e,RD}}$ [Eq.~\eqref{eq:Rabi and Dicke coupling hamiltonian}] represents the unperturbed Hamiltonian of the electronic system, and the second term corresponds to the electron--photon interaction as the coupling between the electric field $\hat{E}$ and the polarization $\hat{P}$.
The spin-$1/2$ operator $\hat{\bm{S}}_j$ represents the $j$th pair of dipole-coupled electronic states in matter.
Examples include a two-level system formed between $s$ and $p$ orbitals in a neutral atomic gas or an excitonic state in a semiconductor.
The parameter $\varDelta$ is the level spacing, and $\mu_{\mathrm{e}}$ is the dipole moment.
The photon field is modeled as a single-mode harmonic oscillator with frequency $\omega$, described by the bosonic operator $\hat{a}$.
The parameter $\gamma$ characterizes the field strength, determined by the spatial mode of the light.

In the following, we set $\varDelta = 1$ as the unit of energy, assume the resonance condition $\omega = \varDelta$, and absorb the dipole moment into the coupling constant by setting $\mu_{\mathrm{e}} = 1$, so that $\gamma$ becomes the dimensionless coupling strength.

For simplicity, we adopt the rotating-wave approximation (RWA), which removes the counter-rotating terms $\hat{a} \hat{S}^{-}_j$ and $\hat{a}^{\dagger} \hat{S}^{+}_j$ oscillating at frequency $2\omega$.
The resulting Hamiltonian is the Tavis--Cummings model~\cite{Tavis1968},
\begin{align}
&\hat{\mathcal{H}}_{\mathrm{TC}} = \hat{\mathcal{H}}_{\mathrm{e,TC}} + \omega \hat{a}^{\dagger} \hat{a}, \label{eq:Tavis--Cummings model Hamiltonian} \\
&\hat{\mathcal{H}}_{\mathrm{e,TC}} = \sum_{j=1}^{N} \varDelta \hat{S}^z_j -\mathrm{i}\gamma \frac{1}{2}\sum_j  \left( \hat{a}\hat{S}^{+}_j - \hat{a}^{\dagger} \hat{S}^{-}_j \right). \label{eq:RWA}
\end{align}
The validity of this approximation for the present work will be examined in Sec.~\ref{sec:bRWA}.

We describe the initial states.
The electronic subsystem is taken to be in the all-spin-down ground state $|{\downarrow\downarrow\cdots\downarrow}\rangle$ of the $\gamma = 0$ Hamiltonian, where $|{\uparrow}\rangle$ and $|{\downarrow}\rangle$ are eigenstates of $\hat{S}_j^z$.
The photonic subsystem is prepared in a general cat state of two coherent states $|\alpha\rangle$ and $|\beta\rangle$:
\begin{equation}
|\mathrm{gcat}\rangle_{\mathrm{p}} = \frac{1}{\sqrt{\mathcal{N}}} \left( |\alpha\rangle + \mathrm{e}^{\mathrm{i} \phi_{\mathrm{cat}} }|\beta\rangle \right). \label{eq:optical cat state}
\end{equation}
Here, $\phi_{\mathrm{cat}}$ is the relative phase between the two coherent states.

The case $\alpha = \alpha_0$ and $\beta = -\alpha_0$ with $\phi_{\mathrm{cat}} = 0$ defines the even cat state in Eq.~\eqref{eq:even cat state}.
This state satisfies $\langle \hat{E}\rangle = 0$, and thus does not induce the electric current or polarization typically observed in optical responses.
Sections~\ref{sec:trace_out} and~\ref{sec:parity} use this even cat-state light to drive the electronic system.

The choice $\alpha = \alpha_0$ and $\beta = 0$ defines the kitten state:
\begin{equation}
|\mathrm{kitten}\rangle_{\mathrm{p}} = \frac{1}{\sqrt{\mathcal{N}}} \left(|\alpha_0\rangle + |0\rangle \right). \label{eq:kitten}
\end{equation}
This state has been experimentally realized in high-harmonic generation~\cite{Lewenstein2021}.
Section~\ref{sec:quadrature} analyzes the electronic dynamics induced by this kitten state of light.

We numerically compute the total-system wave function $|\Psi(t)\rangle_{\mathrm{ep}}$ by expanding the time-evolution operator $\exp(-\mathrm{i}\hat{\mathcal{H}} \updelta t)$ to fourth order in the time step $\updelta t$: $|\Psi(t+\updelta t)\rangle_{\mathrm{ep}} = \exp(-\mathrm{i}\hat{\mathcal{H}} \updelta t)|\Psi(t)\rangle_{\mathrm{ep}} = \sum_{n=0}^{4} (-\mathrm{i}\hat{\mathcal{H}} \updelta t)^n/n!\,|\Psi(t)\rangle_{\mathrm{ep}} + \mathcal{O}(\updelta t^5)$.
The wave function is renormalized at each time step.
We set $\updelta t = 0.001$; for the large-amplitude case $\alpha_0 = 30$, we use $\updelta t = 0.0001$.
The photon Hilbert space is truncated to a maximum photon number for which convergence has been verified: $200$ for $N = 8$, $500$ for $N = 32$, and $1200$ for $\alpha_0 = 30$.
Since the total spin $J = N/2$ is conserved in our models [Eqs.~\eqref{eq:Rabi and Dicke hamiltonian} and~\eqref{eq:Tavis--Cummings model Hamiltonian}], the electronic part of $|\Psi(t)\rangle_{\mathrm{ep}}$ can be expressed using only the $2J+1$ Dicke states $\{ |J,m\rangle \mid m=-J,\ldots,J\}$.
These states are eigenstates of the collective spin operator $\hat{\bm{J}} = \sum_{i} \hat{\bm{S}}_i$, satisfying $\hat{\bm{J}}^2|J,m\rangle = J(J+1)|J,m\rangle$ and $\hat{J}^z|J,m\rangle = m|J,m\rangle$.

\subsection{Photon-state measurement} \label{sec:method:measurement}
We define the projection operators $\mathcal{P}^{\mathrm{M}}_{\mu}$ for the photon-state measurements considered in this study.
Here, $\mathrm{M}$ denotes the type of measurement and $\mu$ its outcome.
The condition for an optimal projector can be formulated as follows.
When the general cat state~\eqref{eq:optical cat state} with large amplitude is irradiated, the total-system wave function in the XFA~\eqref{eq:total state:XFA} reads
\begin{equation}
|\Psi_{\mathrm{XFA}}\rangle_{\mathrm{ep}} \approx |\psi_{\alpha}\rangle |\alpha\rangle +\mathrm{e}^{\mathrm{i}\phi_{\mathrm{cat}}}|\psi_{\beta}\rangle |\beta\rangle. \label{eq:total state:XFA:general cat}
\end{equation}
To project the electronic system onto a cat state, i.e., $|\psi_{\alpha}\rangle + \exp(\mathrm{i}\phi_{\mathrm{cat}})|\psi_{\beta}\rangle$, the optimal photon-state projector $\mathcal{P}^{\mathrm{M}}_{\mu}$ should satisfy
\begin{equation}
\langle \alpha | \mathcal{P}^{\mathrm{M}}_{\mu} |\alpha\rangle = \langle \beta|\mathcal{P}^{\mathrm{M}}_{\mu}|\beta\rangle. \label{eq:optimal projector condition}
\end{equation}

For a superposition of coherent states with equal amplitude [e.g., the even cat state~\eqref{eq:even cat state}], photon-number parity is effective.
The even- and odd-parity projectors, $\mathcal{P}_{\pm}^{\mathrm{P}}$, are defined via the photon-number parity operator $\hat{\Pi}$ as
\begin{equation}
\hat{\Pi} = \mathrm{e}^{\mathrm{i}\pi \hat{n}} = \mathcal{P}_{+}^{\mathrm{P}} - \mathcal{P}_{-}^{\mathrm{P}}, \label{eq:parity operator}
\end{equation}
with
\begin{align}
&\mathcal{P}_{+}^{\mathrm{P}}=\sum_{m =0,1,\ldots} |2m\rangle \langle 2m|, \label{eq:even projector} \\
&\mathcal{P}_{-}^{\mathrm{P}}=\sum_{m =0,1,\ldots} |2m{+}1\rangle \langle 2m{+}1|. \label{eq:odd projector}
\end{align}
Section~\ref{sec:parity} presents an analysis using these parity projectors.

We also consider quadrature projectors, $\mathcal{P}^{\mathrm{Q;\varphi
}}_{x}$, which remain effective even for the kitten state~\eqref{eq:kitten}.
Experimentally, such measurements are implemented via balanced homodyne detection~\cite{Lvovsky2009, DAriano2003, Tyc2004}.
The quadrature operator is defined as
\begin{equation}
\hat{x}_{\varphi} = \left[ \mathrm{e}^{-\mathrm{i}\varphi} \hat{a} + \mathrm{e}^{\mathrm{i}\varphi} \hat{a}^{\dagger} \right]/\sqrt{2}. \label{eq:def:quadrature operator}
\end{equation}
The eigenstates $|x;\varphi\rangle$ with eigenvalue $x \in \mathbb{R}$ are expressed as
\begin{equation}
|x;\varphi\rangle = \sum_{n=0}^{\infty} \frac{\mathrm{e}^{-\mathrm{i}n\varphi}}{\pi^{1/4}\sqrt{2^n n!}}\, H_n(x)\, \mathrm{e}^{-x^2/2}\, |n\rangle, \label{eq:quadrature eigenstate}
\end{equation}
where $H_n(x)$ denotes the $n$th Hermite polynomial.
The ideal projector is
\begin{equation}
\mathcal{P}_{x}^{\mathrm{Q};\varphi} = |x;\varphi\rangle \langle x;\varphi|, \label{eq:ideal quadrature projector}
\end{equation}
as discussed in Refs.~\cite{DAriano2003, Tyc2004}.
A finite-resolution measurement of width $\Delta x$ centered at $x$ is represented by
\begin{equation}
\mathcal{P}_{x}^{\mathrm{Q};\varphi;\Delta x} =  \int_{x-\Delta x}^{x+\Delta x} \mathrm{d}x'\, |x';\varphi\rangle \langle x';\varphi|. \label{eq:finite quadrature projector}
\end{equation}
We employ such quadrature projectors in Sec.~\ref{sec:quadrature}.

Using the photon-state projectors $\mathcal{P}^{\mathrm{M}}_{\mu}$ defined above, the postselected electronic density matrix at time $t$ is given by
\begin{equation}
\hat{\rho}_{\mathrm{e}}^{(\mathrm{M},\mu)}(t) = 
\frac{\mathrm{Tr}_{\mathrm{p}} \left[ \mathcal{P}^{\mathrm{M}}_{\mu}|\Psi(t)\rangle_{\mathrm{ep}}\, {}_{\mathrm{ep}}\langle\Psi(t)|\mathcal{P}^{\mathrm{M}}_{\mu}\right] }{ {}_{\mathrm{ep}}\langle \Psi(t)| \mathcal{P}^{\mathrm{M}}_{\mu}|\Psi(t)\rangle_{\mathrm{ep}} }.  \label{eq:postselected electron state:general}
\end{equation}

\subsection{Quantum Fisher information} \label{sec:quantum Fisher information}
As a measure of quantumness in many-body electronic systems, we adopt the quantum Fisher information (QFI)~\cite{Helstrom1969, Holevo2011, Braunstein1994}.
It is defined as
\begin{equation}
\mathcal{F}_{\mathrm{Q}} = \underset{\hat{A}}{\mathrm{max}}\, \sum_{i,j} 2\frac{(\lambda_i - \lambda_j)^2}{\lambda_i + \lambda_j} |\langle \lambda_i|\hat{A}|\lambda_j \rangle|^2, \label{eq:def quantum Fisher information}
\end{equation}
where $\hat{A}$ is an arbitrary additive observable, and $\lambda_i$ ($|\lambda_i\rangle$) denotes the $i$th eigenvalue (eigenvector) of a given density matrix $\hat{\rho}$.
Although originally developed for parameter estimation and quantum metrology~\cite{Pezze2018}, the QFI has also proven useful for quantifying entanglement depth~\cite{Toth2012, Hyllus2012, Gessner2019, Ren2021c} and macroscopic quantum superpositions~\cite{Frowis2012}.

Entanglement depth~\cite{Toth2012, Hyllus2012, Gessner2019, Ren2021c} quantifies how many particles are genuinely entangled in a given quantum state.
An $N$-particle quantum state $\hat{\rho}$ is $k$-producible if it can be written as $\hat{\rho} = \bigotimes_{i} \hat{\rho}_i$, where each $\hat{\rho}_i$ acts on at most $k$ particles.
If a state is $k$-producible but not $(k{-}1)$-producible, the entanglement depth is $k$.
For $k=1$, the state is separable; for $k=N$, it is genuinely multipartite entangled.
Any $k$-producible state satisfies the inequality~\cite{Toth2012, Hyllus2012}:
\begin{equation}
\mathcal{F}_{\mathrm{Q}} \leq kN. \label{eq:QFI entanglement depth}
\end{equation}
Thus, if $\mathcal{F}_{\mathrm{Q}}/N > k$, the state is not $k$-producible, implying that the entanglement depth is at least $k + 1$, and that there exists a subset of at least $k + 1$ entangled particles.

Macroscopic quantum superposition is a key concept for distinguishing genuine macroscopic quantum coherence (e.g., cat states) from the collective accumulation of microscopic quantum effects (e.g., superconductivity)~\cite{Frowis2018}.
It can be defined as a quantum superposition of states that yield macroscopically distinct values for an additive observable $\hat{A}$.
This property can be quantified by the QFI.
For a pure state, the QFI reads
\begin{equation}
\mathcal{F}_{\mathrm{Q}} = \underset{\hat{A}}{\mathrm{max}} \left[4 \Delta\hat{A} \right], \label{eq:QFI variation}
\end{equation}
where $\Delta\hat{A} \equiv \langle  \hat{A}^2 \rangle - \langle  \hat{A} \rangle^2$ is the variance of $\hat{A}$.
Macroscopic superposition states satisfy $\mathcal{F}_{\mathrm{Q}} = \mathcal{O}(N^2)$~\cite{Frowis2012}.
Thus, when the entanglement depth scales as $k = \mathcal{O}(N)$, the state exhibits both genuine multipartite entanglement and macroscopic superposition, both of which are indicated by an extensive QFI density.

The QFI requires maximization over all additive observables $\hat{A}$.
Since the total spin $\hat{\bm{J}}$ is conserved in our models [Eqs.~\eqref{eq:Rabi and Dicke hamiltonian} and~\eqref{eq:Tavis--Cummings model Hamiltonian}], $\hat{A}$ can be restricted to operators of the form
\begin{equation}
\hat{A} = \bm{n} \cdot \hat{\bm{J}}, \quad \bm{n} = (\sin\theta\cos\phi,\, \sin\theta\sin\phi,\, \cos\theta), \label{eq:additive operator}
\end{equation}
where $\bm{n}$ is a unit vector specifying the measurement direction.
The maximization over $\theta \in [0,\pi)$ and $\phi \in [0, 2\pi)$ then reduces to finding the largest eigenvalue of the $3 \times 3$ matrix $\mathcal{F}_{\mathrm{Q}}^{a,b} = \sum_{i,j} 2(\lambda_i -\lambda_j)^2/(\lambda_i+\lambda_j)\cdot \langle\lambda_i| \hat{J}^a |\lambda_j\rangle \langle \lambda_j| \hat{J}^b |\lambda_i\rangle$ with $a,b \in \{x,y,z\}$.

\subsection{Spin Wigner function} \label{sec:spin Wigner function}
To visualize quantum states of spin systems, we use the spin Wigner function~\cite{Stratonovich1956, Brif1999}.
Given a density matrix $\hat{\rho}$ and a kernel operator $\hat{\Delta}(\Omega)$ defined over the phase space $\Omega$, the Wigner function is defined as
\begin{equation}
W(\Omega)=\mathrm{Tr}[\hat{\rho} \, \hat{\Delta}(\Omega)]. \label{eq:Wigner function}
\end{equation}
The definition is not unique~\cite{Rundle2017}; here, we adopt the form based on the Dicke states $\{ |J,m\rangle\}$, which was analyzed in Ref.~\cite{Davis2021}:
\begin{align}
&\hat{\Delta}(\theta,\phi)= \sum_{m=-J}^{J} \Delta_{J,m} \hat{R}(\theta,\phi)|J,m\rangle \langle J,m|\hat{R}^{\dagger}(\theta,\phi), \label{eq:Wigner func:Delta(Omega)} \\
&\Delta_{J,m} = \sum_{j=0}^{2J} \frac{2j+1}{2J+1} C^{Jm}_{Jm;j0}, \label{eq:Wigner func:Delta_Jm}
\end{align}
with the rotation operator $\hat{R}(\theta,\phi) = \mathrm{e}^{\mathrm{i}\phi \hat{J}^z} \mathrm{e}^{\mathrm{i}\theta \hat{J}^y} = \mathrm{e}^{-\mathrm{i}\theta \bm{l} \cdot \hat{\bm{J}}}$ about the axis $\bm{l} = (-\sin\phi,\cos\phi,0)$.
The coefficients $C^{JM}_{j_1,m_1;j_2,m_2} = \langle j_1,m_1;j_2,m_2|J,M\rangle$ are Clebsch--Gordan coefficients, and the phase-space measure is $\mathrm{d}\Omega = \frac{2J+1}{4\pi} \sin\theta \mathrm{d}\theta \mathrm{d}\phi$.

\section{Electronic dynamics driven by cat-state light without postselection} \label{sec:trace_out}
In this section, we investigate many-body electronic dynamics driven by cat-state light after tracing out the photonic degrees of freedom.
Using the Tavis--Cummings Hamiltonian $\mathcal{H}_{\mathrm{TC}}$ [Eq.~\eqref{eq:Tavis--Cummings model Hamiltonian}], we numerically compute the time-evolved total-system wave function $|\Psi(t)\rangle_{\mathrm{ep}}$, starting from the even cat state of light~\eqref{eq:even cat state} and the all-spin-down electronic state.
We then obtain the reduced electronic density matrix
\begin{equation}
\hat{\rho}_{\mathrm{e}}(t) = \mathrm{Tr}_{\mathrm{p}}\left[|\Psi(t)\rangle_{\mathrm{ep}}\, {}_{\mathrm{ep}} \langle\Psi(t)| \right], \label{eq:trace out:electron density matrix}
\end{equation}
and evaluate the QFI $\mathcal{F}_{\mathrm{Q}}$ [Eq.~\eqref{eq:def quantum Fisher information}] for the electronic system.
The electron--photon coupling strength is set to $\gamma = 0.01$.
Under the resonance condition, the Tavis--Cummings model can be transformed into a rotating frame where $\gamma$ is the only characteristic energy scale.
Consequently, the real-time dynamics scale with $\gamma t$; thus, considering realistic material systems with small $\gamma$ amounts to focusing on shorter timescales.

\begin{figure*}[t]
\centering
\includegraphics[width=2.0\columnwidth]{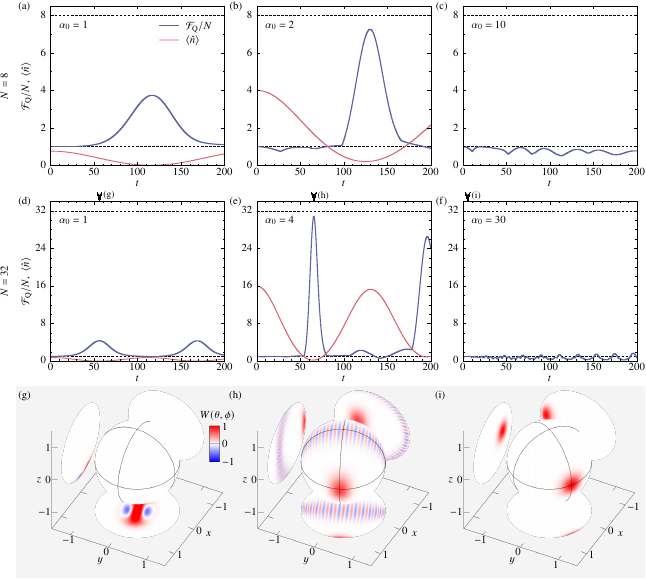}
\caption{
Many-body electronic dynamics driven by the even cat-state light without postselection.
(a)--(f) Time evolution of the QFI density $\mathcal{F}_{\mathrm{Q}}/N$ (blue curves) calculated from $\hat{\rho}_{\mathrm{e}}(t)$ [Eq.~\eqref{eq:trace out:electron density matrix}].
Red curves show the mean photon number $\langle \hat{n} \rangle$.
(a)--(c) $N = 8$; (d)--(f) $N = 32$.
(a),(d) $\alpha_0 = 1$.
(b),(e) $N/2 = \langle \hat{n} \rangle$ ($\alpha_0 = 2$ and $\alpha_0 = 4$, respectively).
(c) $\alpha_0 = 10$; (f) $\alpha_0 = 30$.
(g)--(i) Spin Wigner functions $W(\theta,\phi)$ for $N = 32$, evaluated at the times marked by arrowheads in the corresponding panels.
}
\label{fig:trace_out}
\end{figure*}

Figures~\ref{fig:trace_out}(a)--(f) show the time evolution of the QFI density $\mathcal{F}_{\mathrm{Q}}/N$ for various combinations of electron number $N$ and cat-state amplitude $\alpha_0$.
Figures~\ref{fig:trace_out}(g)--(i) display the corresponding spin Wigner functions $W(\theta, \phi)$ [Eq.~\eqref{eq:Wigner function}], taken at the times marked as arrowheads in the upper panels ($N = 32$).
We classify the dynamics into three regimes according to the relation between the mean photon number $\langle \hat{n} \rangle \approx |\alpha_0|^2$ and the electron number $N$: a small-amplitude regime ($\langle \hat{n} \rangle < N$), discussed in Sec.~\ref{sec:trace out:small} with $\alpha_0 = 1$; an intermediate regime, defined by $|\alpha_0|^2 = N/2$ and treated in Sec.~\ref{sec:trace out:intermediate}; and a large-amplitude regime ($\langle \hat{n} \rangle \gg N$), which is discussed in Sec.~\ref{sec:trace out:large}.

\subsection{Small-amplitude regime} \label{sec:trace out:small}
Figures~\ref{fig:trace_out}(a) and (d) show the QFI for $\alpha_0 = 1$ and $N = 8$ or $32$, respectively.
The QFI density $\mathcal{F}_{\mathrm{Q}}/N$ (blue) exhibits a peak value of $\mathcal{F}_{\mathrm{Q}}/N \gtrsim 4$ in both cases.
This indicates that at least five electrons are mutually entangled ($k \approx 5$).
However, since $\mathcal{F}_{\mathrm{Q}}/N = \mathcal{O}(N^0)$, this QFI scaling does not certify macroscopic entanglement.

Figure~\ref{fig:trace_out}(g) displays the spin Wigner function at the QFI peak time.
In contrast with the simple Gaussian at the south pole expected for the all-spin-down initial state, clear negative regions emerge.
The stretched distribution suggests a quantum superposition of two slightly different spin-polarized states.

To understand the mechanism of this quantum state generation, we examine the dynamics of the photon number $\langle \hat{n} \rangle$.
The red curves in Figs.~\ref{fig:trace_out}(a) and (d) show that the photon number nearly vanishes at the QFI peak time.
This indicates that the initial energy in the photon field has been completely transferred to the electronic system, resulting in perfect absorption.
Such perfect energy absorption is accompanied by the so-called quantum state transfer, previously reported for cat states~\cite{Ling1997, Rundle2021} and for squeezed-vacuum states~\cite{Kuzmich1997, Hald1999, Hald2001}.

This process can be summarized as
\begin{align}
&|\Psi(0)\rangle_{\mathrm{ep}} = |{\downarrow\downarrow\cdots\downarrow}\rangle \otimes (|\alpha_0\rangle + |{-}\alpha_0\rangle) \nonumber \\
&\to |\Psi(t)\rangle_{\mathrm{ep}} \approx (|\psi_{\alpha_0}\rangle_{\mathrm{e}} + |\psi_{-\alpha_0}\rangle_{\mathrm{e}}) \otimes |0\rangle, \label{eq:quantum state transfer}
\end{align}
up to normalization.
Here, $|\psi_{\pm \alpha_0}\rangle$ denote the electronic states driven by $|{\pm} \alpha_0\rangle$, respectively.
Even if each $|\psi_{\pm \alpha_0}\rangle$ has a small QFI density (e.g., $\lesssim 1.1$ for $N = 32$), their quantum superposition results in a finite total QFI.
The electronic states driven by out-of-phase classical light exhibit the out-of-phase spin rotations in the $xy$ plane, which is consistent with the features seen in the spin Wigner function.

\subsection{Intermediate-amplitude regime} \label{sec:trace out:intermediate}
We next ask to what extent large entanglement can be generated via the perfect energy-absorption mechanism~\eqref{eq:quantum state transfer}.
The optimal cat-state amplitude can be estimated from the energy balance for an electronic system driven by $|{\pm}\alpha_0\rangle$.
The opposite-phase coherent-state light induces opposite-phase spin precessions, which are maximally separated when the spins lie in the $xy$ plane, i.e., $\sum_i \langle \hat{S}^z_i \rangle = 0$.
Neglecting the electron--photon interaction energy ($\gamma=0$), the conserved total energy satisfies $\sum_i \langle \hat{S}^z_i \rangle + \langle \hat{n} \rangle = \mathrm{const.}$
Thus, the largest QFI is expected when $-N/2 + \langle \hat{n} \rangle = 0 + 0$, which yields $|\alpha_0| = \sqrt{N/2}$.

Figures~\ref{fig:trace_out}(b) and (e) present the time evolution of the QFI density $\mathcal{F}_{\mathrm{Q}}/N$ for $N = 8$ ($\alpha_0 = 2$) and $N = 32$ ($\alpha_0 = 4$), respectively.
In both cases, the QFI density approaches its theoretical maximum value $N$ at the peak.
This indicates that when $\langle \hat{n}\rangle = N/2$, the even cat-state light can induce nearly genuine multipartite entanglement in the electronic system.
Figure~\ref{fig:trace_out}(h) displays the spin Wigner function at the QFI peak time.
Two Gaussian peaks appear on the $xy$ plane, accompanied by fine but distinct interference fringes along a meridian.
These features are consistent with a superposition of oppositely polarized spin states in the $xy$ plane, $|{\leftarrow\leftarrow \cdots \leftarrow}\rangle + |{\rightarrow\rightarrow\cdots\rightarrow}\rangle$, where $|{\leftarrow}\rangle$ and $|{\rightarrow}\rangle$ denote spin eigenstates in a direction lying in the $xy$ plane.
At the same time, the photon number $\langle \hat{n} \rangle$ also approaches zero, confirming that the electronic system realizes a cat-like state---approximately the GHZ state---via the perfect energy absorption.

Unlike previous works (e.g., Refs.~\cite{Ling1997, Rundle2021}), our results in Figs.~\ref{fig:trace_out}(b) and (e) track the real-time development of entanglement in the electronic system driven by the cat-state light.
The QFI dynamics exhibit features known as the sudden birth and death of entanglement~\cite{Yu2004, Ficek2008, Yu2009}.
These features appear as kinks around $t = 100.0$ and $160.0$ for $N = 8$ [Fig.~\ref{fig:trace_out}(b)], and around $t = 55.0$ and $80.0$ for $N = 32$ [Fig.~\ref{fig:trace_out}(e)].

\begin{figure}[t]
\centering
\includegraphics[width=\columnwidth]{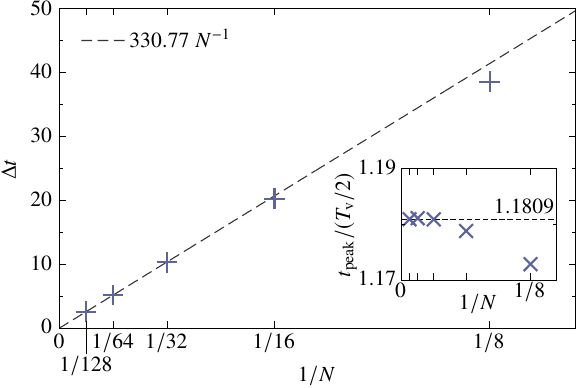}
\caption{
FWHM $\Delta t$ of the QFI-density peak as a function of the electron number $N$.
Dash line is a guide to the eye.
Even cat-state amplitude is $\alpha_0 = \sqrt{N/2}$.
Inset: $N$ dependence of the QFI peak time $t_{\mathrm{peak}}$.
}
\label{fig:fwhm}
\end{figure}

We discuss the $N$ dependence of the quantum-state transfer dynamics.
Figure~\ref{fig:fwhm} shows the full width at half maximum (FWHM) $\Delta t$ of the QFI peak as a function of $N$.
The observed FWHM $\Delta t$ decreases as $N^{-1}$ for large $N$.
As seen in the inset of Fig.~\ref{fig:fwhm}, the QFI peak time $t_{\mathrm{peak}}$ scales as $t_{\mathrm{peak}} \propto T_{\mathrm{v}} = 2\pi/(\gamma \sqrt{N})$ for large $N$, where $T_{\mathrm{v}}$ is the period of the vacuum Rabi oscillation, reflecting the collective coupling $\gamma \to \gamma\sqrt{N}$.
These $N$ dependencies indicate that, in the thermodynamic limit $N \to \infty$ with $\gamma \sqrt{\langle \hat{n} \rangle} = \gamma \sqrt{N/2} \equiv C = \mathrm{const.}$, the QFI develops a delta-function-like spike at time $t \sim \pi/\sqrt{2}C$.
From $\mathcal{F}_{\mathrm{Q}} \propto \Delta \hat{A}^2$ [Eq.~\eqref{eq:QFI variation}], this spike corresponds to the ultrafast growth of macroscopic fluctuations in the electronic system.

\subsection{Large-amplitude regime} \label{sec:trace out:large}
Figures~\ref{fig:trace_out}(c) and (f) show the time evolution of the QFI density $\mathcal{F}_{\mathrm{Q}}/N$ in the large-amplitude regime, where $\langle \hat{n} \rangle \gg N$.
The values stay below unity for most of the evolution; consequently, the QFI does not witness macroscopic entanglement.
Here, the large photon number prevents the photon field from collapsing to the vacuum via energy absorption, unlike in the small- and intermediate-amplitude regimes.

The XFA description (Sec.~\ref{sec:XFA}) accounts for the negligible increase in the QFI under large-amplitude cat-state light irradiation without measurement.
The XFA total-system wave function $|\Psi_{\mathrm{XFA}}\rangle_{\mathrm{ep}} \sim |\psi_{\alpha_0}\rangle_{\mathrm{e}} |\alpha_0\rangle + |\psi_{-\alpha_0}\rangle_{\mathrm{e}} |{-}\alpha_0\rangle$ [Eq.~\eqref{eq:derive:superposition total state}] leads, after tracing out photons, to the reduced electronic density matrix~\eqref{eq:trace out:electron density matrix}
\begin{align}
\hat{\rho}_{\mathrm{e}} 
&\sim 
|\psi_{\alpha_0}\rangle \langle \psi_{\alpha_0}|
+|\psi_{-\alpha_0}\rangle \langle \psi_{-\alpha_0}| \nonumber \\
&+\langle \alpha_0|{-}\alpha_0\rangle (|\psi_{\alpha_0}\rangle \langle \psi_{-\alpha_0}|
+ |\psi_{-\alpha_0}\rangle \langle \psi_{\alpha_0}|). \label{eq:trace_out density matrix}
\end{align}
Here, the overlap $\langle \alpha_0|{-}\alpha_0\rangle = \exp(-2|\alpha_0|^2)$ exponentially suppresses the electronic interference terms $|\psi_{\pm\alpha_0}\rangle \langle \psi_{\mp\alpha_0}|$, leaving a classical mixture of the two classically driven states $|\psi_{\pm\alpha_0}\rangle$ [see Fig.~\ref{fig:trace_out}(i)].
With no electron--electron interactions~\eqref{eq:RWA}, each $|\psi_{\pm\alpha_0}\rangle$ remains unentangled, and tracing out the light does not create entanglement.
Thus, large-amplitude cat-state light alone does not generate electronic entanglement.

Quantum control of condensed-matter systems requires an extensive QFI density in the thermodynamic limit $\gamma \sqrt{\langle \hat{n} \rangle} = \mathrm{const.},\ \gamma \to 0$ [Eq.~\eqref{eq:thermodynamics limit}].
Our results show that, in this regime, cat-state light acts effectively as a classical drive and, by itself, does not produce macroscopic entanglement in matter.
Sections~\ref{sec:parity} and~\ref{sec:quadrature} demonstrate that appropriate projective measurements on the light overcome this limitation.

\section{Electronic dynamics conditioned on photon-number-parity measurements} \label{sec:parity}
In this section, we analyze postselected many-body electronic dynamics driven by the even cat-state light~\eqref{eq:even cat state}, conditioned on a projective measurement of the photon-number parity~\eqref{eq:parity operator}. 
Section~\ref{sec:parity:numerical} presents numerical results using the same setup as Sec.~\ref{sec:trace_out}.
Section~\ref{sec:parity:analytical} compares these results with the XFA state $|\Psi_{\mathrm{XFA}}\rangle_{\mathrm{ep}}$ [Eq.~\eqref{eq:derive:superposition total state}].
Section~\ref{sec:overcomplete} discusses the role of coherent-state overcompleteness within the XFA.

\subsection{Numerical simulation} \label{sec:parity:numerical}
We numerically compute the time-evolved total-system wave function $|\Psi(t)\rangle_{\mathrm{ep}}$ with the Tavis--Cummings Hamiltonian $\mathcal{H}_{\mathrm{TC}}$ [Eq.~\eqref{eq:Tavis--Cummings model Hamiltonian}], starting from the even cat state of light~\eqref{eq:even cat state} and the all-spin-down electronic state.
We construct the postselected electronic density matrices $\hat{\rho}_{\mathrm{e}}^{(\mathrm{P},\pm)}(t)$ via Eq.~\eqref{eq:postselected electron state:general} using the even or odd photon-number-parity projectors $\mathcal{P}_{\pm}^{\mathrm{P}}$ [Eqs.~\eqref{eq:even projector} and~\eqref{eq:odd projector}].
We then evaluate the corresponding QFI $\mathcal{F}_{\mathrm{Q}}$ [Eq.~\eqref{eq:def quantum Fisher information}].
The electron--photon coupling strength is set to $\gamma = 0.01$.

\begin{figure*}[t]
\centering
\includegraphics[width=2.0\columnwidth]{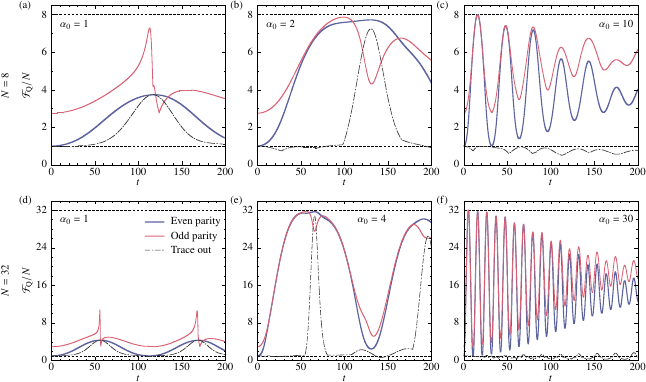}
\caption{
Time evolution of the QFI density $\mathcal{F}_{\mathrm{Q}}/N$ for postselected electronic states $\hat{\rho}_{\mathrm{e}}^{(\mathrm{P},\pm)}(t)$ conditioned on photon-number-parity measurements, under driving by even cat-state light.
Blue: even parity; red: odd parity; dash-dotted: trace-out baseline (same as in Fig.~\ref{fig:trace_out}).
(a)--(c) $N = 8$; (d)--(f) $N = 32$.
(a),(d) $\alpha_0 = 1$.
(b),(e) $N/2 = \langle \hat{n} \rangle$ ($\alpha_0 = 2$ and $\alpha_0 = 4$, respectively).
(c) $\alpha_0 = 10$; (f) $\alpha_0 = 30$.}
\label{fig:parity}
\end{figure*}

Figure~\ref{fig:parity} shows the time evolution of the QFI density $\mathcal{F}_{\mathrm{Q}}/N$ for several combinations of the electron number $N$ and the cat-state amplitude $\alpha_0$.
For even-parity postselection (blue curves), the QFI is enhanced relative to the trace-out cases (dash-dotted curves) across all parameters examined.
The postselection also removes the nonmonotonic, kink-like behavior associated with the sudden birth and death of entanglement, as shown in Figs.~\ref{fig:parity}(b) and (e), enabling a rapid rise in QFI.
Such acceleration is advantageous for realistic systems subject to finite relaxation and decoherence.

In contrast, the odd-parity case (red curves) exhibits some distinctive features.
A nonzero QFI appears already at $t = 0^+$, and for small cat amplitudes [e.g., $\alpha_0 = 1$; Figs.~\ref{fig:parity}(a) and (d)] it can even exceed that of the even-parity outcome.
This behavior originates from postselection onto the single-excitation Dicke state $|J,-J+1\rangle$, whose QFI is $\mathcal{F}_\mathrm{Q}/N=3-2/N$, consistent with the initial nonzero QFI observed in Fig.~\ref{fig:parity}.
This mechanism is understood from first-order perturbation theory: $|\Psi(t)\rangle \approx |\Psi(0)\rangle - \mathrm{i} \hat{\mathcal{H}}_{\mathrm{TC}} t |\Psi(0)\rangle = |{\downarrow\downarrow\cdots\downarrow}\rangle |\mathrm{ecat}\rangle -(\gamma t /2)\sqrt{N} |J,-J+1\rangle \hat{a}|\mathrm{ecat}\rangle$, where $\hat{a}|\mathrm{ecat}\rangle$ has support only on odd-photon-number Fock states, since the even cat state $|\mathrm{ecat}\rangle$ [Eq.~\eqref{eq:even cat state}] contains only even-photon-number Fock states.
However, this odd-parity-specific behavior is accompanied by a vanishing success probability at $t=0$ and remains strongly suppressed near photon depletion, making it difficult to observe in practice.

We consider how parity postselection affects the quantum-state transfer process~\eqref{eq:quantum state transfer} in the intermediate-amplitude regime ($N/2 = |\alpha_0|^2$).
Figures~\ref{fig:parity}(b) and (e) show that significant QFI can be generated even at times when $\mathcal{F}_{\mathrm{Q}}/N < 1$ without photon-state measurements.
Moreover, in Fig.~\ref{fig:parity}(e) ($N = 32$), the QFI becomes nearly identical for both even- and odd-parity outcomes over $t \in [10,50]$.
This does not mean that resolving the parity outcome is unnecessary; if the outcome is not recorded, averaging over parities is equivalent to tracing out the photonic degrees of freedom (dash-dotted curves).

Parity postselection qualitatively changes the QFI dynamics in the large-amplitude regime.
Despite the trace-out baseline showing no QFI enhancement, Figs.~\ref{fig:parity}(c) and (f) show coherent oscillations of the QFI density once the parity is resolved.
During the initial cycles, the QFI density reaches its theoretical maximum, $\mathcal{F}_{\mathrm{Q}}/N \simeq N$.
This demonstrates that genuine multipartite entanglement and macroscopic superposition can be realized in matter by combining large-amplitude cat-state light with photon-number-parity measurements.
Section~\ref{sec:parity:analytical} links these oscillations to a Rabi-oscillation cat state predicted by the XFA and briefly comments on backaction effects beyond the XFA~\cite{footnote_squeezing_in_Rabi_collapse}.

\subsection{Comparison with Rabi-oscillation cat state} \label{sec:parity:analytical}
We apply the XFA (Sec.~\ref{sec:XFA}) to the Tavis--Cummings electronic Hamiltonian $\hat{\mathcal{H}}_{\mathrm{e,TC}}$ [Eq.~\eqref{eq:RWA}].
Within the XFA, the photon mode acts as the prescribed $c$ number $\alpha(t) = \alpha\, \exp(-\mathrm{i}\omega t)$ [Eq.~\eqref{eq:alpha(t)}] and the electronic state $|\psi_{\alpha}(t)\rangle_{\mathrm{e}}$ obeys
\begin{equation}
\mathrm{i} \partial_t |\psi_{\alpha}(t)\rangle_{\mathrm{e}} = \hat{\mathcal{H}}_{\mathrm{e,TC}}[\alpha(t)]\, |\psi_{\alpha}(t)\rangle_{\mathrm{e}}. \label{eq:electron state with alpha:TC}
\end{equation}
For driving with even cat-state light, the total-system wave function takes the superposition form of Eq.~\eqref{eq:derive:superposition total state} using the solutions $|\psi_{\pm\alpha_0}(t)\rangle_{\mathrm{e}}$.

In the Tavis--Cummings model, Eq.~\eqref{eq:electron state with alpha:TC} admits an analytical solution that exhibits Rabi oscillations.
For the all-down initial state ($J = N/2$), the solution reads
\begin{equation}
|\psi_{\alpha}(t)\rangle_{\mathrm{e}} = \sum_{m=-J}^{J} \sqrt{ \binom{2J}{J+m}}\, a(t)^{J-m}\, b(t)^{J+m} \mathrm{e}^{-\mathrm{i}m\omega t} |J,m\rangle, \label{eq:N qubit Rabi solution}
\end{equation}
with the single-qubit amplitudes (in a frame rotating at frequency $\omega$)
\begin{align}
&a(t) = \cos\left( \Omega_{\mathrm{eff}}t/2 \right) + \mathrm{i} [(\Delta - \omega)/\Omega_{\mathrm{eff}}] \sin\left( \Omega_{\mathrm{eff}}t/2 \right), \label{eq:1 qubit Rabi solution:a} \\
&b(t) = \mathrm{i} (E_{\alpha}/\Omega_{\mathrm{eff}}) \sin\left( \Omega_{\mathrm{eff}}t/2 \right), \label{eq:1 qubit Rabi solution:b} \\
&\Omega_{\mathrm{eff}}=\sqrt{(\Delta - \omega)^2 + |E_{\alpha}|^2},\ \ \ \ E_{\alpha} = \mathrm{i} \gamma \omega \alpha. \label{eq:1 qubit Rabi solution:parameters}
\end{align}
Here, $\Omega_{\mathrm{eff}}$ is the effective Rabi frequency, and $E_{\alpha}$ is the field strength.

\begin{figure}[t]
\centering
\includegraphics[width=\columnwidth]{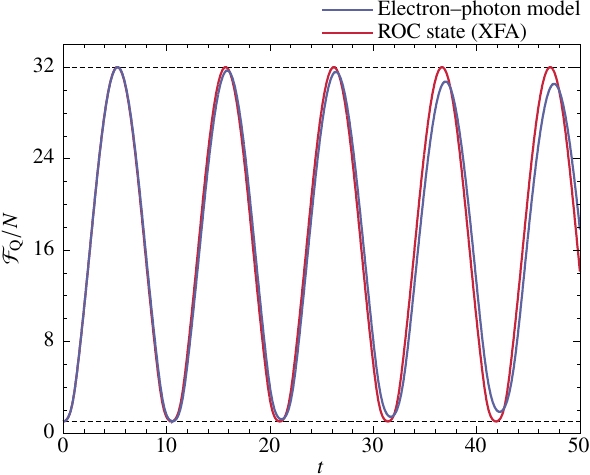}
\caption{
Comparison of QFI dynamics for the ROC state within the XFA.
The blue and red curves show the QFI density $\mathcal{F}_{\mathrm{Q}}/N$ calculated from the Tavis--Cummings model [same data as Fig.~\ref{fig:parity}(f)] and from the ROC state [Eq.~\eqref{eq:derive:superposition of Rabi oscillations}], respectively.
The photon-number parity is even.
Parameters are set to $N = 32$, $\gamma = 0.01$, and $\alpha_0 = 30$.}
\label{fig:superposition of Rabi oscillations}
\end{figure}

By applying the even-parity projector $\mathcal{P}_{+}^{\mathrm{P}}$ [Eq.~\eqref{eq:even projector}] to the XFA state~\eqref{eq:derive:superposition total state}, the postselected electronic state becomes a superposition of the two Rabi oscillations $|\psi_{\pm\alpha_0}\rangle_{\mathrm{e}}$,
\begin{equation}
|\psi_{\mathrm{ROC}}\rangle_{\mathrm{e}} =  \frac{1}{\sqrt{\mathcal{N}}}\ (|\psi_{+\alpha_0}\rangle_{\mathrm{e}}  + |\psi_{{-}\alpha_0}\rangle_{\mathrm{e}} ), \label{eq:derive:superposition of Rabi oscillations}
\end{equation}
where $\mathcal{N}=2(1+ \mathrm{Re}[\langle\psi_{\alpha_0}|\psi_{-\alpha_0}\rangle] )$.
We refer to this state as the Rabi-oscillation cat (ROC) state.
Substituting the analytical solution~\eqref{eq:N qubit Rabi solution} into Eq.~\eqref{eq:derive:superposition of Rabi oscillations}, we compute the QFI density $\mathcal{F}_{\mathrm{Q}}/N$ and show its time evolution as the red curve in Fig.~\ref{fig:superposition of Rabi oscillations}.
The ROC prediction agrees quantitatively with the numerical result of the Tavis--Cummings model (blue) during the first few cycles.
At longer times, however, a deviation gradually develops, which can be understood as the collapse of Rabi oscillations~\cite{Eberly1980, Gea-Banacloche1991}.
Within the present iterative framework, this collapse arises from the backaction of the Rabi-oscillating electronic subsystem onto the photon field, which enters at the next Born iteration: the photon field acquires an electronic-state--conditioned displacement and becomes entangled with the electrons even for a coherent-state drive~\cite{footnote_collapse_of_Rabi}.
Consequently, the otherwise persistent Rabi oscillations acquire a Gaussian envelope $\exp [-(\gamma \omega t)^2/8 ]$, consistent with the collapse time scale observed in Fig.~\ref{fig:parity}(f).
Including higher-order Born iterations would systematically improve the approximation, whereas the thermodynamic limit~\eqref{eq:thermodynamics limit} extends the time window where the XFA remains valid (see Sec.~\ref{sec:overcomplete}).

\begin{figure}[t]
\centering
\includegraphics[width=\columnwidth]{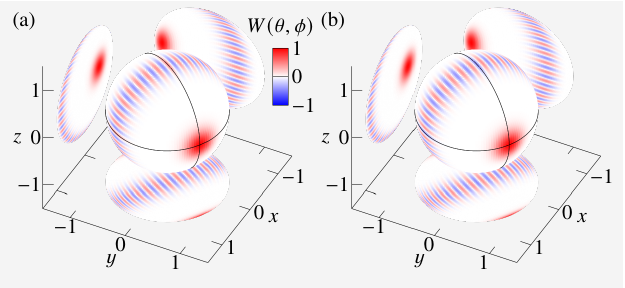}
\caption{
Spin Wigner functions $W(\theta,\phi)$ after driving with even cat-state light and postselection on photon-number parity: (a) even parity; (b) odd parity.
Time is $t = \pi/(2\gamma \alpha_0)$.
Parameters are $N = 32$, $\gamma = 0.01$, and $\alpha_0 = 30$.}
\label{fig:spin Wigner function}
\end{figure}

We visualize the ROC state using the spin Wigner function $W(\theta, \phi)$ [Eq.~\eqref{eq:Wigner function}].
Figure~\ref{fig:spin Wigner function} shows the spin Wigner functions calculated from $\hat{\rho}_{\mathrm{e}}^{(\mathrm{P},\pm)}(t)$ at a quarter period of the Rabi oscillation, $t = \pi/(2\gamma |\alpha_0|$).
Figures~\ref{fig:spin Wigner function}(a) and (b) correspond to even and odd parity, respectively.
The other settings are the same as Fig.~\ref{fig:parity}(f), with $N = 32$ and $\alpha_0 = 30$.
In both cases, two Gaussian intensities appear at antipodal positions on the equator, as in the trace-out case [Fig.~\ref{fig:trace_out}(i)].
In addition, distinct interference fringes run along the meridional direction, proving the emergence of macroscopic quantum superpositions in the electronic system.
We overlay meridional and equatorial guide lines through the Gaussian centers: comparing Figs.~\ref{fig:spin Wigner function}(a) and (b) reveals a half-period shift in the fringe patterns.
This indicates that the many-body electronic state becomes an even or odd cat state $|\psi_{\alpha_0}\rangle \pm |\psi_{-\alpha_0}\rangle$ heralded by the photon-number parity (even or odd).

\begin{figure}[t]
\centering
\includegraphics[width=\columnwidth]{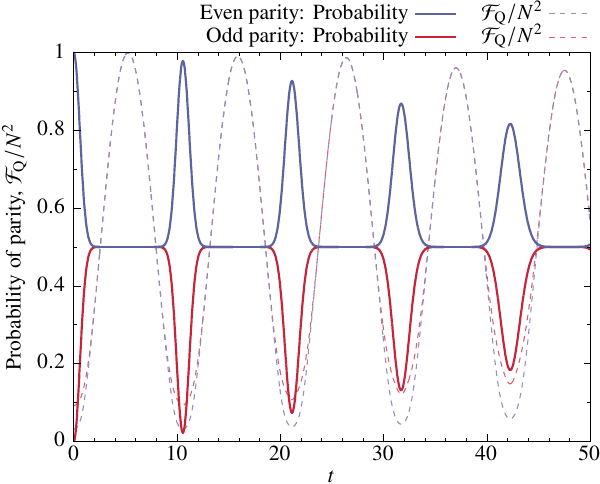}
\caption{
Probabilities of photon-number parity outcomes.
Blue (red) solid curve shows even (odd) parity case.
Dashed curves represent the corresponding QFI $\mathcal{F}_{\mathrm{Q}}/N^2$ for reference.
Parameters are the same as in Fig.~\ref{fig:parity}(f).}
\label{fig:probability}
\end{figure}

Finally, we examine the statistics of the parity outcomes.
Figure~\ref{fig:probability} shows the measurement probabilities of even (blue solid curve) and odd (red solid curve) photon-number parity for $N = 32$ and $\alpha_0 = 30$.
As a reference, the QFIs normalized to the Heisenberg limit, $\mathcal{F}_{\mathrm{Q}}/N^2$, are shown as dashed curves.
When the QFI reaches its maximum, the two outcomes become nearly equiprobable ($\approx 0.5$).
This behavior can be understood from the XFA wave function $|\Psi_{\mathrm{XFA}}\rangle_{\mathrm{ep}}$ [Eq.~\eqref{eq:derive:superposition total state}].
When the electronic states $|\psi_{\pm\alpha_0}\rangle$ are nearly orthogonal (i.e., $\mathcal{F}_{\mathrm{Q}} \approx N^2$), tracing out the electronic system results in a photonic reduced density matrix that approximates a classical statistical mixture of two coherent states: $|\alpha_0\rangle \langle \alpha_0| + |{-}\alpha_0\rangle \langle {-}\alpha_0|$.
In this mixture, the even and odd probabilities approach $1/2$ for large $|\alpha_0|$, consistent with Fig.~\ref{fig:probability}.

\subsection{Overcompleteness of coherent-state basis} \label{sec:overcomplete}
In this section, we examine how the overcompleteness of the coherent-state basis affects predictions of the XFA.
We compare two different coherent-state expansions of the even cat state~\eqref{eq:even cat state}.
The first expansion coefficient is given by $f_{\mathrm{ecat}}(\alpha) = [\delta^2(\alpha- \alpha_0) + \delta^2(\alpha+\alpha_0)]/ \sqrt{\mathcal{N}}$, as in Eq.~\eqref{eq:even cat state:coherent state rep}.
An alternative coefficient $\tilde{f}_{\mathrm{ecat}}(\alpha)$ is obtained by inserting the identity operator $\mathbb{1} = \int \mathrm{d}^2\alpha\, |\alpha\rangle \langle \alpha|/\pi$ into Eq.~\eqref{eq:even cat state}:
\begin{equation}
\tilde{f}_{\mathrm{ecat}}(\alpha) = \frac{1}{\pi \sqrt{\mathcal{N}}} \mathrm{e}^{-|\alpha|^2/2} \mathrm{e}^{-|\alpha_0|^2/2} \left( \mathrm{e}^{\overline{\alpha}\alpha_0} + \mathrm{e}^{-\overline{\alpha}\alpha_0} \right). \label{eq:even cat state:overcomplete:coefficients}
\end{equation}
The photonic state itself is unchanged as $\int \mathrm{d}^2\alpha f_{\mathrm{ecat}}(\alpha)|\alpha\rangle = \int \mathrm{d}^2\alpha\, \tilde{f}_{\mathrm{ecat}}(\alpha) |\alpha\rangle$.

However, within the XFA, the total-system state $|\Psi_{\mathrm{XFA}}(t)\rangle_{\mathrm{ep}}$ in Eq.~\eqref{eq:total state:XFA} is generally representation-dependent.
Using $f_{\mathrm{ecat}}(\alpha)$ yields Eq.~\eqref{eq:derive:superposition total state}, whereas $\tilde{f}_{\mathrm{ecat}}(\alpha)$ gives 
\begin{equation}
|\tilde{\Psi}_{\mathrm{XFA}}(t)\rangle_{\mathrm{ep}} = \int \mathrm{d}^2\alpha\, \tilde{f}_{\mathrm{ecat}}(\alpha) |\psi_{\alpha}(t)\rangle_{\mathrm{e}}|\alpha(t)\rangle.  \label{eq:derive:superposition total state:overcomplete}
\end{equation}
The discrepancy arises from $|\psi_{\alpha}(t)\rangle_{\mathrm{e}}$ for $\alpha \neq \pm \alpha_0$.
Since the electronic dynamics in the XFA depend on the intensive field $\gamma \alpha$ rather than $\alpha$ itself, this difference is expected to vanish in the thermodynamic limit~\eqref{eq:thermodynamics limit}.

\begin{figure}[t]
\centering
\includegraphics[width=\columnwidth]{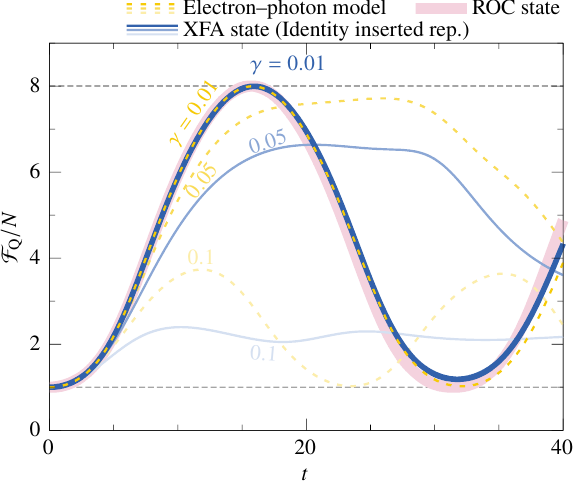}
\caption{
Effect of coherent-state overcompleteness on the XFA framework in the QFI dynamics.
Yellow dashed curves show exact results obtained by solving the Tavis--Cummings model~\eqref{eq:Tavis--Cummings model Hamiltonian}.
Thick red curve shows the XFA result for the ROC state~\eqref{eq:derive:superposition of Rabi oscillations}.
Solid blue curves show the XFA results for the alternative expansion $|\tilde{\Psi}_{\mathrm{XFA}}\rangle_{\mathrm{ep}}$ [Eq.~\eqref{eq:derive:superposition total state:overcomplete}] obtained by inserting the identity.
Field strength is fixed as $|E_{\alpha_0}| = \gamma \alpha_0 = 0.1$, with $\gamma = 0.01, 0.05, 0.1$.
Electron number is $N = 8$.
}
\label{fig:overcompleteness}
\end{figure}

To test this expectation, we numerically analyze the $\gamma$ dependence of the results from Eqs.~\eqref{eq:derive:superposition total state} and~\eqref{eq:derive:superposition total state:overcomplete}.
Figure~\ref{fig:overcompleteness} shows the time evolution of the QFI density $\mathcal{F}_{\mathrm{Q}}/N$ for the postselected electronic state (even photon-number parity) driven by the even cat-state light.
The yellow dashed curves are the numerical results for the Tavis--Cummings model~\eqref{eq:Tavis--Cummings model Hamiltonian}; the thick red curve is obtained from $|\Psi_{\mathrm{XFA}}\rangle_{\mathrm{ep}}$ with $f_{\mathrm{ecat}}$ [Eq.~\eqref{eq:derive:superposition total state}], and the solid blue curves are given by $|\tilde{\Psi}_{\mathrm{XFA}}\rangle_{\mathrm{ep}}$ with $\tilde{f}_{\mathrm{ecat}}$ [Eq.~\eqref{eq:derive:superposition total state:overcomplete}].
The field strength is fixed at $|E_{\alpha_0}| = \gamma \alpha_0 = 0.1$, and we vary $\gamma = 0.01$, $0.05$, and $0.1$.
Note that the ROC result (red) does not depend explicitly on $\gamma$, since the Rabi oscillation is governed solely by $\gamma \alpha_0$ [see Eq.~\eqref{eq:N qubit Rabi solution}].
As $\gamma$ decreases at fixed $\gamma |\alpha_0|$, it is found that both the yellow dashed curves and the blue solid curves converge toward the red curve, indicating thermodynamic convergence.

We consider the mechanism behind this convergence.
The results from the Tavis--Cummings model (yellow) approach that of the ROC state (red) because the XFA validity condition~\eqref{eq:photon num >> excited electron num} is increasingly satisfied as $\gamma \to 0$ while $\langle \hat{n} \rangle \approx |\alpha_0|^2 \to \infty$ (with $\gamma |\alpha_0|$ fixed).
This suggests that the time window over which the XFA remains valid, $t \ll 1/\gamma$, becomes longer.

We next discuss the thermodynamic convergence of the QFI dynamics obtained from $\tilde{f}_{\mathrm{ecat}}(\alpha)$ (blue) and $f_{\mathrm{ecat}}(\alpha)$ (red).
The two terms of $\tilde{f}_{\mathrm{ecat}}(\alpha)$ in Eq.~\eqref{eq:even cat state:overcomplete:coefficients} originate from the overlaps $\langle \alpha |{\pm} \alpha_0 \rangle$.
These absolute values read $|\langle \alpha| {\pm}\alpha_0\rangle|^2 = \exp(-|\gamma\alpha - \gamma\alpha_0|^2/\gamma^2)$, which approach a delta function $\delta^2(\gamma\alpha \pm \gamma\alpha_0)$ in the thermodynamic limit.
Hence the total-system state $|\tilde{\Psi}_{\mathrm{XFA}}(t)\rangle_{\mathrm{ep}}$ [Eq.~\eqref{eq:derive:superposition total state:overcomplete}] collapses onto two contributions at $\gamma \alpha = \pm \gamma \alpha_0$ and becomes equivalent to $|\Psi_{\mathrm{XFA}}(t)\rangle_{\mathrm{ep}}$ [Eq.~\eqref{eq:derive:superposition total state}], explaining the observed convergence.
This observation suggests that XFA predictions can be representation-dependent for the choice of $f(\alpha)$ in Eq.~\eqref{eq:photon state:coherent state rep} at finite $\gamma$, and that one should verify thermodynamic convergence.
A full proof that any coherent-state expansion yields the same XFA state in the thermodynamic limit~\eqref{eq:thermodynamics limit} is left for future work.

\section{Electronic dynamics conditioned on optical-quadrature measurements} \label{sec:quadrature}
In this section, we analyze postselected many-body electronic dynamics driven by the kitten state of light~\eqref{eq:kitten}, conditioned on projective measurements of the optical quadrature $\hat{x}_{\varphi}$~\eqref{eq:def:quadrature operator}.
Section~\ref{sec:quadrature:ideal} presents numerical results for the ideal projection.
Section~\ref{sec:quadrature:finite} examines the effects of finite measurement resolution.

\subsection{Ideal quadrature projection} \label{sec:quadrature:ideal}
We numerically compute the time-evolved total-system wave function $|\Psi(t)\rangle_{\mathrm{ep}}$ with the Tavis--Cummings model $\mathcal{H}_{\mathrm{TC}}$ [Eq.~\eqref{eq:Tavis--Cummings model Hamiltonian}], starting from the kitten state of light~\eqref{eq:kitten} and the all-spin-down electronic state.
We construct the postselected electronic density matrix $\hat{\rho}_{\mathrm{e}}^{(\mathrm{Q};\varphi,x)}(t)$ via Eq.~\eqref{eq:postselected electron state:general} using the quadrature projector $\mathcal{P}_x^{\mathrm{Q};\varphi}$ [Eq.~\eqref{eq:ideal quadrature projector}] for $\hat{x}_{\varphi}$.

To optimally induce macroscopic quantum states in the electronic system, we set the quadrature parameters to
\begin{equation}
x=0, \ \ \varphi(t) = \frac{\pi}{2} + \mathrm{arg}[\alpha_0(t)]=\frac{\pi}{2}-\omega t, \label{eq:optimal x and varphi}
\end{equation}
for $\alpha_0(t) = \alpha_0 \exp(-\mathrm{i}\omega t)$ [$\alpha_0 \in \mathbb{R}$, Eq.~\eqref{eq:alpha(t)}].
These values are determined from the optimal projector criterion in Eq.~\eqref{eq:optimal projector condition}, specifically $\langle x;\varphi| \alpha_0(t)\rangle=\langle x;\varphi| 0\rangle$.
The overlap is
\begin{equation}
\langle x;\varphi| \alpha\rangle = \pi^{-1/4} \, \mathrm{e}^{-\mathrm{i}x_0 p_0 + \mathrm{i} \sqrt{2} p_0 x} \, \mathrm{e}^{-(x-\sqrt{2}x_0)^2/2}, \label{eq:<x|alpha>}
\end{equation}
with $\alpha \exp(-\mathrm{i}\varphi)= x_0 + \mathrm{i} p_0$.
With this optimal measurement, we then evaluate the corresponding QFI $\mathcal{F}_{\mathrm{Q}}$ [Eq.~\eqref{eq:def quantum Fisher information}].
The electron--photon coupling strength is set to $\gamma = 0.01$.

\begin{figure}[t]
\centering
\includegraphics[width=\columnwidth]{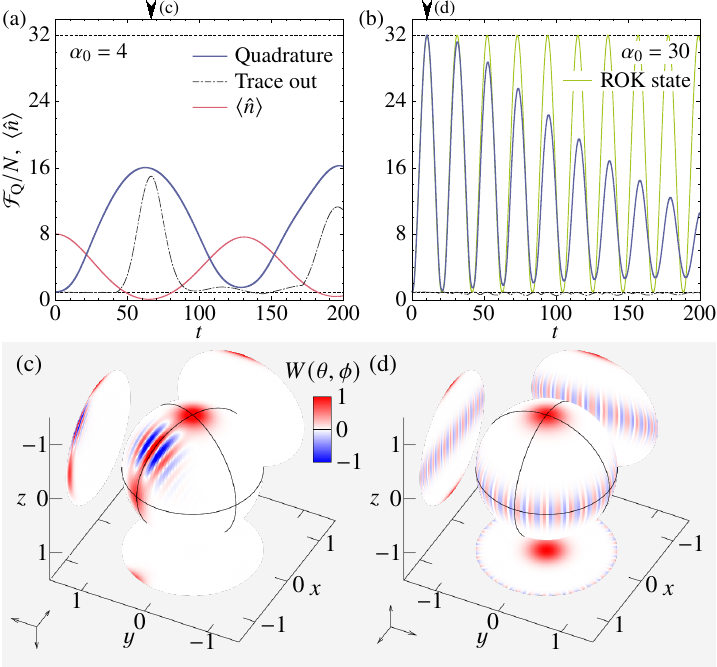}
\caption{
Many-body electronic dynamics driven by the kitten-state light with quadrature postselection.
(a),(b) Time evolution of the QFI density $\mathcal{F}_{\mathrm{Q}}/N$.
Blue: postselected states $\hat{\rho}_{\mathrm{e}}^{(\mathrm{Q};\varphi(t),0)}$.
Dash-dotted: trace-out baseline.
Red: photon number $\langle \hat{n} \rangle$.
Green: XFA-based ROK state~\eqref{eq:derive:superposition of Rabi oscillations:kitten}.
The kitten-state amplitude is (a)~$\alpha_0=4$ and (b)~$\alpha_0 = 30$.
(c),(d) Spin Wigner functions $W(\theta,\phi)$ [Eq.~\eqref{eq:Wigner function}] at the times marked as arrowheads in panels (a) and (b).
Electron number is $N = 32$.}
\label{fig:quadrature_ideal}
\end{figure}

Figure~\ref{fig:quadrature_ideal} shows the time evolution of the QFI density $\mathcal{F}_{\mathrm{Q}}/N$ for $N = 32$.
Figures~\ref{fig:quadrature_ideal}(a) and (b) correspond to the intermediate-amplitude regime ($\alpha_0 = 4$) and the large-amplitude regime ($\alpha_0 = 30$), respectively.
The blue curves represent the QFI calculated from $\hat{\rho}_{\mathrm{e}}^{(\mathrm{Q};\varphi(t),0)}$.
The dash-dotted curves show the trace-out baseline calculated from the reduced electronic density matrix $\hat{\rho}_{\mathrm{e}}(t)$ [Eq.~\eqref{eq:trace out:electron density matrix}], i.e., without postselection.

We first discuss the case without photon-state measurement, focusing on quantum-state transfer induced by kitten-state light.
Coherent-state light $|\alpha_0\rangle$ can fully excite the all-spin-down state into an $xy$ plane-polarized state via perfect energy absorption.
Neglecting the interaction energy ($\gamma = 0$), energy conservation requires $\sum_i \langle S^z_i\rangle + \langle\alpha_0| \hat{n} |\alpha_0\rangle = -N/2 + |\alpha_0|^2 = 0$, which yields $|\alpha_0| = \sqrt{N/2}$.
Figure~\ref{fig:quadrature_ideal}(a) shows the QFI dynamics under this condition.
When the photon number $\langle \hat{n} \rangle$ vanishes (red curve), the QFI of $\hat{\rho}_{\mathrm{e}}(t)$ (dash-dotted curve) reaches its peak.
The peak value is $\mathcal{F}_{\mathrm{Q}}/N \approx N/2$, which is the largest QFI achievable without measurement under kitten-state optical driving.

Figure~\ref{fig:quadrature_ideal}(c) shows the spin Wigner function $W(\theta,\phi)$ [Eq.~\eqref{eq:Wigner function}] at the QFI peak time  (with all axes inverted for visibility).
A Gaussian peak remains at the south pole (representing the initial state), while another appears on the equator.
Distinct interference fringes between these peaks evidence a macroscopic superposition of the form $|{\downarrow\downarrow \cdots \downarrow}\rangle + |{\rightarrow\rightarrow\cdots\rightarrow}\rangle$, where $|{\rightarrow}\rangle$ denotes a spin eigenstate oriented in the $xy$ plane.
Although quadrature projection indeed enhances QFI generation across the entire time region, shown as the blue curve in Fig.~\ref{fig:quadrature_ideal}(a), it has little impact on the quantum-state transfer due to photon depletion.

We examine the dynamics in the large-amplitude regime.
The QFI density stays below unity without postselection [Fig.~\ref{fig:quadrature_ideal}(b), dash-dotted curve], indicating that the QFI does not witness macroscopic entanglement under large-amplitude kitten-state driving.
In contrast, quadrature projection raises the QFI density to its maximum value, $\mathcal{F}_{\mathrm{Q}}/N = N$ (blue curve).
The corresponding spin Wigner function $W(\theta, \phi)$ at the QFI peak time is shown in Fig.~\ref{fig:quadrature_ideal}(d).
Two Gaussians appear at the poles with pronounced quantum interference fringes along the equator, indicating the formation of a macroscopic cat state $|{\downarrow\downarrow \cdots \downarrow}\rangle + |{\uparrow\uparrow \cdots \uparrow}\rangle$.

This generation of the macroscopic electronic cat state can be interpreted within the XFA.
After quadrature projection, the postselected electronic state is approximately given by
\begin{equation}
|\psi_{\mathrm{ROK}}\rangle_{\mathrm{e}} = \frac{1}{\sqrt{\mathcal{N}}}\ (|\psi_{\alpha_0}\rangle_{\mathrm{e}} + |{\downarrow\downarrow \cdots \downarrow}\rangle_{\mathrm{e}} ), \label{eq:derive:superposition of Rabi oscillations:kitten}
\end{equation}
which we term the Rabi-oscillation kitten (ROK) state.
By substituting the analytical solution for $|\psi_{\alpha_0}\rangle_{\mathrm{e}}$ [Eq.~\eqref{eq:N qubit Rabi solution}], the resulting QFI is plotted as the green curve in Fig.~\ref{fig:quadrature_ideal}(b).
The XFA result agrees quantitatively with the numerical result (blue) during the initial coherent-oscillation stage ($t \approx 0$–$40$), before the collapse of Rabi oscillations due to electronic backaction on the photons.

\subsection{Resolution dependence} \label{sec:quadrature:finite}
To assess the role of finite measurement resolution, we consider the finite-resolution quadrature projector $\mathcal{P}_{x}^{(\mathrm{Q};\varphi;\Delta x)}$ [Eq.~\eqref{eq:finite quadrature projector}], instead of the ideal projector used above.
The corresponding postselected electronic density matrix $\hat{\rho}_{\mathrm{e}}^{(\mathrm{Q};\varphi;\Delta x, x)}$ is computed via Eq.~\eqref{eq:postselected electron state:general}, from which we evaluate the QFI $\mathcal{F}_{\mathrm{Q}}$ [Eq.~\eqref{eq:def quantum Fisher information}] of the electronic system.

\begin{figure}[t]
\centering
\includegraphics[width=\columnwidth]{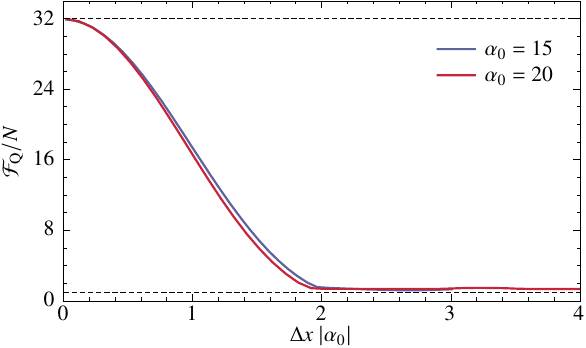}
\caption{
Dependence of the maximum QFI density $\mathcal{F}_{\mathrm{Q}}/N$ on quadrature measurement resolution $\Delta x$.
Blue (red) curve corresponds to a kitten-state amplitude $\alpha_0 = 15$ ($\alpha_0 = 20$).
Electron number is $N = 32$, and the simulation setting is identical to that in Fig.~\ref{fig:quadrature_ideal}.}
\label{fig:quadrature window}
\end{figure}

Figure~\ref{fig:quadrature window} shows the dependence of the maximum QFI density on the quadrature measurement resolution $\Delta x$.
We examine two values of the kitten-state amplitude~\eqref{eq:kitten}, $\alpha_0 = 15$ and $\alpha_0 = 20$, while keeping all other simulation parameters identical to those in Sec.~\ref{sec:quadrature:ideal}.
We find that the resulting curves collapse when rescaled by $\Delta x \cdot |\alpha_0|$.
This scaling implies that, to achieve a given QFI density---or equivalently, a certain entanglement depth---the required measurement resolution $\Delta x$ must decrease inversely with the kitten-state amplitude $\alpha_0$.

Such a requirement is not unique to material responses; it also arises when verifying the quantum properties of the incident light itself.
Therefore, advances in quantum measurement technologies are essential to realize macroscopic quantum responses in matter using multiphoton quantum light.

Finally, we note that the postselection success probability, evaluated at the QFI-peak time, scales approximately as $P_{\mathrm{succ}} \approx 1.065 \Delta x$ for $\Delta x |\alpha_0| \lesssim 2$.
This highlights a practical trade-off between achievable QFI enhancement and the event rate.

\section{Discussion} \label{sec:discussion}

\subsection{Effects of rotating-wave approximation} \label{sec:bRWA}
To assess the validity of the RWA used in the previous sections, we repeat the analysis of Sec.~\ref{sec:parity:numerical} without applying the RWA.
We use the $N$-qubit quantum Rabi and Dicke model $\mathcal{H}_{\mathrm{RD}}$ [Eq.~\eqref{eq:Rabi and Dicke hamiltonian}], and compute the time-evolved total-system wave function $|\Psi(t)\rangle_{\mathrm{ep}}$, starting from the even cat-state light~\eqref{eq:even cat state} and the all-spin-down electronic state.
We construct the postselected electronic density matrix $\hat{\rho}_{\mathrm{e}}^{(\mathrm{P},+)}(t)$ via Eq.~\eqref{eq:postselected electron state:general} using the even photon-number parity projector $\mathcal{P}_{+}^{\mathrm{P}}$ [Eq.~\eqref{eq:even projector}], from which we evaluate the electronic QFI $\mathcal{F}_{\mathrm{Q}}$ [Eq.~\eqref{eq:def quantum Fisher information}].

\begin{figure}[t]
\centering
\includegraphics[width=\columnwidth]{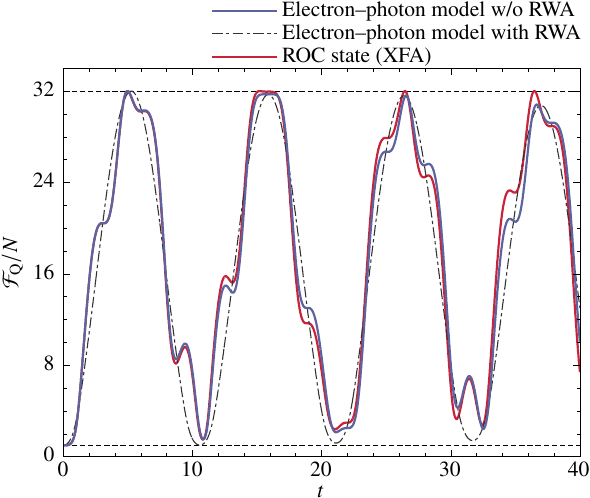}
\caption{
Effect of the RWA on the QFI dynamics under even cat-state optical driving and photon-number parity measurement.
Blue: simulation using the full electron--photon Hamiltonian $\hat{\mathcal{H}}_{\mathrm{RD}}$ [Eq.~\eqref{eq:Rabi and Dicke hamiltonian}] (no RWA).
Dash-dotted: RWA result [same as Fig.~\ref{fig:parity}(f)].
Red: XFA result obtained from the ROC state~\eqref{eq:derive:superposition of Rabi oscillations}.
Parameters: $N = 32$, $\alpha_0 = 30$.
}
\label{fig:without RWA}
\end{figure}

Figure~\ref{fig:without RWA} compares the time evolution of the QFI density $\mathcal{F}_{\mathrm{Q}}/N$ from simulations with and without the RWA.
Without the RWA (blue), the dynamics exhibit small $4\pi$-period modulations, corresponding to $2\omega$ oscillations ($\omega = 1$) induced by counter-rotating terms.
These high-frequency modulations are absent in the RWA result (dash-dotted) but do not qualitatively affect the overall QFI dynamics.
Thus, our main conclusion holds regardless of the RWA.

We examine the effect of the RWA on the XFA framework.
Without the RWA, no analytical solution exists for the semiclassical equation of motion for the electronic system in Eq.~\eqref{eq:electron state with alpha}.
Instead, we numerically solve Eq.~\eqref{eq:electron state with alpha} with the electronic Rabi and Dicke Hamiltonian $\hat{\mathcal{H}}_{\mathrm{e,RD}}[\pm \alpha_0(t)]$ [Eq.~\eqref{eq:Rabi and Dicke coupling hamiltonian}], construct the ROC state from Eq.~\eqref{eq:derive:superposition of Rabi oscillations}, and evaluate the QFI~\eqref{eq:def quantum Fisher information}.
The result is shown as the red curve in Fig.~\ref{fig:without RWA}.
We find that the ROC state reproduces the initial coherent oscillations of the QFI, including the small, rapid oscillations originating from the counter-rotating terms.
This confirms that the ROC-state picture based on the XFA remains valid, irrespective of whether the RWA is applied.

\subsection{Experimental demonstration} \label{sec:experiment}
We now discuss the experimental feasibility of the present proposal.
The theoretical models employed in this work [the Tavis--Cummings model~\eqref{eq:Tavis--Cummings model Hamiltonian} and the $N$-qubit Rabi--Dicke model~\eqref{eq:Rabi and Dicke hamiltonian}] provide effective descriptions of, e.g., cavity-QED or cold-atom platforms and band insulators.
In cavity-QED and cold-atom systems, the photon field represents a cavity-confined mode, while the $N$ two-level systems are realized by a dipole-allowed transition between two internal states of $N$ trapped atoms (e.g., Rb, Cs, Sr, or Yb).
In band insulating systems, the photon field can be interpreted as a propagating optical mode (with other continuous modes assumed to be in vacuum states), while the effective two-level systems correspond to interband transitions (valence band $\leftrightarrow$ conduction band), i.e., electron--hole pair excitations resonant with the photon energy.

In realistic systems, dissipation is unavoidable and may reduce the visibility of the ROC state.
The relevant processes include photon loss of the light mode, characterized by a decay rate $\kappa$, as well as relaxation and dephasing of the matter subsystem, characterized by $T_1$ and $T_2$ (including, e.g., electron--electron interaction-induced dephasing and inhomogeneous broadening).
When the dissipation times are much longer than the Rabi period $T_{\mathrm{Rabi}} = 2\pi/|E_{\alpha_0}|$, i.e., $T_{\mathrm{Rabi}} \ll T_1,T_2,\kappa^{-1}$, the ROC state is expected to be observable; otherwise, dissipation washes out the coherent Rabi oscillations and thereby degrades the ROC-state coherence.
Accordingly, promising experimental platforms are those where coherent Rabi oscillations have already been demonstrated, for example, cavity-QED~\cite{Dominici2014}, cold-atom~\cite{Lim2014}, and semiconductor systems~\cite{Furst1997}.
We also note that Sec.~\ref{sec:POVM} discusses the backaction associated with weak detection, which can often be formulated in a way analogous to coupling to a (Markovian) bath.

The QFI, which plays a central role in our analysis, can be indirectly probed through measurements of dynamic susceptibility~\cite{Hauke2016}.
Recent advances have extended such techniques to transient systems~\cite{Hales2023}, suggesting that they could be applied to probe the ultrafast dynamics driven by cat-state light, as explored in this work.

While our work has focused on generating quantum states of matter using quantum light, it is natural to expect that quantum light may also be emitted via backaction from the resulting nonequilibrium quantum material~\cite{Pizzi2022}.
Such emission constitutes nonlinear optical phenomena driven by quantum light, as explored in recent studies on squeezed vacuum states~\cite{Gorlach2022b, Tzur2022a, Tzur2024, Rasputnyi2024, EvenTzur2024, Wang2024h, Lemieux2025, Gothelf2025} and large-amplitude kitten-state light~\cite{Lamprou2025}.
Our results suggest that projective measurements on either the matter or the incident light could be used to extract and amplify the quantum properties of the emitted light.

We outline possible experimental implementations of the photon-number parity measurement used in Sec.~\ref{sec:parity}.
Photon-number parity detection is a fundamental tool in quantum information processing and continues to be actively developed~\cite{Gerry2010, Birrittella2021}.
The most direct approach is to explicitly measure the photon number.
In the optical frequency regime, photon-number-resolving detection has been demonstrated for up to $55$ photons~\cite{Xu2024a}.
This benchmark allows one to estimate the Rabi-oscillation period $T_{\mathrm{R}}=2\pi/(\gamma\sqrt{\langle \hat{n}\rangle})$ as $T_{\mathrm{R}}\lesssim 10~\mu\mathrm{s}$ for Sr atoms [with $\gamma/(2\pi)\sim 15~\mathrm{kHz}$~\cite{Song2025} (standard notation: $\gamma \to 2g$)] and $T_{\mathrm{R}} \sim 0.1~\mu\mathrm{s}$ for rare-earth solid-state systems (with $\gamma/(2\pi)\sim 1~\mathrm{MHz}$~\cite{Shen2025}).
Comparing these estimates with the dephasing time, $T_2\lesssim 42~\mu\mathrm{s}$ for the Sr 689-nm transition (inferred from the linewidth in Ref.~\cite{Song2025}) and $T_2\sim 1~\mathrm{ms}$ for rare-earth-doped systems~\cite{Askarani2021}, we expect that the ROC state can be realized within $T_2$ in both platforms.
Several theoretical proposals not relying on photon counting have also been made, including QND-type schemes based on nonlinear optical interactions~\cite{Gerry2005, Munro2005}.
In addition, interference-based approaches using beam splitters with two photon-number-resolving detectors can implement an even-parity measurement~\cite{Thekkadath2020}.

We also consider possible implementations of the quadrature projective measurements used in Sec.~\ref{sec:quadrature}.
The standard method is balanced homodyne detection~\cite{Lvovsky2009}, which realizes a projective measurement onto quadrature eigenstates in the strong local oscillator limit~\cite{DAriano2003, Tyc2004}.
A quadrature resolution can be estimated as $\Delta x\lesssim 0.05$ (in shot-noise units) based on, e.g., $8$-bit ADC~\cite{Bruynsteen2023} and $28$-dB shot-noise clearance~\cite{Bruynsteen2021}.
Using the condition for achieving a large QFI, $|\alpha_0|\sim \Delta x^{-1}$, we obtain $T_{\mathrm{R}}\lesssim 3~\mu\mathrm{s}$ for Sr and $T_{\mathrm{R}}\lesssim 30~\mathrm{ns}$ for rare-earth solid-state systems, again well within the available coherence time $T_2$.
Recent developments in high-speed, real-time homodyne detection~\cite{Lordi2024, Hubenschmid2024, Yang2023b} further open the way to applying these measurements in ultrafast quantum dynamics considered in this work.

\subsection{Weak and continuous measurements} \label{sec:POVM}
We briefly discuss how our conclusions for single-shot projective measurements extend to more general positive operator-valued measurements (POVMs), including weak measurements and continuous monitoring~\cite{Kraus1983, Wiseman2009}.

We first consider a single-shot weak measurement. 
Conditioned on a given outcome, the corresponding POVM element can be taken close to the identity, $\hat{E}\propto \mathbb{1}+\epsilon\hat{e}$ with $\epsilon \ll 1$ (up to an overall normalization)~\cite{Aharonov1988}.
For an even cat-state drive, postselection multiplies the electronic interference term by $\langle -\alpha_0|\hat{E}|\alpha_0\rangle \propto \exp(-2|\alpha_0|^2)+\epsilon\langle -\alpha_0|\hat{e}|\alpha_0\rangle$.
Hence, the interference is restored at order $\epsilon$ only for weak measurements for which $\langle -\alpha_0|\hat{e}|\alpha_0\rangle=\mathcal{O}(|\alpha_0|^0)$, such as a weak photon-number-parity measurement.
This is expected to improve upon the exponential suppression observed without postselection.

Next, we consider a continuous, weak measurement.
In this case, the free harmonic evolution of the field~\eqref{eq:alpha(t)} is replaced by a stochastic evolution conditioned on the measurement record, i.e., a quantum trajectory~\cite{Caves1987, Dalibard1992, Wiseman1993}.
The conditional photonic state gradually collapses toward an eigenstate of the measured operator.

The resulting dynamics depend on whether the monitoring process acquires which-branch information about the two coherent-state branches of the cat state.
When the measurement distinguishes the branches, it quickly collapses the photonic state onto one branch, yielding the corresponding classically driven electronic state.
In contrast, when the measurement does not distinguish the branches, both branch states can collapse onto the same eigenstate of the measurement, allowing the electronic system to exhibit a large QFI, as in the results above.
For continuous measurement of the $\hat{x}$ quadrature and the even cat state of light $|\alpha\rangle + |{-}\alpha\rangle$, these two cases correspond to $\alpha \notin \mathrm{i}\mathbb{R}$ and $\alpha \in \mathrm{i}\mathbb{R}$, respectively.

Furthermore, under continuous monitoring during optical drive, measurement backaction changes each coherent state into nonclassical light.
For example, under continuous monitoring of $\hat{x}$, the conditioned photonic state becomes gradually squeezed in $\hat{x}$ (and antisqueezed in $\hat{p}$).
Consequently, the electronic states in each branch are driven by squeezed light; this situation may be described by applying the XFA method to a squeezed-light drive.

Finally, we briefly comment on repeated projective measurements.
After applying the cat-state light for some duration, repeated projective measurements, e.g., of the photon-number parity, can freeze the subsequent electron--photon dynamics via the quantum Zeno effect~\cite{Misra1977}.
Since light--matter absorption and emission flip the photon-number parity, frequent parity measurements suppress these processes.
This Zeno suppression could help stabilize the resulting ROC or ROK state.

\section{Summary} \label{sec:summary}
We studied the generation of macroscopic quantum states in many-body electronic systems driven by Schr\"{o}dinger cat and kitten states of light.
We found that projective measurements on the light field are essential to induce macroscopic cat states in matter in the thermodynamic limit.
Using the Tavis--Cummings model or quantum Rabi or Dicke model, we identified the conditions for achieving large quantum Fisher information (QFI), and classified the dynamics into small-, intermediate-, and large-amplitude regimes.

Without postselection, macroscopic entanglement appears only for an ultrashort time, and only in the intermediate regime, via quantum state transfer.
However, even in the thermodynamic limit (the large-amplitude regime), projective measurements on the light---photon-number parity and quadrature---restore macroscopic quantum superpositions: the Rabi-oscillation cat (ROC) and kitten (ROK) states, which we derive within the external-field approximation (XFA) framework.

Our results highlight that combining multiphoton quantum light with advanced quantum measurements is key to the macroscopic quantum control of matter.

\begin{acknowledgments}
The author thanks Atsushi Ono for fruitful discussions.
The author thanks Zongping Gong for helpful comments regarding QFI and entanglement.
This work was supported by JST FOREST (Grant No.~JPMJFR2131) and JSPS KAKENHI (Grants No.~JP23K19030 and No.~JP25K17343).
\end{acknowledgments}

\bibliography{ref}

\begin{thebibliography}{115}%
\makeatletter
\providecommand \@ifxundefined [1]{%
 \@ifx{#1\undefined}
}%
\providecommand \@ifnum [1]{%
 \ifnum #1\expandafter \@firstoftwo
 \else \expandafter \@secondoftwo
 \fi
}%
\providecommand \@ifx [1]{%
 \ifx #1\expandafter \@firstoftwo
 \else \expandafter \@secondoftwo
 \fi
}%
\providecommand \natexlab [1]{#1}%
\providecommand \enquote  [1]{``#1''}%
\providecommand \bibnamefont  [1]{#1}%
\providecommand \bibfnamefont [1]{#1}%
\providecommand \citenamefont [1]{#1}%
\providecommand \href@noop [0]{\@secondoftwo}%
\providecommand \href [0]{\begingroup \@sanitize@url \@href}%
\providecommand \@href[1]{\@@startlink{#1}\@@href}%
\providecommand \@@href[1]{\endgroup#1\@@endlink}%
\providecommand \@sanitize@url [0]{\catcode `\\12\catcode `\$12\catcode `\&12\catcode `\#12\catcode `\^12\catcode `\_12\catcode `\%12\relax}%
\providecommand \@@startlink[1]{}%
\providecommand \@@endlink[0]{}%
\providecommand \url  [0]{\begingroup\@sanitize@url \@url }%
\providecommand \@url [1]{\endgroup\@href {#1}{\urlprefix }}%
\providecommand \urlprefix  [0]{URL }%
\providecommand \Eprint [0]{\href }%
\providecommand \doibase [0]{https://doi.org/}%
\providecommand \selectlanguage [0]{\@gobble}%
\providecommand \bibinfo  [0]{\@secondoftwo}%
\providecommand \bibfield  [0]{\@secondoftwo}%
\providecommand \translation [1]{[#1]}%
\providecommand \BibitemOpen [0]{}%
\providecommand \bibitemStop [0]{}%
\providecommand \bibitemNoStop [0]{.\EOS\space}%
\providecommand \EOS [0]{\spacefactor3000\relax}%
\providecommand \BibitemShut  [1]{\csname bibitem#1\endcsname}%
\let\auto@bib@innerbib\@empty
\bibitem [{\citenamefont {Oka}\ and\ \citenamefont {Aoki}(2009)}]{Oka2009}%
  \BibitemOpen
  \bibfield  {author} {\bibinfo {author} {\bibfnamefont {T.}~\bibnamefont {Oka}}\ and\ \bibinfo {author} {\bibfnamefont {H.}~\bibnamefont {Aoki}},\ }\bibfield  {title} {\bibinfo {title} {{Photovoltaic Hall effect in graphene}},\ }\href {https://doi.org/10.1103/PhysRevB.79.081406} {\bibfield  {journal} {\bibinfo  {journal} {Phys. Rev. B}\ }\textbf {\bibinfo {volume} {79}},\ \bibinfo {pages} {081406(R)} (\bibinfo {year} {2009})}\BibitemShut {NoStop}%
\bibitem [{\citenamefont {McIver}\ \emph {et~al.}(2020)\citenamefont {McIver}, \citenamefont {Schulte}, \citenamefont {Stein}, \citenamefont {Matsuyama}, \citenamefont {Jotzu}, \citenamefont {Meier},\ and\ \citenamefont {Cavalleri}}]{McIver2020}%
  \BibitemOpen
  \bibfield  {author} {\bibinfo {author} {\bibfnamefont {J.~W.}\ \bibnamefont {McIver}}, \bibinfo {author} {\bibfnamefont {B.}~\bibnamefont {Schulte}}, \bibinfo {author} {\bibfnamefont {F.-U.}\ \bibnamefont {Stein}}, \bibinfo {author} {\bibfnamefont {T.}~\bibnamefont {Matsuyama}}, \bibinfo {author} {\bibfnamefont {G.}~\bibnamefont {Jotzu}}, \bibinfo {author} {\bibfnamefont {G.}~\bibnamefont {Meier}},\ and\ \bibinfo {author} {\bibfnamefont {A.}~\bibnamefont {Cavalleri}},\ }\bibfield  {title} {\bibinfo {title} {{Light-induced anomalous Hall effect in graphene}},\ }\href {https://doi.org/10.1038/s41567-019-0698-y} {\bibfield  {journal} {\bibinfo  {journal} {Nat. Phys.}\ }\textbf {\bibinfo {volume} {16}},\ \bibinfo {pages} {38} (\bibinfo {year} {2020})}\BibitemShut {NoStop}%
\bibitem [{\citenamefont {Lewenstein}\ \emph {et~al.}(2021)\citenamefont {Lewenstein}, \citenamefont {Ciappina}, \citenamefont {Pisanty}, \citenamefont {Rivera-Dean}, \citenamefont {Stammer}, \citenamefont {Lamprou},\ and\ \citenamefont {Tzallas}}]{Lewenstein2021}%
  \BibitemOpen
  \bibfield  {author} {\bibinfo {author} {\bibfnamefont {M.}~\bibnamefont {Lewenstein}}, \bibinfo {author} {\bibfnamefont {M.~F.}\ \bibnamefont {Ciappina}}, \bibinfo {author} {\bibfnamefont {E.}~\bibnamefont {Pisanty}}, \bibinfo {author} {\bibfnamefont {J.}~\bibnamefont {Rivera-Dean}}, \bibinfo {author} {\bibfnamefont {P.}~\bibnamefont {Stammer}}, \bibinfo {author} {\bibfnamefont {T.}~\bibnamefont {Lamprou}},\ and\ \bibinfo {author} {\bibfnamefont {P.}~\bibnamefont {Tzallas}},\ }\bibfield  {title} {\bibinfo {title} {{Generation of optical Schr{\"{o}}dinger cat states in intense laser^^e2^^80^^93matter interactions}},\ }\href {https://doi.org/10.1038/s41567-021-01317-w} {\bibfield  {journal} {\bibinfo  {journal} {Nat. Phys.}\ }\textbf {\bibinfo {volume} {17}},\ \bibinfo {pages} {1104} (\bibinfo {year} {2021})}\BibitemShut {NoStop}%
\bibitem [{\citenamefont {Rivera-Dean}\ \emph {et~al.}(2022)\citenamefont {Rivera-Dean}, \citenamefont {Lamprou}, \citenamefont {Pisanty}, \citenamefont {Stammer}, \citenamefont {Ord{\'{o}}{\~{n}}ez}, \citenamefont {Maxwell}, \citenamefont {Ciappina}, \citenamefont {Lewenstein},\ and\ \citenamefont {Tzallas}}]{Rivera-Dean2022}%
  \BibitemOpen
  \bibfield  {author} {\bibinfo {author} {\bibfnamefont {J.}~\bibnamefont {Rivera-Dean}}, \bibinfo {author} {\bibfnamefont {T.}~\bibnamefont {Lamprou}}, \bibinfo {author} {\bibfnamefont {E.}~\bibnamefont {Pisanty}}, \bibinfo {author} {\bibfnamefont {P.}~\bibnamefont {Stammer}}, \bibinfo {author} {\bibfnamefont {A.~F.}\ \bibnamefont {Ord{\'{o}}{\~{n}}ez}}, \bibinfo {author} {\bibfnamefont {A.~S.}\ \bibnamefont {Maxwell}}, \bibinfo {author} {\bibfnamefont {M.~F.}\ \bibnamefont {Ciappina}}, \bibinfo {author} {\bibfnamefont {M.}~\bibnamefont {Lewenstein}},\ and\ \bibinfo {author} {\bibfnamefont {P.}~\bibnamefont {Tzallas}},\ }\bibfield  {title} {\bibinfo {title} {{Strong laser fields and their power to generate controllable high-photon-number coherent-state superpositions}},\ }\href {https://doi.org/10.1103/PhysRevA.105.033714} {\bibfield  {journal} {\bibinfo  {journal} {Phys. Rev. A}\ }\textbf {\bibinfo {volume} {105}},\ \bibinfo {pages} {033714} (\bibinfo {year} {2022})}\BibitemShut {NoStop}%
\bibitem [{\citenamefont {Lamprou}\ \emph {et~al.}(2025)\citenamefont {Lamprou}, \citenamefont {Rivera-Dean}, \citenamefont {Stammer}, \citenamefont {Lewenstein},\ and\ \citenamefont {Tzallas}}]{Lamprou2025}%
  \BibitemOpen
  \bibfield  {author} {\bibinfo {author} {\bibfnamefont {T.}~\bibnamefont {Lamprou}}, \bibinfo {author} {\bibfnamefont {J.}~\bibnamefont {Rivera-Dean}}, \bibinfo {author} {\bibfnamefont {P.}~\bibnamefont {Stammer}}, \bibinfo {author} {\bibfnamefont {M.}~\bibnamefont {Lewenstein}},\ and\ \bibinfo {author} {\bibfnamefont {P.}~\bibnamefont {Tzallas}},\ }\bibfield  {title} {\bibinfo {title} {{Nonlinear Optics Using Intense Optical Coherent State Superpositions}},\ }\href {https://doi.org/10.1103/PhysRevLett.134.013601} {\bibfield  {journal} {\bibinfo  {journal} {Phys. Rev. Lett.}\ }\textbf {\bibinfo {volume} {134}},\ \bibinfo {pages} {013601} (\bibinfo {year} {2025})}\BibitemShut {NoStop}%
\bibitem [{\citenamefont {Stammer}(2022)}]{Stammer2022a}%
  \BibitemOpen
  \bibfield  {author} {\bibinfo {author} {\bibfnamefont {P.}~\bibnamefont {Stammer}},\ }\bibfield  {title} {\bibinfo {title} {{Theory of entanglement and measurement in high-order harmonic generation}},\ }\href {https://doi.org/10.1103/PhysRevA.106.L050402} {\bibfield  {journal} {\bibinfo  {journal} {Phys. Rev. A}\ }\textbf {\bibinfo {volume} {106}},\ \bibinfo {pages} {L050402} (\bibinfo {year} {2022})}\BibitemShut {NoStop}%
\bibitem [{\citenamefont {Stammer}\ \emph {et~al.}(2022)\citenamefont {Stammer}, \citenamefont {Rivera-Dean}, \citenamefont {Lamprou}, \citenamefont {Pisanty}, \citenamefont {Ciappina}, \citenamefont {Tzallas},\ and\ \citenamefont {Lewenstein}}]{Stammer2022}%
  \BibitemOpen
  \bibfield  {author} {\bibinfo {author} {\bibfnamefont {P.}~\bibnamefont {Stammer}}, \bibinfo {author} {\bibfnamefont {J.}~\bibnamefont {Rivera-Dean}}, \bibinfo {author} {\bibfnamefont {T.}~\bibnamefont {Lamprou}}, \bibinfo {author} {\bibfnamefont {E.}~\bibnamefont {Pisanty}}, \bibinfo {author} {\bibfnamefont {M.~F.}\ \bibnamefont {Ciappina}}, \bibinfo {author} {\bibfnamefont {P.}~\bibnamefont {Tzallas}},\ and\ \bibinfo {author} {\bibfnamefont {M.}~\bibnamefont {Lewenstein}},\ }\bibfield  {title} {\bibinfo {title} {{High Photon Number Entangled States and Coherent State Superposition from the Extreme Ultraviolet to the Far Infrared}},\ }\href {https://doi.org/10.1103/PhysRevLett.128.123603} {\bibfield  {journal} {\bibinfo  {journal} {Phys. Rev. Lett.}\ }\textbf {\bibinfo {volume} {128}},\ \bibinfo {pages} {123603} (\bibinfo {year} {2022})}\BibitemShut {NoStop}%
\bibitem [{\citenamefont {Rivera-Dean}\ \emph {et~al.}(2025)\citenamefont {Rivera-Dean}, \citenamefont {Lamprou}, \citenamefont {Pisanty}, \citenamefont {Ciappina}, \citenamefont {Tzallas}, \citenamefont {Lewenstein},\ and\ \citenamefont {Stammer}}]{Anonymous2025a}%
  \BibitemOpen
  \bibfield  {author} {\bibinfo {author} {\bibfnamefont {J.}~\bibnamefont {Rivera-Dean}}, \bibinfo {author} {\bibfnamefont {T.}~\bibnamefont {Lamprou}}, \bibinfo {author} {\bibfnamefont {E.}~\bibnamefont {Pisanty}}, \bibinfo {author} {\bibfnamefont {M.~F.}\ \bibnamefont {Ciappina}}, \bibinfo {author} {\bibfnamefont {P.}~\bibnamefont {Tzallas}}, \bibinfo {author} {\bibfnamefont {M.}~\bibnamefont {Lewenstein}},\ and\ \bibinfo {author} {\bibfnamefont {P.}~\bibnamefont {Stammer}},\ }\bibfield  {title} {\bibinfo {title} {{Quantum state engineering of light using intensity measurements and postselection}},\ }\href {https://doi.org/10.1103/11vz-9gcz} {\bibfield  {journal} {\bibinfo  {journal} {Phys. Rev. A}\ }\textbf {\bibinfo {volume} {112}},\ \bibinfo {pages} {013110} (\bibinfo {year} {2025})}\BibitemShut {NoStop}%
\bibitem [{\citenamefont {Stammer}\ \emph {et~al.}(2023)\citenamefont {Stammer}, \citenamefont {Rivera-Dean}, \citenamefont {Maxwell}, \citenamefont {Lamprou}, \citenamefont {Ord{\'{o}}{\~{n}}ez}, \citenamefont {Ciappina}, \citenamefont {Tzallas},\ and\ \citenamefont {Lewenstein}}]{Stammer2023}%
  \BibitemOpen
  \bibfield  {author} {\bibinfo {author} {\bibfnamefont {P.}~\bibnamefont {Stammer}}, \bibinfo {author} {\bibfnamefont {J.}~\bibnamefont {Rivera-Dean}}, \bibinfo {author} {\bibfnamefont {A.}~\bibnamefont {Maxwell}}, \bibinfo {author} {\bibfnamefont {T.}~\bibnamefont {Lamprou}}, \bibinfo {author} {\bibfnamefont {A.}~\bibnamefont {Ord{\'{o}}{\~{n}}ez}}, \bibinfo {author} {\bibfnamefont {M.~F.}\ \bibnamefont {Ciappina}}, \bibinfo {author} {\bibfnamefont {P.}~\bibnamefont {Tzallas}},\ and\ \bibinfo {author} {\bibfnamefont {M.}~\bibnamefont {Lewenstein}},\ }\bibfield  {title} {\bibinfo {title} {{Quantum Electrodynamics of Intense Laser-Matter Interactions: A Tool for Quantum State Engineering}},\ }\href {https://doi.org/10.1103/PRXQuantum.4.010201} {\bibfield  {journal} {\bibinfo  {journal} {PRX Quantum}\ }\textbf {\bibinfo {volume} {4}},\ \bibinfo {pages} {010201} (\bibinfo {year} {2023})}\BibitemShut {NoStop}%
\bibitem [{\citenamefont {Bhattacharya}\ \emph {et~al.}(2023)\citenamefont {Bhattacharya}, \citenamefont {Lamprou}, \citenamefont {Maxwell}, \citenamefont {Ord{\'{o}}{\~{n}}ez}, \citenamefont {Pisanty}, \citenamefont {Rivera-Dean}, \citenamefont {Stammer}, \citenamefont {Ciappina}, \citenamefont {Lewenstein},\ and\ \citenamefont {Tzallas}}]{Helversen2023}%
  \BibitemOpen
  \bibfield  {author} {\bibinfo {author} {\bibfnamefont {U.}~\bibnamefont {Bhattacharya}}, \bibinfo {author} {\bibfnamefont {T.}~\bibnamefont {Lamprou}}, \bibinfo {author} {\bibfnamefont {A.~S.}\ \bibnamefont {Maxwell}}, \bibinfo {author} {\bibfnamefont {A.}~\bibnamefont {Ord{\'{o}}{\~{n}}ez}}, \bibinfo {author} {\bibfnamefont {E.}~\bibnamefont {Pisanty}}, \bibinfo {author} {\bibfnamefont {J.}~\bibnamefont {Rivera-Dean}}, \bibinfo {author} {\bibfnamefont {P.}~\bibnamefont {Stammer}}, \bibinfo {author} {\bibfnamefont {M.~F.}\ \bibnamefont {Ciappina}}, \bibinfo {author} {\bibfnamefont {M.}~\bibnamefont {Lewenstein}},\ and\ \bibinfo {author} {\bibfnamefont {P.}~\bibnamefont {Tzallas}},\ }\bibfield  {title} {\bibinfo {title} {{Strong^^e2^^80^^93laser^^e2^^80^^93field physics, non^^e2^^80^^93classical light states and quantum information science}},\ }\href {https://doi.org/10.1088/1361-6633/acea31} {\bibfield  {journal} {\bibinfo  {journal} {Reports Prog. Phys.}\ }\textbf {\bibinfo {volume} {86}},\ \bibinfo
  {pages} {094401} (\bibinfo {year} {2023})}\BibitemShut {NoStop}%
\bibitem [{\citenamefont {Lewenstein}\ \emph {et~al.}(2024)\citenamefont {Lewenstein}, \citenamefont {Baldelli}, \citenamefont {Bhattacharya}, \citenamefont {Biegert}, \citenamefont {Ciappina}, \citenamefont {Grass}, \citenamefont {Grochowski}, \citenamefont {Johnson}, \citenamefont {Lamprou}, \citenamefont {Maxwell}, \citenamefont {Ord{\'{o}}{\~{n}}ez}, \citenamefont {Pisanty}, \citenamefont {Rivera-Dean}, \citenamefont {Stammer},\ and\ \citenamefont {Tzallas}}]{Lewenstein2024}%
  \BibitemOpen
  \bibfield  {author} {\bibinfo {author} {\bibfnamefont {M.}~\bibnamefont {Lewenstein}}, \bibinfo {author} {\bibfnamefont {N.}~\bibnamefont {Baldelli}}, \bibinfo {author} {\bibfnamefont {U.}~\bibnamefont {Bhattacharya}}, \bibinfo {author} {\bibfnamefont {J.}~\bibnamefont {Biegert}}, \bibinfo {author} {\bibfnamefont {M.~F.}\ \bibnamefont {Ciappina}}, \bibinfo {author} {\bibfnamefont {T.}~\bibnamefont {Grass}}, \bibinfo {author} {\bibfnamefont {P.~T.}\ \bibnamefont {Grochowski}}, \bibinfo {author} {\bibfnamefont {A.~S.}\ \bibnamefont {Johnson}}, \bibinfo {author} {\bibfnamefont {T.}~\bibnamefont {Lamprou}}, \bibinfo {author} {\bibfnamefont {A.~S.}\ \bibnamefont {Maxwell}}, \bibinfo {author} {\bibfnamefont {A.}~\bibnamefont {Ord{\'{o}}{\~{n}}ez}}, \bibinfo {author} {\bibfnamefont {E.}~\bibnamefont {Pisanty}}, \bibinfo {author} {\bibfnamefont {J.}~\bibnamefont {Rivera-Dean}}, \bibinfo {author} {\bibfnamefont {P.}~\bibnamefont {Stammer}},\ and\ \bibinfo {author} {\bibfnamefont {P.}~\bibnamefont {Tzallas}},\
  }\bibfield  {title} {\bibinfo {title} {{Attosecond Physics and Quantum Information Science}},\ }in\ \href {https://doi.org/10.1007/978-3-031-47938-0_4} {\emph {\bibinfo {booktitle} {Springer Proc. Phys.}}}\ (\bibinfo {year} {2024})\ pp.\ \bibinfo {pages} {27--44}\BibitemShut {NoStop}%
\bibitem [{\citenamefont {Cruz-Rodriguez}\ \emph {et~al.}(2024)\citenamefont {Cruz-Rodriguez}, \citenamefont {Dey}, \citenamefont {Freibert},\ and\ \citenamefont {Stammer}}]{Cruz-Rodriguez2024a}%
  \BibitemOpen
  \bibfield  {author} {\bibinfo {author} {\bibfnamefont {L.}~\bibnamefont {Cruz-Rodriguez}}, \bibinfo {author} {\bibfnamefont {D.}~\bibnamefont {Dey}}, \bibinfo {author} {\bibfnamefont {A.}~\bibnamefont {Freibert}},\ and\ \bibinfo {author} {\bibfnamefont {P.}~\bibnamefont {Stammer}},\ }\bibfield  {title} {\bibinfo {title} {{Quantum phenomena in attosecond science}},\ }\href {https://doi.org/10.1038/s42254-024-00769-2} {\bibfield  {journal} {\bibinfo  {journal} {Nat. Rev. Phys.}\ }\textbf {\bibinfo {volume} {6}},\ \bibinfo {pages} {691} (\bibinfo {year} {2024})}\BibitemShut {NoStop}%
\bibitem [{\citenamefont {Vidiella-Barranco}\ \emph {et~al.}(1992)\citenamefont {Vidiella-Barranco}, \citenamefont {Moya-Cessa},\ and\ \citenamefont {Bu{\v{z}}ek}}]{Vidiella-Barranco1992}%
  \BibitemOpen
  \bibfield  {author} {\bibinfo {author} {\bibfnamefont {A.}~\bibnamefont {Vidiella-Barranco}}, \bibinfo {author} {\bibfnamefont {H.}~\bibnamefont {Moya-Cessa}},\ and\ \bibinfo {author} {\bibfnamefont {V.}~\bibnamefont {Bu{\v{z}}ek}},\ }\bibfield  {title} {\bibinfo {title} {{Interaction of Superpositions of Coherent States of Light with Two-level Atoms}},\ }\href {https://doi.org/10.1080/09500349214551481} {\bibfield  {journal} {\bibinfo  {journal} {J. Mod. Opt.}\ }\textbf {\bibinfo {volume} {39}},\ \bibinfo {pages} {1441} (\bibinfo {year} {1992})}\BibitemShut {NoStop}%
\bibitem [{\citenamefont {Gerry}\ and\ \citenamefont {Hach}(1993)}]{Gerry1993}%
  \BibitemOpen
  \bibfield  {author} {\bibinfo {author} {\bibfnamefont {C.~C.}\ \bibnamefont {Gerry}}\ and\ \bibinfo {author} {\bibfnamefont {E.~E.}\ \bibnamefont {Hach}},\ }\bibfield  {title} {\bibinfo {title} {{Interaction of a two-level atom with an even coherent state}},\ }\href {https://doi.org/10.1016/0375-9601(93)91081-F} {\bibfield  {journal} {\bibinfo  {journal} {Phys. Lett. A}\ }\textbf {\bibinfo {volume} {179}},\ \bibinfo {pages} {1} (\bibinfo {year} {1993})}\BibitemShut {NoStop}%
\bibitem [{\citenamefont {Moya-Cessa}\ and\ \citenamefont {Vidiella-Barranco}(1995)}]{Moya-Cessa1995}%
  \BibitemOpen
  \bibfield  {author} {\bibinfo {author} {\bibfnamefont {H.}~\bibnamefont {Moya-Cessa}}\ and\ \bibinfo {author} {\bibfnamefont {A.}~\bibnamefont {Vidiella-Barranco}},\ }\bibfield  {title} {\bibinfo {title} {{On the Interaction of Two-level Atoms with Superpositions of Coherent States of Light}},\ }\href {https://doi.org/10.1080/09500349514551341} {\bibfield  {journal} {\bibinfo  {journal} {J. Mod. Opt.}\ }\textbf {\bibinfo {volume} {42}},\ \bibinfo {pages} {1547} (\bibinfo {year} {1995})}\BibitemShut {NoStop}%
\bibitem [{\citenamefont {Joshi}\ and\ \citenamefont {Singh}(1995)}]{Joshi1995}%
  \BibitemOpen
  \bibfield  {author} {\bibinfo {author} {\bibfnamefont {A.}~\bibnamefont {Joshi}}\ and\ \bibinfo {author} {\bibfnamefont {M.}~\bibnamefont {Singh}},\ }\bibfield  {title} {\bibinfo {title} {{Effects of Even and Odd Coherent States on the Evolution of the Two-photon Jaynes-Cummings model}},\ }\href {https://doi.org/10.1080/713824410} {\bibfield  {journal} {\bibinfo  {journal} {J. Mod. Opt.}\ }\textbf {\bibinfo {volume} {42}},\ \bibinfo {pages} {775} (\bibinfo {year} {1995})}\BibitemShut {NoStop}%
\bibitem [{\citenamefont {Bocanegra-Garay}\ \emph {et~al.}(2024)\citenamefont {Bocanegra-Garay}, \citenamefont {Castillo-Celeita}, \citenamefont {Negro}, \citenamefont {Nieto},\ and\ \citenamefont {G{\'{o}}mez-Ruiz}}]{Bocanegra-Garay2024}%
  \BibitemOpen
  \bibfield  {author} {\bibinfo {author} {\bibfnamefont {I.~A.}\ \bibnamefont {Bocanegra-Garay}}, \bibinfo {author} {\bibfnamefont {M.}~\bibnamefont {Castillo-Celeita}}, \bibinfo {author} {\bibfnamefont {J.}~\bibnamefont {Negro}}, \bibinfo {author} {\bibfnamefont {L.~M.}\ \bibnamefont {Nieto}},\ and\ \bibinfo {author} {\bibfnamefont {F.~J.}\ \bibnamefont {G{\'{o}}mez-Ruiz}},\ }\bibfield  {title} {\bibinfo {title} {{Exploring supersymmetry: Interchangeability between Jaynes-Cummings and anti-Jaynes-Cummings models}},\ }\href {https://doi.org/10.1103/PhysRevResearch.6.043218} {\bibfield  {journal} {\bibinfo  {journal} {Phys. Rev. Res.}\ }\textbf {\bibinfo {volume} {6}},\ \bibinfo {pages} {043218} (\bibinfo {year} {2024})}\BibitemShut {NoStop}%
\bibitem [{\citenamefont {Tang-Kun}(2006)}]{Liu2006}%
  \BibitemOpen
  \bibfield  {author} {\bibinfo {author} {\bibfnamefont {L.}~\bibnamefont {Tang-Kun}},\ }\bibfield  {title} {\bibinfo {title} {{Entropy evolvement properties in a system of Schr{\"{o}}dinger cat state light field interacting with two entangled atoms}},\ }\href {https://doi.org/10.1088/1009-1963/15/3/016} {\bibfield  {journal} {\bibinfo  {journal} {Chinese Phys.}\ }\textbf {\bibinfo {volume} {15}},\ \bibinfo {pages} {542} (\bibinfo {year} {2006})}\BibitemShut {NoStop}%
\bibitem [{\citenamefont {Mohamed}\ \emph {et~al.}(2019)\citenamefont {Mohamed}, \citenamefont {Eleuch},\ and\ \citenamefont {Ooi}}]{Mohamed2019a}%
  \BibitemOpen
  \bibfield  {author} {\bibinfo {author} {\bibfnamefont {A.~B.~A.}\ \bibnamefont {Mohamed}}, \bibinfo {author} {\bibfnamefont {H.}~\bibnamefont {Eleuch}},\ and\ \bibinfo {author} {\bibfnamefont {C.~H.~R.}\ \bibnamefont {Ooi}},\ }\bibfield  {title} {\bibinfo {title} {{Non-locality Correlation in Two Driven Qubits Inside an Open Coherent Cavity: Trace Norm Distance and Maximum Bell Function}},\ }\href {https://doi.org/10.1038/s41598-019-55548-2} {\bibfield  {journal} {\bibinfo  {journal} {Sci. Rep.}\ }\textbf {\bibinfo {volume} {9}},\ \bibinfo {pages} {19632} (\bibinfo {year} {2019})}\BibitemShut {NoStop}%
\bibitem [{\citenamefont {Mohamed}\ \emph {et~al.}(2021)\citenamefont {Mohamed}, \citenamefont {Khalil}, \citenamefont {Selim},\ and\ \citenamefont {Eleuch}}]{Mohamed2021d}%
  \BibitemOpen
  \bibfield  {author} {\bibinfo {author} {\bibfnamefont {A.-B.~A.}\ \bibnamefont {Mohamed}}, \bibinfo {author} {\bibfnamefont {E.~M.}\ \bibnamefont {Khalil}}, \bibinfo {author} {\bibfnamefont {M.~M.}\ \bibnamefont {Selim}},\ and\ \bibinfo {author} {\bibfnamefont {H.}~\bibnamefont {Eleuch}},\ }\bibfield  {title} {\bibinfo {title} {{Quantum Fisher Information and Bures Distance Correlations of Coupled Two Charge-Qubits Inside a Coherent Cavity with the Intrinsic Decoherence}},\ }\href {https://doi.org/10.3390/sym13020352} {\bibfield  {journal} {\bibinfo  {journal} {Symmetry (Basel).}\ }\textbf {\bibinfo {volume} {13}},\ \bibinfo {pages} {352} (\bibinfo {year} {2021})}\BibitemShut {NoStop}%
\bibitem [{\citenamefont {Abdel-Khalek}\ \emph {et~al.}(2021)\citenamefont {Abdel-Khalek}, \citenamefont {Berrada}, \citenamefont {Khalil}, \citenamefont {Eleuch}, \citenamefont {Obada},\ and\ \citenamefont {Reda}}]{Abdel-Khalek2021}%
  \BibitemOpen
  \bibfield  {author} {\bibinfo {author} {\bibfnamefont {S.}~\bibnamefont {Abdel-Khalek}}, \bibinfo {author} {\bibfnamefont {K.}~\bibnamefont {Berrada}}, \bibinfo {author} {\bibfnamefont {E.~M.}\ \bibnamefont {Khalil}}, \bibinfo {author} {\bibfnamefont {H.}~\bibnamefont {Eleuch}}, \bibinfo {author} {\bibfnamefont {A.-S.~F.}\ \bibnamefont {Obada}},\ and\ \bibinfo {author} {\bibfnamefont {E.}~\bibnamefont {Reda}},\ }\bibfield  {title} {\bibinfo {title} {{Tavis^^e2^^80^^93Cummings Model with Moving Atoms}},\ }\href {https://doi.org/10.3390/e23040452} {\bibfield  {journal} {\bibinfo  {journal} {Entropy}\ }\textbf {\bibinfo {volume} {23}},\ \bibinfo {pages} {452} (\bibinfo {year} {2021})}\BibitemShut {NoStop}%
\bibitem [{\citenamefont {Movahedi}\ \emph {et~al.}(2023)\citenamefont {Movahedi}, \citenamefont {Afshar},\ and\ \citenamefont {Jafarpour}}]{Movahedi2023}%
  \BibitemOpen
  \bibfield  {author} {\bibinfo {author} {\bibfnamefont {R.}~\bibnamefont {Movahedi}}, \bibinfo {author} {\bibfnamefont {D.}~\bibnamefont {Afshar}},\ and\ \bibinfo {author} {\bibfnamefont {M.}~\bibnamefont {Jafarpour}},\ }\bibfield  {title} {\bibinfo {title} {{Improvement of the entanglement generation in atomic states using a single-mode field in the Tavis^^e2^^80^^93Cummings model}},\ }\href {https://doi.org/10.1140/epjd/s10053-023-00647-z} {\bibfield  {journal} {\bibinfo  {journal} {Eur. Phys. J. D}\ }\textbf {\bibinfo {volume} {77}},\ \bibinfo {pages} {59} (\bibinfo {year} {2023})}\BibitemShut {NoStop}%
\bibitem [{\citenamefont {Imai}\ \emph {et~al.}()\citenamefont {Imai}, \citenamefont {Ono},\ and\ \citenamefont {Tsuji}}]{Imai2025b}%
  \BibitemOpen
  \bibfield  {author} {\bibinfo {author} {\bibfnamefont {S.}~\bibnamefont {Imai}}, \bibinfo {author} {\bibfnamefont {A.}~\bibnamefont {Ono}},\ and\ \bibinfo {author} {\bibfnamefont {N.}~\bibnamefont {Tsuji}},\ }\bibfield  {title} {\bibinfo {title} {{Electron dynamics induced by quantum cat-state light}},\ }\Eprint {https://arxiv.org/abs/2501.16801} {arXiv:2501.16801} \BibitemShut {NoStop}%
\bibitem [{\citenamefont {Leman}\ \emph {et~al.}(1997)\citenamefont {Leman}, \citenamefont {Yiwen},\ and\ \citenamefont {Molin}}]{Kuang1997}%
  \BibitemOpen
  \bibfield  {author} {\bibinfo {author} {\bibfnamefont {K.}~\bibnamefont {Leman}}, \bibinfo {author} {\bibfnamefont {W.}~\bibnamefont {Yiwen}},\ and\ \bibinfo {author} {\bibfnamefont {G.}~\bibnamefont {Molin}},\ }\bibfield  {title} {\bibinfo {title} {{Supercurrent and Its Quantum Statistical Properties in Mesoscopic Josephson Junction in the Presence of Nonclassical Light Fields}},\ }\href {https://doi.org/10.1088/0253-6102/28/4/391} {\bibfield  {journal} {\bibinfo  {journal} {Commun. Theor. Phys.}\ }\textbf {\bibinfo {volume} {28}},\ \bibinfo {pages} {391} (\bibinfo {year} {1997})}\BibitemShut {NoStop}%
\bibitem [{\citenamefont {Horoshko}\ and\ \citenamefont {{Ya Kilin}}(2000)}]{Horoshko2000}%
  \BibitemOpen
  \bibfield  {author} {\bibinfo {author} {\bibfnamefont {D.~B.}\ \bibnamefont {Horoshko}}\ and\ \bibinfo {author} {\bibfnamefont {S.}~\bibnamefont {{Ya Kilin}}},\ }\bibfield  {title} {\bibinfo {title} {{Resonance fluorescence excited by macroscopic superposition in a feedback loop}},\ }\href {https://doi.org/10.1134/1.559158} {\bibfield  {journal} {\bibinfo  {journal} {J. Exp. Theor. Phys.}\ }\textbf {\bibinfo {volume} {90}},\ \bibinfo {pages} {733} (\bibinfo {year} {2000})}\BibitemShut {NoStop}%
\bibitem [{\citenamefont {Tomilin}\ and\ \citenamefont {Il'ichov}(2016)}]{Tomilin2016}%
  \BibitemOpen
  \bibfield  {author} {\bibinfo {author} {\bibfnamefont {V.}~\bibnamefont {Tomilin}}\ and\ \bibinfo {author} {\bibfnamefont {L.}~\bibnamefont {Il'ichov}},\ }\bibfield  {title} {\bibinfo {title} {{The stationary resonance fluorescence of a two-level atom in a cat-state field}},\ }\href {https://doi.org/10.1016/j.optcom.2016.04.067} {\bibfield  {journal} {\bibinfo  {journal} {Opt. Commun.}\ }\textbf {\bibinfo {volume} {375}},\ \bibinfo {pages} {38} (\bibinfo {year} {2016})}\BibitemShut {NoStop}%
\bibitem [{\citenamefont {Tomilin}\ and\ \citenamefont {Il'ichov}(2017{\natexlab{a}})}]{Tomilin2017}%
  \BibitemOpen
  \bibfield  {author} {\bibinfo {author} {\bibfnamefont {V.~A.}\ \bibnamefont {Tomilin}}\ and\ \bibinfo {author} {\bibfnamefont {L.~V.}\ \bibnamefont {Il'ichov}},\ }\bibfield  {title} {\bibinfo {title} {{Correlations of photoemissions in a multiatomic ensemble driven by a cat-state field}},\ }\href {https://doi.org/10.1103/PhysRevA.96.063805} {\bibfield  {journal} {\bibinfo  {journal} {Phys. Rev. A}\ }\textbf {\bibinfo {volume} {96}},\ \bibinfo {pages} {063805} (\bibinfo {year} {2017}{\natexlab{a}})}\BibitemShut {NoStop}%
\bibitem [{\citenamefont {Tomilin}\ and\ \citenamefont {Il'ichov}(2017{\natexlab{b}})}]{Tomilin2017a}%
  \BibitemOpen
  \bibfield  {author} {\bibinfo {author} {\bibfnamefont {V.~A.}\ \bibnamefont {Tomilin}}\ and\ \bibinfo {author} {\bibfnamefont {L.~V.}\ \bibnamefont {Il'ichov}},\ }\bibfield  {title} {\bibinfo {title} {{Lambda-scheme spectroscopy in the cat-state field}},\ }\href {https://doi.org/10.1134/S1063776117030086} {\bibfield  {journal} {\bibinfo  {journal} {J. Exp. Theor. Phys.}\ }\textbf {\bibinfo {volume} {124}},\ \bibinfo {pages} {707} (\bibinfo {year} {2017}{\natexlab{b}})}\BibitemShut {NoStop}%
\bibitem [{\citenamefont {Bertassoli}\ and\ \citenamefont {Vidiella-Barranco}(2024)}]{Bertassoli2024}%
  \BibitemOpen
  \bibfield  {author} {\bibinfo {author} {\bibfnamefont {J.~L.~T.}\ \bibnamefont {Bertassoli}}\ and\ \bibinfo {author} {\bibfnamefont {A.}~\bibnamefont {Vidiella-Barranco}},\ }\bibfield  {title} {\bibinfo {title} {{Note on the emission spectrum and trapping states in the Jaynes^^e2^^80^^93Cummings model}},\ }\href {https://doi.org/10.1364/JOSAB.524429} {\bibfield  {journal} {\bibinfo  {journal} {J. Opt. Soc. Am. B}\ }\textbf {\bibinfo {volume} {41}},\ \bibinfo {pages} {C199} (\bibinfo {year} {2024})}\BibitemShut {NoStop}%
\bibitem [{\citenamefont {Ling}\ and\ \citenamefont {Guo}(1997)}]{Ling1997}%
  \BibitemOpen
  \bibfield  {author} {\bibinfo {author} {\bibfnamefont {T.}~\bibnamefont {Ling}}\ and\ \bibinfo {author} {\bibfnamefont {G.-C.}\ \bibnamefont {Guo}},\ }\bibfield  {title} {\bibinfo {title} {{Superposition of the atomic Bloch state: preparation method}},\ }\href {https://doi.org/10.1364/JOSAB.14.001537} {\bibfield  {journal} {\bibinfo  {journal} {J. Opt. Soc. Am. B}\ }\textbf {\bibinfo {volume} {14}},\ \bibinfo {pages} {1537} (\bibinfo {year} {1997})}\BibitemShut {NoStop}%
\bibitem [{\citenamefont {Rundle}\ and\ \citenamefont {Everitt}(2021)}]{Rundle2021}%
  \BibitemOpen
  \bibfield  {author} {\bibinfo {author} {\bibfnamefont {R.~P.}\ \bibnamefont {Rundle}}\ and\ \bibinfo {author} {\bibfnamefont {M.~J.}\ \bibnamefont {Everitt}},\ }\bibfield  {title} {\bibinfo {title} {{An informationally complete Wigner function for the Tavis^^e2^^80^^93Cummings model}},\ }\href {https://doi.org/10.1007/s10825-021-01777-6} {\bibfield  {journal} {\bibinfo  {journal} {J. Comput. Electron.}\ }\textbf {\bibinfo {volume} {20}},\ \bibinfo {pages} {2180} (\bibinfo {year} {2021})}\BibitemShut {NoStop}%
\bibitem [{\citenamefont {Zurek}(2003)}]{Zurek2003}%
  \BibitemOpen
  \bibfield  {author} {\bibinfo {author} {\bibfnamefont {W.~H.}\ \bibnamefont {Zurek}},\ }\bibfield  {title} {\bibinfo {title} {{Decoherence, einselection, and the quantum origins of the classical}},\ }\href {https://doi.org/10.1103/RevModPhys.75.715} {\bibfield  {journal} {\bibinfo  {journal} {Rev. Mod. Phys.}\ }\textbf {\bibinfo {volume} {75}},\ \bibinfo {pages} {715} (\bibinfo {year} {2003})}\BibitemShut {NoStop}%
\bibitem [{\citenamefont {Kuzmich}\ \emph {et~al.}(1997)\citenamefont {Kuzmich}, \citenamefont {M{\o}lmer},\ and\ \citenamefont {Polzik}}]{Kuzmich1997}%
  \BibitemOpen
  \bibfield  {author} {\bibinfo {author} {\bibfnamefont {A.}~\bibnamefont {Kuzmich}}, \bibinfo {author} {\bibfnamefont {K.}~\bibnamefont {M{\o}lmer}},\ and\ \bibinfo {author} {\bibfnamefont {E.~S.}\ \bibnamefont {Polzik}},\ }\bibfield  {title} {\bibinfo {title} {{Spin Squeezing in an Ensemble of Atoms Illuminated with Squeezed Light}},\ }\href {https://doi.org/10.1103/PhysRevLett.79.4782} {\bibfield  {journal} {\bibinfo  {journal} {Phys. Rev. Lett.}\ }\textbf {\bibinfo {volume} {79}},\ \bibinfo {pages} {4782} (\bibinfo {year} {1997})}\BibitemShut {NoStop}%
\bibitem [{\citenamefont {Hald}\ \emph {et~al.}(1999)\citenamefont {Hald}, \citenamefont {S{\o}rensen}, \citenamefont {Schori},\ and\ \citenamefont {Polzik}}]{Hald1999}%
  \BibitemOpen
  \bibfield  {author} {\bibinfo {author} {\bibfnamefont {J.}~\bibnamefont {Hald}}, \bibinfo {author} {\bibfnamefont {J.~L.}\ \bibnamefont {S{\o}rensen}}, \bibinfo {author} {\bibfnamefont {C.}~\bibnamefont {Schori}},\ and\ \bibinfo {author} {\bibfnamefont {E.~S.}\ \bibnamefont {Polzik}},\ }\bibfield  {title} {\bibinfo {title} {{Spin Squeezed Atoms: A Macroscopic Entangled Ensemble Created by Light}},\ }\href {https://doi.org/10.1103/PhysRevLett.83.1319} {\bibfield  {journal} {\bibinfo  {journal} {Phys. Rev. Lett.}\ }\textbf {\bibinfo {volume} {83}},\ \bibinfo {pages} {1319} (\bibinfo {year} {1999})}\BibitemShut {NoStop}%
\bibitem [{\citenamefont {Hald}\ and\ \citenamefont {Polzik}(2001)}]{Hald2001}%
  \BibitemOpen
  \bibfield  {author} {\bibinfo {author} {\bibfnamefont {J.}~\bibnamefont {Hald}}\ and\ \bibinfo {author} {\bibfnamefont {E.~S.}\ \bibnamefont {Polzik}},\ }\bibfield  {title} {\bibinfo {title} {{Mapping a quantum state of light onto atoms}},\ }\href {https://doi.org/10.1088/1464-4266/3/1/365} {\bibfield  {journal} {\bibinfo  {journal} {J. Opt. B Quantum Semiclassical Opt.}\ }\textbf {\bibinfo {volume} {3}},\ \bibinfo {pages} {S83} (\bibinfo {year} {2001})}\BibitemShut {NoStop}%
\bibitem [{\citenamefont {Ma}\ \emph {et~al.}(2011)\citenamefont {Ma}, \citenamefont {Wang}, \citenamefont {Sun},\ and\ \citenamefont {Nori}}]{Ma2011c}%
  \BibitemOpen
  \bibfield  {author} {\bibinfo {author} {\bibfnamefont {J.}~\bibnamefont {Ma}}, \bibinfo {author} {\bibfnamefont {X.}~\bibnamefont {Wang}}, \bibinfo {author} {\bibfnamefont {C.~P.}\ \bibnamefont {Sun}},\ and\ \bibinfo {author} {\bibfnamefont {F.}~\bibnamefont {Nori}},\ }\bibfield  {title} {\bibinfo {title} {{Quantum spin squeezing}},\ }\href {https://doi.org/10.1016/j.physrep.2011.08.003} {\bibfield  {journal} {\bibinfo  {journal} {Phys. Rep.}\ }\textbf {\bibinfo {volume} {509}},\ \bibinfo {pages} {89} (\bibinfo {year} {2011})}\BibitemShut {NoStop}%
\bibitem [{\citenamefont {Fr{\"{o}}wis}\ \emph {et~al.}(2018)\citenamefont {Fr{\"{o}}wis}, \citenamefont {Sekatski}, \citenamefont {D{\"{u}}r}, \citenamefont {Gisin},\ and\ \citenamefont {Sangouard}}]{Frowis2018}%
  \BibitemOpen
  \bibfield  {author} {\bibinfo {author} {\bibfnamefont {F.}~\bibnamefont {Fr{\"{o}}wis}}, \bibinfo {author} {\bibfnamefont {P.}~\bibnamefont {Sekatski}}, \bibinfo {author} {\bibfnamefont {W.}~\bibnamefont {D{\"{u}}r}}, \bibinfo {author} {\bibfnamefont {N.}~\bibnamefont {Gisin}},\ and\ \bibinfo {author} {\bibfnamefont {N.}~\bibnamefont {Sangouard}},\ }\bibfield  {title} {\bibinfo {title} {{Macroscopic quantum states: Measures, fragility, and implementations}},\ }\href {https://doi.org/10.1103/RevModPhys.90.025004} {\bibfield  {journal} {\bibinfo  {journal} {Rev. Mod. Phys.}\ }\textbf {\bibinfo {volume} {90}},\ \bibinfo {pages} {025004} (\bibinfo {year} {2018})}\BibitemShut {NoStop}%
\bibitem [{\citenamefont {Shimizu}\ and\ \citenamefont {Miyadera}(2002)}]{Shimizu2002a}%
  \BibitemOpen
  \bibfield  {author} {\bibinfo {author} {\bibfnamefont {A.}~\bibnamefont {Shimizu}}\ and\ \bibinfo {author} {\bibfnamefont {T.}~\bibnamefont {Miyadera}},\ }\bibfield  {title} {\bibinfo {title} {{Stability of Quantum States of Finite Macroscopic Systems against Classical Noises, Perturbations from Environments, and Local Measurements}},\ }\href {https://doi.org/10.1103/PhysRevLett.89.270403} {\bibfield  {journal} {\bibinfo  {journal} {Phys. Rev. Lett.}\ }\textbf {\bibinfo {volume} {89}},\ \bibinfo {pages} {270403} (\bibinfo {year} {2002})}\BibitemShut {NoStop}%
\bibitem [{\citenamefont {Fr{\"{o}}wis}\ and\ \citenamefont {D{\"{u}}r}(2012)}]{Frowis2012}%
  \BibitemOpen
  \bibfield  {author} {\bibinfo {author} {\bibfnamefont {F.}~\bibnamefont {Fr{\"{o}}wis}}\ and\ \bibinfo {author} {\bibfnamefont {W.}~\bibnamefont {D{\"{u}}r}},\ }\bibfield  {title} {\bibinfo {title} {{Measures of macroscopicity for quantum spin systems}},\ }\href {https://doi.org/10.1088/1367-2630/14/9/093039} {\bibfield  {journal} {\bibinfo  {journal} {New J. Phys.}\ }\textbf {\bibinfo {volume} {14}},\ \bibinfo {pages} {093039} (\bibinfo {year} {2012})}\BibitemShut {NoStop}%
\bibitem [{\citenamefont {T{\'{o}}th}(2012)}]{Toth2012}%
  \BibitemOpen
  \bibfield  {author} {\bibinfo {author} {\bibfnamefont {G.}~\bibnamefont {T{\'{o}}th}},\ }\bibfield  {title} {\bibinfo {title} {{Multipartite entanglement and high-precision metrology}},\ }\href {https://doi.org/10.1103/PhysRevA.85.022322} {\bibfield  {journal} {\bibinfo  {journal} {Phys. Rev. A}\ }\textbf {\bibinfo {volume} {85}},\ \bibinfo {pages} {022322} (\bibinfo {year} {2012})}\BibitemShut {NoStop}%
\bibitem [{\citenamefont {Hyllus}\ \emph {et~al.}(2012)\citenamefont {Hyllus}, \citenamefont {Laskowski}, \citenamefont {Krischek}, \citenamefont {Schwemmer}, \citenamefont {Wieczorek}, \citenamefont {Weinfurter}, \citenamefont {Pezz{\'{e}}},\ and\ \citenamefont {Smerzi}}]{Hyllus2012}%
  \BibitemOpen
  \bibfield  {author} {\bibinfo {author} {\bibfnamefont {P.}~\bibnamefont {Hyllus}}, \bibinfo {author} {\bibfnamefont {W.}~\bibnamefont {Laskowski}}, \bibinfo {author} {\bibfnamefont {R.}~\bibnamefont {Krischek}}, \bibinfo {author} {\bibfnamefont {C.}~\bibnamefont {Schwemmer}}, \bibinfo {author} {\bibfnamefont {W.}~\bibnamefont {Wieczorek}}, \bibinfo {author} {\bibfnamefont {H.}~\bibnamefont {Weinfurter}}, \bibinfo {author} {\bibfnamefont {L.}~\bibnamefont {Pezz{\'{e}}}},\ and\ \bibinfo {author} {\bibfnamefont {A.}~\bibnamefont {Smerzi}},\ }\bibfield  {title} {\bibinfo {title} {{Fisher information and multiparticle entanglement}},\ }\href {https://doi.org/10.1103/PhysRevA.85.022321} {\bibfield  {journal} {\bibinfo  {journal} {Phys. Rev. A}\ }\textbf {\bibinfo {volume} {85}},\ \bibinfo {pages} {022321} (\bibinfo {year} {2012})}\BibitemShut {NoStop}%
\bibitem [{\citenamefont {Agarwal}\ \emph {et~al.}(1997)\citenamefont {Agarwal}, \citenamefont {Puri},\ and\ \citenamefont {Singh}}]{Agarwal1997}%
  \BibitemOpen
  \bibfield  {author} {\bibinfo {author} {\bibfnamefont {G.~S.}\ \bibnamefont {Agarwal}}, \bibinfo {author} {\bibfnamefont {R.~R.}\ \bibnamefont {Puri}},\ and\ \bibinfo {author} {\bibfnamefont {R.~P.}\ \bibnamefont {Singh}},\ }\bibfield  {title} {\bibinfo {title} {{Atomic Schr{\"{o}}dinger cat states}},\ }\href {https://doi.org/10.1103/PhysRevA.56.2249} {\bibfield  {journal} {\bibinfo  {journal} {Phys. Rev. A}\ }\textbf {\bibinfo {volume} {56}},\ \bibinfo {pages} {2249} (\bibinfo {year} {1997})}\BibitemShut {NoStop}%
\bibitem [{\citenamefont {Gerry}\ and\ \citenamefont {Grobe}(1997)}]{Gerry1997}%
  \BibitemOpen
  \bibfield  {author} {\bibinfo {author} {\bibfnamefont {C.~C.}\ \bibnamefont {Gerry}}\ and\ \bibinfo {author} {\bibfnamefont {R.}~\bibnamefont {Grobe}},\ }\bibfield  {title} {\bibinfo {title} {{Generation and properties of collective atomic Schr{\"{o}}dinger-cat states}},\ }\href {https://doi.org/10.1103/PhysRevA.56.2390} {\bibfield  {journal} {\bibinfo  {journal} {Phys. Rev. A}\ }\textbf {\bibinfo {volume} {56}},\ \bibinfo {pages} {2390} (\bibinfo {year} {1997})}\BibitemShut {NoStop}%
\bibitem [{\citenamefont {Massar}\ and\ \citenamefont {Polzik}(2003)}]{Massar2003}%
  \BibitemOpen
  \bibfield  {author} {\bibinfo {author} {\bibfnamefont {S.}~\bibnamefont {Massar}}\ and\ \bibinfo {author} {\bibfnamefont {E.~S.}\ \bibnamefont {Polzik}},\ }\bibfield  {title} {\bibinfo {title} {{Generating a Superposition of Spin States in an Atomic Ensemble}},\ }\href {https://doi.org/10.1103/PhysRevLett.91.060401} {\bibfield  {journal} {\bibinfo  {journal} {Phys. Rev. Lett.}\ }\textbf {\bibinfo {volume} {91}},\ \bibinfo {pages} {060401} (\bibinfo {year} {2003})}\BibitemShut {NoStop}%
\bibitem [{\citenamefont {Genes}\ and\ \citenamefont {Berman}(2006)}]{Genes2006}%
  \BibitemOpen
  \bibfield  {author} {\bibinfo {author} {\bibfnamefont {C.}~\bibnamefont {Genes}}\ and\ \bibinfo {author} {\bibfnamefont {P.~R.}\ \bibnamefont {Berman}},\ }\bibfield  {title} {\bibinfo {title} {{Generating conditional atomic entanglement by measuring photon number in a single output channel}},\ }\href {https://doi.org/10.1103/PhysRevA.73.013801} {\bibfield  {journal} {\bibinfo  {journal} {Phys. Rev. A}\ }\textbf {\bibinfo {volume} {73}},\ \bibinfo {pages} {013801} (\bibinfo {year} {2006})}\BibitemShut {NoStop}%
\bibitem [{\citenamefont {Filip}(2008)}]{Filip2008}%
  \BibitemOpen
  \bibfield  {author} {\bibinfo {author} {\bibfnamefont {R.}~\bibnamefont {Filip}},\ }\bibfield  {title} {\bibinfo {title} {{Excess-noise-free recording and uploading of nonclassical states to continuous-variable quantum memory}},\ }\href {https://doi.org/10.1103/PhysRevA.78.012329} {\bibfield  {journal} {\bibinfo  {journal} {Phys. Rev. A}\ }\textbf {\bibinfo {volume} {78}},\ \bibinfo {pages} {012329} (\bibinfo {year} {2008})}\BibitemShut {NoStop}%
\bibitem [{\citenamefont {Lemr}\ and\ \citenamefont {Fiur{\'{a}}{\v{s}}ek}(2009)}]{Lemr2009}%
  \BibitemOpen
  \bibfield  {author} {\bibinfo {author} {\bibfnamefont {K.}~\bibnamefont {Lemr}}\ and\ \bibinfo {author} {\bibfnamefont {J.}~\bibnamefont {Fiur{\'{a}}{\v{s}}ek}},\ }\bibfield  {title} {\bibinfo {title} {{Conditional preparation of arbitrary superpositions of atomic Dicke states}},\ }\href {https://doi.org/10.1103/PhysRevA.79.043808} {\bibfield  {journal} {\bibinfo  {journal} {Phys. Rev. A}\ }\textbf {\bibinfo {volume} {79}},\ \bibinfo {pages} {043808} (\bibinfo {year} {2009})}\BibitemShut {NoStop}%
\bibitem [{\citenamefont {Nielsen}\ \emph {et~al.}(2009)\citenamefont {Nielsen}, \citenamefont {Poulsen}, \citenamefont {Negretti},\ and\ \citenamefont {M{\o}lmer}}]{Nielsen2009}%
  \BibitemOpen
  \bibfield  {author} {\bibinfo {author} {\bibfnamefont {A.~E.~B.}\ \bibnamefont {Nielsen}}, \bibinfo {author} {\bibfnamefont {U.~V.}\ \bibnamefont {Poulsen}}, \bibinfo {author} {\bibfnamefont {A.}~\bibnamefont {Negretti}},\ and\ \bibinfo {author} {\bibfnamefont {K.}~\bibnamefont {M{\o}lmer}},\ }\bibfield  {title} {\bibinfo {title} {{Atomic quantum superposition state generation via optical probing}},\ }\href {https://doi.org/10.1103/PhysRevA.79.023841} {\bibfield  {journal} {\bibinfo  {journal} {Phys. Rev. A}\ }\textbf {\bibinfo {volume} {79}},\ \bibinfo {pages} {023841} (\bibinfo {year} {2009})}\BibitemShut {NoStop}%
\bibitem [{\citenamefont {Christensen}\ \emph {et~al.}(2013)\citenamefont {Christensen}, \citenamefont {B{\'{e}}guin}, \citenamefont {S{\o}rensen}, \citenamefont {Bookjans}, \citenamefont {Oblak}, \citenamefont {M{\"{u}}ller}, \citenamefont {Appel},\ and\ \citenamefont {Polzik}}]{Christensen2013}%
  \BibitemOpen
  \bibfield  {author} {\bibinfo {author} {\bibfnamefont {S.~L.}\ \bibnamefont {Christensen}}, \bibinfo {author} {\bibfnamefont {J.~B.}\ \bibnamefont {B{\'{e}}guin}}, \bibinfo {author} {\bibfnamefont {H.~L.}\ \bibnamefont {S{\o}rensen}}, \bibinfo {author} {\bibfnamefont {E.}~\bibnamefont {Bookjans}}, \bibinfo {author} {\bibfnamefont {D.}~\bibnamefont {Oblak}}, \bibinfo {author} {\bibfnamefont {J.~H.}\ \bibnamefont {M{\"{u}}ller}}, \bibinfo {author} {\bibfnamefont {J.}~\bibnamefont {Appel}},\ and\ \bibinfo {author} {\bibfnamefont {E.~S.}\ \bibnamefont {Polzik}},\ }\bibfield  {title} {\bibinfo {title} {{Toward quantum state tomography of a single polariton state of an atomic ensemble}},\ }\href {https://doi.org/10.1088/1367-2630/15/1/015002} {\bibfield  {journal} {\bibinfo  {journal} {New J. Phys.}\ }\textbf {\bibinfo {volume} {15}},\ \bibinfo {pages} {015002} (\bibinfo {year} {2013})}\BibitemShut {NoStop}%
\bibitem [{\citenamefont {McConnell}\ \emph {et~al.}(2013)\citenamefont {McConnell}, \citenamefont {Zhang}, \citenamefont {{\'{C}}uk}, \citenamefont {Hu}, \citenamefont {Schleier-Smith},\ and\ \citenamefont {Vuleti{\'{c}}}}]{McConnell2013}%
  \BibitemOpen
  \bibfield  {author} {\bibinfo {author} {\bibfnamefont {R.}~\bibnamefont {McConnell}}, \bibinfo {author} {\bibfnamefont {H.}~\bibnamefont {Zhang}}, \bibinfo {author} {\bibfnamefont {S.}~\bibnamefont {{\'{C}}uk}}, \bibinfo {author} {\bibfnamefont {J.}~\bibnamefont {Hu}}, \bibinfo {author} {\bibfnamefont {M.~H.}\ \bibnamefont {Schleier-Smith}},\ and\ \bibinfo {author} {\bibfnamefont {V.}~\bibnamefont {Vuleti{\'{c}}}},\ }\bibfield  {title} {\bibinfo {title} {{Generating entangled spin states for quantum metrology by single-photon detection}},\ }\href {https://doi.org/10.1103/PhysRevA.88.063802} {\bibfield  {journal} {\bibinfo  {journal} {Phys. Rev. A}\ }\textbf {\bibinfo {volume} {88}},\ \bibinfo {pages} {063802} (\bibinfo {year} {2013})}\BibitemShut {NoStop}%
\bibitem [{\citenamefont {McConnell}\ \emph {et~al.}(2015)\citenamefont {McConnell}, \citenamefont {Zhang}, \citenamefont {Hu}, \citenamefont {{\'{C}}uk},\ and\ \citenamefont {Vuleti{\'{c}}}}]{McConnell2015}%
  \BibitemOpen
  \bibfield  {author} {\bibinfo {author} {\bibfnamefont {R.}~\bibnamefont {McConnell}}, \bibinfo {author} {\bibfnamefont {H.}~\bibnamefont {Zhang}}, \bibinfo {author} {\bibfnamefont {J.}~\bibnamefont {Hu}}, \bibinfo {author} {\bibfnamefont {S.}~\bibnamefont {{\'{C}}uk}},\ and\ \bibinfo {author} {\bibfnamefont {V.}~\bibnamefont {Vuleti{\'{c}}}},\ }\bibfield  {title} {\bibinfo {title} {{Entanglement with negative Wigner function of almost 3,000 atoms heralded by one photon}},\ }\href {https://doi.org/10.1038/nature14293} {\bibfield  {journal} {\bibinfo  {journal} {Nature}\ }\textbf {\bibinfo {volume} {519}},\ \bibinfo {pages} {439} (\bibinfo {year} {2015})}\BibitemShut {NoStop}%
\bibitem [{\citenamefont {Huang}\ and\ \citenamefont {Agarwal}(2015)}]{Huang2015}%
  \BibitemOpen
  \bibfield  {author} {\bibinfo {author} {\bibfnamefont {S.}~\bibnamefont {Huang}}\ and\ \bibinfo {author} {\bibfnamefont {G.~S.}\ \bibnamefont {Agarwal}},\ }\bibfield  {title} {\bibinfo {title} {{Weak value amplification of atomic cat states}},\ }\href {https://doi.org/10.1088/1367-2630/17/9/093032} {\bibfield  {journal} {\bibinfo  {journal} {New J. Phys.}\ }\textbf {\bibinfo {volume} {17}},\ \bibinfo {pages} {093032} (\bibinfo {year} {2015})}\BibitemShut {NoStop}%
\bibitem [{\citenamefont {Rabi}(1937)}]{Rabi1937}%
  \BibitemOpen
  \bibfield  {author} {\bibinfo {author} {\bibfnamefont {I.~I.}\ \bibnamefont {Rabi}},\ }\bibfield  {title} {\bibinfo {title} {{Space Quantization in a Gyrating Magnetic Field}},\ }\href {https://doi.org/10.1103/PhysRev.51.652} {\bibfield  {journal} {\bibinfo  {journal} {Phys. Rev.}\ }\textbf {\bibinfo {volume} {51}},\ \bibinfo {pages} {652} (\bibinfo {year} {1937})}\BibitemShut {NoStop}%
\bibitem [{\citenamefont {Dicke}(1954)}]{Dicke1954}%
  \BibitemOpen
  \bibfield  {author} {\bibinfo {author} {\bibfnamefont {R.~H.}\ \bibnamefont {Dicke}},\ }\bibfield  {title} {\bibinfo {title} {{Coherence in Spontaneous Radiation Processes}},\ }\href {https://doi.org/10.1103/PhysRev.93.99} {\bibfield  {journal} {\bibinfo  {journal} {Phys. Rev.}\ }\textbf {\bibinfo {volume} {93}},\ \bibinfo {pages} {99} (\bibinfo {year} {1954})}\BibitemShut {NoStop}%
\bibitem [{\citenamefont {Tavis}\ and\ \citenamefont {Cummings}(1968)}]{Tavis1968}%
  \BibitemOpen
  \bibfield  {author} {\bibinfo {author} {\bibfnamefont {M.}~\bibnamefont {Tavis}}\ and\ \bibinfo {author} {\bibfnamefont {F.~W.}\ \bibnamefont {Cummings}},\ }\bibfield  {title} {\bibinfo {title} {{Exact Solution for an $N$-Molecule---Radiation-Field Hamiltonian}},\ }\href {https://doi.org/10.1103/PhysRev.170.379} {\bibfield  {journal} {\bibinfo  {journal} {Phys. Rev.}\ }\textbf {\bibinfo {volume} {170}},\ \bibinfo {pages} {379} (\bibinfo {year} {1968})}\BibitemShut {NoStop}%
\bibitem [{\citenamefont {Helstrom}(1976)}]{Helstrom1969}%
  \BibitemOpen
  \bibfield  {author} {\bibinfo {author} {\bibfnamefont {C.~W.}\ \bibnamefont {Helstrom}},\ }\href {https://doi.org/10.1007/BF01007479} {\emph {\bibinfo {title} {{Quantum detection and estimation theory}}}}\ (\bibinfo  {publisher} {Academic Press},\ \bibinfo {address} {New York},\ \bibinfo {year} {1976})\BibitemShut {NoStop}%
\bibitem [{\citenamefont {Holevo}(1982)}]{Holevo2011}%
  \BibitemOpen
  \bibfield  {author} {\bibinfo {author} {\bibfnamefont {A.~S.}\ \bibnamefont {Holevo}},\ }\href {https://doi.org/10.1007/978-88-7642-378-9} {\emph {\bibinfo {title} {{Probabilistic and Statistical Aspects of Quantum Theory}}}}\ (\bibinfo  {publisher} {North-Holland},\ \bibinfo {address} {Amsterdam},\ \bibinfo {year} {1982})\BibitemShut {NoStop}%
\bibitem [{\citenamefont {Braunstein}\ and\ \citenamefont {Caves}(1994)}]{Braunstein1994}%
  \BibitemOpen
  \bibfield  {author} {\bibinfo {author} {\bibfnamefont {S.~L.}\ \bibnamefont {Braunstein}}\ and\ \bibinfo {author} {\bibfnamefont {C.~M.}\ \bibnamefont {Caves}},\ }\bibfield  {title} {\bibinfo {title} {{Statistical distance and the geometry of quantum states}},\ }\href {https://doi.org/10.1103/PhysRevLett.72.3439} {\bibfield  {journal} {\bibinfo  {journal} {Phys. Rev. Lett.}\ }\textbf {\bibinfo {volume} {72}},\ \bibinfo {pages} {3439} (\bibinfo {year} {1994})}\BibitemShut {NoStop}%
\bibitem [{\citenamefont {Greenberger}\ \emph {et~al.}(1989)\citenamefont {Greenberger}, \citenamefont {Horne},\ and\ \citenamefont {Zeilinger}}]{Greenberger1989}%
  \BibitemOpen
  \bibfield  {author} {\bibinfo {author} {\bibfnamefont {D.~M.}\ \bibnamefont {Greenberger}}, \bibinfo {author} {\bibfnamefont {M.~A.}\ \bibnamefont {Horne}},\ and\ \bibinfo {author} {\bibfnamefont {A.}~\bibnamefont {Zeilinger}},\ }\bibinfo {title} {Going beyond bell's theorem},\ in\ \href {https://doi.org/10.1007/978-94-017-0849-4_10} {\emph {\bibinfo {booktitle} {Bell's Theorem, Quantum Theory and Conceptions of the Universe}}}\ (\bibinfo  {publisher} {Springer Netherlands},\ \bibinfo {address} {Dordrecht},\ \bibinfo {year} {1989})\ pp.\ \bibinfo {pages} {69--72}\BibitemShut {NoStop}%
\bibitem [{foo({\natexlab{a}})}]{footnote_Fock_Bargmann}%
  \BibitemOpen
  \bibinfo {note} {A related representation is the Fock--Bargmann representation~\cite{Fock1928, Bargmann1961}, where a quantum state is uniquely specified by a holomorphic (Bargmann) function, which in turn defines a canonical coefficient function $f(\alpha)$ in the coherent-state expansion. We also allow more general (non-canonical) $f(\alpha)$ functions that are strongly localized in phase space.}\BibitemShut {Stop}%
\bibitem [{\citenamefont {Gorlach}\ \emph {et~al.}(2023)\citenamefont {Gorlach}, \citenamefont {Tzur}, \citenamefont {Birk}, \citenamefont {Kr{\"{u}}ger}, \citenamefont {Rivera}, \citenamefont {Cohen},\ and\ \citenamefont {Kaminer}}]{Gorlach2022b}%
  \BibitemOpen
  \bibfield  {author} {\bibinfo {author} {\bibfnamefont {A.}~\bibnamefont {Gorlach}}, \bibinfo {author} {\bibfnamefont {M.~E.}\ \bibnamefont {Tzur}}, \bibinfo {author} {\bibfnamefont {M.}~\bibnamefont {Birk}}, \bibinfo {author} {\bibfnamefont {M.}~\bibnamefont {Kr{\"{u}}ger}}, \bibinfo {author} {\bibfnamefont {N.}~\bibnamefont {Rivera}}, \bibinfo {author} {\bibfnamefont {O.}~\bibnamefont {Cohen}},\ and\ \bibinfo {author} {\bibfnamefont {I.}~\bibnamefont {Kaminer}},\ }\bibfield  {title} {\bibinfo {title} {{High-harmonic generation driven by quantum light}},\ }\href {https://doi.org/10.1038/s41567-023-02127-y} {\bibfield  {journal} {\bibinfo  {journal} {Nat. Phys.}\ }\textbf {\bibinfo {volume} {19}},\ \bibinfo {pages} {1689} (\bibinfo {year} {2023})}\BibitemShut {NoStop}%
\bibitem [{\citenamefont {Lvovsky}\ and\ \citenamefont {Raymer}(2009)}]{Lvovsky2009}%
  \BibitemOpen
  \bibfield  {author} {\bibinfo {author} {\bibfnamefont {A.~I.}\ \bibnamefont {Lvovsky}}\ and\ \bibinfo {author} {\bibfnamefont {M.~G.}\ \bibnamefont {Raymer}},\ }\bibfield  {title} {\bibinfo {title} {{Continuous-variable optical quantum-state tomography}},\ }\href {https://doi.org/10.1103/RevModPhys.81.299} {\bibfield  {journal} {\bibinfo  {journal} {Rev. Mod. Phys.}\ }\textbf {\bibinfo {volume} {81}},\ \bibinfo {pages} {299} (\bibinfo {year} {2009})}\BibitemShut {NoStop}%
\bibitem [{\citenamefont {{Mauro D'Ariano}}\ \emph {et~al.}(2003)\citenamefont {{Mauro D'Ariano}}, \citenamefont {Paris},\ and\ \citenamefont {Sacchi}}]{DAriano2003}%
  \BibitemOpen
  \bibfield  {author} {\bibinfo {author} {\bibfnamefont {G.}~\bibnamefont {{Mauro D'Ariano}}}, \bibinfo {author} {\bibfnamefont {M.~G.}\ \bibnamefont {Paris}},\ and\ \bibinfo {author} {\bibfnamefont {M.~F.}\ \bibnamefont {Sacchi}},\ }\bibfield  {title} {\bibinfo {title} {{Quantum Tomography}},\ }in\ \href {https://doi.org/10.1016/S1076-5670(03)80065-4} {\emph {\bibinfo {booktitle} {Adv. Imaging Electron Phys.}}},\ Vol.\ \bibinfo {volume} {128}\ (\bibinfo {year} {2003})\ pp.\ \bibinfo {pages} {205--308}\BibitemShut {NoStop}%
\bibitem [{\citenamefont {Tyc}\ and\ \citenamefont {Sanders}(2004)}]{Tyc2004}%
  \BibitemOpen
  \bibfield  {author} {\bibinfo {author} {\bibfnamefont {T.}~\bibnamefont {Tyc}}\ and\ \bibinfo {author} {\bibfnamefont {B.~C.}\ \bibnamefont {Sanders}},\ }\bibfield  {title} {\bibinfo {title} {{Operational formulation of homodyne detection}},\ }\href {https://doi.org/10.1088/0305-4470/37/29/010} {\bibfield  {journal} {\bibinfo  {journal} {J. Phys. A. Math. Gen.}\ }\textbf {\bibinfo {volume} {37}},\ \bibinfo {pages} {7341} (\bibinfo {year} {2004})}\BibitemShut {NoStop}%
\bibitem [{\citenamefont {Pezz{\`{e}}}\ \emph {et~al.}(2018)\citenamefont {Pezz{\`{e}}}, \citenamefont {Smerzi}, \citenamefont {Oberthaler}, \citenamefont {Schmied},\ and\ \citenamefont {Treutlein}}]{Pezze2018}%
  \BibitemOpen
  \bibfield  {author} {\bibinfo {author} {\bibfnamefont {L.}~\bibnamefont {Pezz{\`{e}}}}, \bibinfo {author} {\bibfnamefont {A.}~\bibnamefont {Smerzi}}, \bibinfo {author} {\bibfnamefont {M.~K.}\ \bibnamefont {Oberthaler}}, \bibinfo {author} {\bibfnamefont {R.}~\bibnamefont {Schmied}},\ and\ \bibinfo {author} {\bibfnamefont {P.}~\bibnamefont {Treutlein}},\ }\bibfield  {title} {\bibinfo {title} {{Quantum metrology with nonclassical states of atomic ensembles}},\ }\href {https://doi.org/10.1103/RevModPhys.90.035005} {\bibfield  {journal} {\bibinfo  {journal} {Rev. Mod. Phys.}\ }\textbf {\bibinfo {volume} {90}},\ \bibinfo {pages} {035005} (\bibinfo {year} {2018})}\BibitemShut {NoStop}%
\bibitem [{\citenamefont {Gessner}\ \emph {et~al.}(2019)\citenamefont {Gessner}, \citenamefont {Smerzi},\ and\ \citenamefont {Pezz{\`{e}}}}]{Gessner2019}%
  \BibitemOpen
  \bibfield  {author} {\bibinfo {author} {\bibfnamefont {M.}~\bibnamefont {Gessner}}, \bibinfo {author} {\bibfnamefont {A.}~\bibnamefont {Smerzi}},\ and\ \bibinfo {author} {\bibfnamefont {L.}~\bibnamefont {Pezz{\`{e}}}},\ }\bibfield  {title} {\bibinfo {title} {{Metrological Nonlinear Squeezing Parameter}},\ }\href {https://doi.org/10.1103/PhysRevLett.122.090503} {\bibfield  {journal} {\bibinfo  {journal} {Phys. Rev. Lett.}\ }\textbf {\bibinfo {volume} {122}},\ \bibinfo {pages} {090503} (\bibinfo {year} {2019})}\BibitemShut {NoStop}%
\bibitem [{\citenamefont {Ren}\ \emph {et~al.}(2021)\citenamefont {Ren}, \citenamefont {Li}, \citenamefont {Smerzi},\ and\ \citenamefont {Gessner}}]{Ren2021c}%
  \BibitemOpen
  \bibfield  {author} {\bibinfo {author} {\bibfnamefont {Z.}~\bibnamefont {Ren}}, \bibinfo {author} {\bibfnamefont {W.}~\bibnamefont {Li}}, \bibinfo {author} {\bibfnamefont {A.}~\bibnamefont {Smerzi}},\ and\ \bibinfo {author} {\bibfnamefont {M.}~\bibnamefont {Gessner}},\ }\bibfield  {title} {\bibinfo {title} {{Metrological Detection of Multipartite Entanglement from Young Diagrams}},\ }\href {https://doi.org/10.1103/PhysRevLett.126.080502} {\bibfield  {journal} {\bibinfo  {journal} {Phys. Rev. Lett.}\ }\textbf {\bibinfo {volume} {126}},\ \bibinfo {pages} {080502} (\bibinfo {year} {2021})}\BibitemShut {NoStop}%
\bibitem [{\citenamefont {Stratonovich}(1956)}]{Stratonovich1956}%
  \BibitemOpen
  \bibfield  {author} {\bibinfo {author} {\bibfnamefont {R.~L.}\ \bibnamefont {Stratonovich}},\ }\bibfield  {title} {\bibinfo {title} {{On Distributions in Representation Space}},\ }\href {http://jetp.ras.ru/cgi-bin/e/index/e/4/6/p891?a=list} {\bibfield  {journal} {\bibinfo  {journal} {Zh. Eksp. Teor. Fiz.}\ }\textbf {\bibinfo {volume} {31}},\ \bibinfo {pages} {1012} (\bibinfo {year} {1956})},\ \bibinfo {note} {[Sov. Phys. JETP 4, 891 (1957)]}\BibitemShut {NoStop}%
\bibitem [{\citenamefont {Brif}\ and\ \citenamefont {Mann}(1999)}]{Brif1999}%
  \BibitemOpen
  \bibfield  {author} {\bibinfo {author} {\bibfnamefont {C.}~\bibnamefont {Brif}}\ and\ \bibinfo {author} {\bibfnamefont {A.}~\bibnamefont {Mann}},\ }\bibfield  {title} {\bibinfo {title} {{Phase-space formulation of quantum mechanics and quantum-state reconstruction for physical systems with Lie-group symmetries}},\ }\href {https://doi.org/10.1103/PhysRevA.59.971} {\bibfield  {journal} {\bibinfo  {journal} {Phys. Rev. A}\ }\textbf {\bibinfo {volume} {59}},\ \bibinfo {pages} {971} (\bibinfo {year} {1999})}\BibitemShut {NoStop}%
\bibitem [{\citenamefont {Rundle}\ \emph {et~al.}(2017)\citenamefont {Rundle}, \citenamefont {Mills}, \citenamefont {Tilma}, \citenamefont {Samson},\ and\ \citenamefont {Everitt}}]{Rundle2017}%
  \BibitemOpen
  \bibfield  {author} {\bibinfo {author} {\bibfnamefont {R.~P.}\ \bibnamefont {Rundle}}, \bibinfo {author} {\bibfnamefont {P.~W.}\ \bibnamefont {Mills}}, \bibinfo {author} {\bibfnamefont {T.}~\bibnamefont {Tilma}}, \bibinfo {author} {\bibfnamefont {J.~H.}\ \bibnamefont {Samson}},\ and\ \bibinfo {author} {\bibfnamefont {M.~J.}\ \bibnamefont {Everitt}},\ }\bibfield  {title} {\bibinfo {title} {{Simple procedure for phase-space measurement and entanglement validation}},\ }\href {https://doi.org/10.1103/PhysRevA.96.022117} {\bibfield  {journal} {\bibinfo  {journal} {Phys. Rev. A}\ }\textbf {\bibinfo {volume} {96}},\ \bibinfo {pages} {022117} (\bibinfo {year} {2017})}\BibitemShut {NoStop}%
\bibitem [{\citenamefont {Davis}\ \emph {et~al.}(2021)\citenamefont {Davis}, \citenamefont {Kumari}, \citenamefont {Mann},\ and\ \citenamefont {Ghose}}]{Davis2021}%
  \BibitemOpen
  \bibfield  {author} {\bibinfo {author} {\bibfnamefont {J.}~\bibnamefont {Davis}}, \bibinfo {author} {\bibfnamefont {M.}~\bibnamefont {Kumari}}, \bibinfo {author} {\bibfnamefont {R.~B.}\ \bibnamefont {Mann}},\ and\ \bibinfo {author} {\bibfnamefont {S.}~\bibnamefont {Ghose}},\ }\bibfield  {title} {\bibinfo {title} {{Wigner negativity in spin-$j$ systems}},\ }\href {https://doi.org/10.1103/PhysRevResearch.3.033134} {\bibfield  {journal} {\bibinfo  {journal} {Phys. Rev. Res.}\ }\textbf {\bibinfo {volume} {3}},\ \bibinfo {pages} {033134} (\bibinfo {year} {2021})}\BibitemShut {NoStop}%
\bibitem [{\citenamefont {Yu}\ and\ \citenamefont {Eberly}(2004)}]{Yu2004}%
  \BibitemOpen
  \bibfield  {author} {\bibinfo {author} {\bibfnamefont {T.}~\bibnamefont {Yu}}\ and\ \bibinfo {author} {\bibfnamefont {J.~H.}\ \bibnamefont {Eberly}},\ }\bibfield  {title} {\bibinfo {title} {{Finite-Time Disentanglement Via Spontaneous Emission}},\ }\href {https://doi.org/10.1103/PhysRevLett.93.140404} {\bibfield  {journal} {\bibinfo  {journal} {Phys. Rev. Lett.}\ }\textbf {\bibinfo {volume} {93}},\ \bibinfo {pages} {140404} (\bibinfo {year} {2004})}\BibitemShut {NoStop}%
\bibitem [{\citenamefont {Ficek}\ and\ \citenamefont {Tana{\'{s}}}(2008)}]{Ficek2008}%
  \BibitemOpen
  \bibfield  {author} {\bibinfo {author} {\bibfnamefont {Z.}~\bibnamefont {Ficek}}\ and\ \bibinfo {author} {\bibfnamefont {R.}~\bibnamefont {Tana{\'{s}}}},\ }\bibfield  {title} {\bibinfo {title} {{Delayed sudden birth of entanglement}},\ }\href {https://doi.org/10.1103/PhysRevA.77.054301} {\bibfield  {journal} {\bibinfo  {journal} {Phys. Rev. A}\ }\textbf {\bibinfo {volume} {77}},\ \bibinfo {pages} {054301} (\bibinfo {year} {2008})}\BibitemShut {NoStop}%
\bibitem [{\citenamefont {Yu}\ and\ \citenamefont {Eberly}(2009)}]{Yu2009}%
  \BibitemOpen
  \bibfield  {author} {\bibinfo {author} {\bibfnamefont {T.}~\bibnamefont {Yu}}\ and\ \bibinfo {author} {\bibfnamefont {J.~H.}\ \bibnamefont {Eberly}},\ }\bibfield  {title} {\bibinfo {title} {{Sudden Death of Entanglement}},\ }\href {https://doi.org/10.1126/science.1167343} {\bibfield  {journal} {\bibinfo  {journal} {Science (80-. ).}\ }\textbf {\bibinfo {volume} {323}},\ \bibinfo {pages} {598} (\bibinfo {year} {2009})}\BibitemShut {NoStop}%
\bibitem [{foo({\natexlab{b}})}]{footnote_squeezing_in_Rabi_collapse}%
  \BibitemOpen
  \bibinfo {note} {As an example of the backaction effect, Figs.~\ref{fig:parity}(c) and (f) show that the QFI density gradually separates for the even- and odd-parity outcomes. A plausible interpretation is that a squeezed single-excitation Dicke-like state appears only for odd-parity outcomes, exhibiting greater quantumness than the unsqueezed one at $t=0^+$. Such spin squeezing is known to develop in coherent-field^^e2^^80^^93driven Tavis^^e2^^80^^93Cummings dynamics~\cite{Genes2003}.}\BibitemShut {Stop}%
\bibitem [{\citenamefont {Eberly}\ \emph {et~al.}(1980)\citenamefont {Eberly}, \citenamefont {Narozhny},\ and\ \citenamefont {Sanchez-Mondragon}}]{Eberly1980}%
  \BibitemOpen
  \bibfield  {author} {\bibinfo {author} {\bibfnamefont {J.~H.}\ \bibnamefont {Eberly}}, \bibinfo {author} {\bibfnamefont {N.~B.}\ \bibnamefont {Narozhny}},\ and\ \bibinfo {author} {\bibfnamefont {J.~J.}\ \bibnamefont {Sanchez-Mondragon}},\ }\bibfield  {title} {\bibinfo {title} {{Periodic Spontaneous Collapse and Revival in a Simple Quantum Model}},\ }\href {https://doi.org/10.1103/PhysRevLett.44.1323} {\bibfield  {journal} {\bibinfo  {journal} {Phys. Rev. Lett.}\ }\textbf {\bibinfo {volume} {44}},\ \bibinfo {pages} {1323} (\bibinfo {year} {1980})}\BibitemShut {NoStop}%
\bibitem [{\citenamefont {Gea-Banacloche}(1991)}]{Gea-Banacloche1991}%
  \BibitemOpen
  \bibfield  {author} {\bibinfo {author} {\bibfnamefont {J.}~\bibnamefont {Gea-Banacloche}},\ }\bibfield  {title} {\bibinfo {title} {{Atom- and field-state evolution in the Jaynes-Cummings model for large initial fields}},\ }\href {https://doi.org/10.1103/PhysRevA.44.5913} {\bibfield  {journal} {\bibinfo  {journal} {Phys. Rev. A}\ }\textbf {\bibinfo {volume} {44}},\ \bibinfo {pages} {5913} (\bibinfo {year} {1991})}\BibitemShut {NoStop}%
\bibitem [{foo({\natexlab{c}})}]{footnote_collapse_of_Rabi}%
  \BibitemOpen
  \bibinfo {note} {For $N=1$ and an initial coherent state $|\alpha_0\rangle$ ($\alpha_0 \in \mathbb{R}$), the next Born iteration (first backaction correction) yields the Heisenberg-picture field operator as $\hat{a}_{\mathrm{H,1BA}}(t)\approx \left[ \hat{a} - \mathrm{i}(\gamma t/4)\,\hat{\sigma}^{y} \right]\,\mathrm{e}^{-\mathrm{i}\omega t}$ (keeping only the secular term linear in $t$). Equivalently, the Schr{\"{o}}dinger-picture state is $|\Psi_{\mathrm{1BA}}(t)\rangle \approx \hat{U}_{\mathrm{XFA}}(t)\, \hat{D}\left(-\mathrm{i}(\gamma t/4)\hat{\sigma}^{y}\right)\, |{\downarrow}\rangle|\alpha_0\rangle$, with $\hat{U}_{\mathrm{XFA}}(t)$ the XFA time-evolution operator and $\hat{D}(\alpha)=\exp(\alpha \hat{a}^{\dagger}-\overline{\alpha} \hat{a})$ the displacement operator. The additional factor $\hat{D}\left(-\mathrm{i}(\gamma t/4)\hat\sigma^{y} \right)$ produces a qubit-state--conditioned displacement of the light mode, implying light--matter entanglement even for a coherent-state drive and yielding a
  Gaussian collapse for the population dynamics, $\langle \hat{\sigma}^{z}(t)\rangle=-\cos(\Omega_{\mathrm{eff}} t)\exp[-(\gamma\omega t)^2/8]$, consistent with the $1/|\alpha_0|$ expansion of the exact solution in Ref.~\cite{Gea-Banacloche1991}.}\BibitemShut {Stop}%
\bibitem [{\citenamefont {Dominici}\ \emph {et~al.}(2014)\citenamefont {Dominici}, \citenamefont {Colas}, \citenamefont {Donati}, \citenamefont {{Restrepo Cuartas}}, \citenamefont {{De Giorgi}}, \citenamefont {Ballarini}, \citenamefont {Guirales}, \citenamefont {{L{\'{o}}pez Carre{\~{n}}o}}, \citenamefont {Bramati}, \citenamefont {Gigli}, \citenamefont {del Valle}, \citenamefont {Laussy},\ and\ \citenamefont {Sanvitto}}]{Dominici2014}%
  \BibitemOpen
  \bibfield  {author} {\bibinfo {author} {\bibfnamefont {L.}~\bibnamefont {Dominici}}, \bibinfo {author} {\bibfnamefont {D.}~\bibnamefont {Colas}}, \bibinfo {author} {\bibfnamefont {S.}~\bibnamefont {Donati}}, \bibinfo {author} {\bibfnamefont {J.~P.}\ \bibnamefont {{Restrepo Cuartas}}}, \bibinfo {author} {\bibfnamefont {M.}~\bibnamefont {{De Giorgi}}}, \bibinfo {author} {\bibfnamefont {D.}~\bibnamefont {Ballarini}}, \bibinfo {author} {\bibfnamefont {G.}~\bibnamefont {Guirales}}, \bibinfo {author} {\bibfnamefont {J.~C.}\ \bibnamefont {{L{\'{o}}pez Carre{\~{n}}o}}}, \bibinfo {author} {\bibfnamefont {A.}~\bibnamefont {Bramati}}, \bibinfo {author} {\bibfnamefont {G.}~\bibnamefont {Gigli}}, \bibinfo {author} {\bibfnamefont {E.}~\bibnamefont {del Valle}}, \bibinfo {author} {\bibfnamefont {F.~P.}\ \bibnamefont {Laussy}},\ and\ \bibinfo {author} {\bibfnamefont {D.}~\bibnamefont {Sanvitto}},\ }\bibfield  {title} {\bibinfo {title} {{Ultrafast Control and Rabi Oscillations of Polaritons}},\ }\href
  {https://doi.org/10.1103/PhysRevLett.113.226401} {\bibfield  {journal} {\bibinfo  {journal} {Phys. Rev. Lett.}\ }\textbf {\bibinfo {volume} {113}},\ \bibinfo {pages} {226401} (\bibinfo {year} {2014})}\BibitemShut {NoStop}%
\bibitem [{\citenamefont {Lim}\ \emph {et~al.}(2014)\citenamefont {Lim}, \citenamefont {Lee}, \citenamefont {Lee}, \citenamefont {Park},\ and\ \citenamefont {Ahn}}]{Lim2014}%
  \BibitemOpen
  \bibfield  {author} {\bibinfo {author} {\bibfnamefont {J.}~\bibnamefont {Lim}}, \bibinfo {author} {\bibfnamefont {H.-g.}\ \bibnamefont {Lee}}, \bibinfo {author} {\bibfnamefont {S.}~\bibnamefont {Lee}}, \bibinfo {author} {\bibfnamefont {C.-Y.}\ \bibnamefont {Park}},\ and\ \bibinfo {author} {\bibfnamefont {J.}~\bibnamefont {Ahn}},\ }\bibfield  {title} {\bibinfo {title} {{Ultrafast Ramsey interferometry to implement cold atomic qubit gates}},\ }\href {https://doi.org/10.1038/srep05867} {\bibfield  {journal} {\bibinfo  {journal} {Sci. Rep.}\ }\textbf {\bibinfo {volume} {4}},\ \bibinfo {pages} {5867} (\bibinfo {year} {2014})}\BibitemShut {NoStop}%
\bibitem [{\citenamefont {F{\"{u}}rst}\ \emph {et~al.}(1997)\citenamefont {F{\"{u}}rst}, \citenamefont {Leitenstorfer}, \citenamefont {Nutsch}, \citenamefont {Tr{\"{a}}nkle},\ and\ \citenamefont {Zrenner}}]{Furst1997}%
  \BibitemOpen
  \bibfield  {author} {\bibinfo {author} {\bibfnamefont {C.}~\bibnamefont {F{\"{u}}rst}}, \bibinfo {author} {\bibfnamefont {A.}~\bibnamefont {Leitenstorfer}}, \bibinfo {author} {\bibfnamefont {A.}~\bibnamefont {Nutsch}}, \bibinfo {author} {\bibfnamefont {G.}~\bibnamefont {Tr{\"{a}}nkle}},\ and\ \bibinfo {author} {\bibfnamefont {A.}~\bibnamefont {Zrenner}},\ }\bibfield  {title} {\bibinfo {title} {{Ultrafast Rabi Oscillations of Free-Carrier Transitions in InP}},\ }\href {https://doi.org/10.1002/1521-3951(199711)204:1<20::AID-PSSB20>3.0.CO;2-4} {\bibfield  {journal} {\bibinfo  {journal} {phys. stat. sol. (b)}\ }\textbf {\bibinfo {volume} {204}},\ \bibinfo {pages} {20} (\bibinfo {year} {1997})}\BibitemShut {NoStop}%
\bibitem [{\citenamefont {Hauke}\ \emph {et~al.}(2016)\citenamefont {Hauke}, \citenamefont {Heyl}, \citenamefont {Tagliacozzo},\ and\ \citenamefont {Zoller}}]{Hauke2016}%
  \BibitemOpen
  \bibfield  {author} {\bibinfo {author} {\bibfnamefont {P.}~\bibnamefont {Hauke}}, \bibinfo {author} {\bibfnamefont {M.}~\bibnamefont {Heyl}}, \bibinfo {author} {\bibfnamefont {L.}~\bibnamefont {Tagliacozzo}},\ and\ \bibinfo {author} {\bibfnamefont {P.}~\bibnamefont {Zoller}},\ }\bibfield  {title} {\bibinfo {title} {{Measuring multipartite entanglement through dynamic susceptibilities}},\ }\href {https://doi.org/10.1038/nphys3700} {\bibfield  {journal} {\bibinfo  {journal} {Nat. Phys.}\ }\textbf {\bibinfo {volume} {12}},\ \bibinfo {pages} {778} (\bibinfo {year} {2016})}\BibitemShut {NoStop}%
\bibitem [{\citenamefont {Hales}\ \emph {et~al.}(2023)\citenamefont {Hales}, \citenamefont {Bajpai}, \citenamefont {Liu}, \citenamefont {Baykusheva}, \citenamefont {Li}, \citenamefont {Mitrano},\ and\ \citenamefont {Wang}}]{Hales2023}%
  \BibitemOpen
  \bibfield  {author} {\bibinfo {author} {\bibfnamefont {J.}~\bibnamefont {Hales}}, \bibinfo {author} {\bibfnamefont {U.}~\bibnamefont {Bajpai}}, \bibinfo {author} {\bibfnamefont {T.}~\bibnamefont {Liu}}, \bibinfo {author} {\bibfnamefont {D.~R.}\ \bibnamefont {Baykusheva}}, \bibinfo {author} {\bibfnamefont {M.}~\bibnamefont {Li}}, \bibinfo {author} {\bibfnamefont {M.}~\bibnamefont {Mitrano}},\ and\ \bibinfo {author} {\bibfnamefont {Y.}~\bibnamefont {Wang}},\ }\bibfield  {title} {\bibinfo {title} {{Witnessing light-driven entanglement using time-resolved resonant inelastic X-ray scattering}},\ }\href {https://doi.org/10.1038/s41467-023-38540-3} {\bibfield  {journal} {\bibinfo  {journal} {Nat. Commun.}\ }\textbf {\bibinfo {volume} {14}},\ \bibinfo {pages} {3512} (\bibinfo {year} {2023})}\BibitemShut {NoStop}%
\bibitem [{\citenamefont {Pizzi}\ \emph {et~al.}(2023)\citenamefont {Pizzi}, \citenamefont {Gorlach}, \citenamefont {Rivera}, \citenamefont {Nunnenkamp},\ and\ \citenamefont {Kaminer}}]{Pizzi2022}%
  \BibitemOpen
  \bibfield  {author} {\bibinfo {author} {\bibfnamefont {A.}~\bibnamefont {Pizzi}}, \bibinfo {author} {\bibfnamefont {A.}~\bibnamefont {Gorlach}}, \bibinfo {author} {\bibfnamefont {N.}~\bibnamefont {Rivera}}, \bibinfo {author} {\bibfnamefont {A.}~\bibnamefont {Nunnenkamp}},\ and\ \bibinfo {author} {\bibfnamefont {I.}~\bibnamefont {Kaminer}},\ }\bibfield  {title} {\bibinfo {title} {{Light emission from strongly driven many-body systems}},\ }\href {https://doi.org/10.1038/s41567-022-01910-7} {\bibfield  {journal} {\bibinfo  {journal} {Nat. Phys.}\ }\textbf {\bibinfo {volume} {19}},\ \bibinfo {pages} {551} (\bibinfo {year} {2023})}\BibitemShut {NoStop}%
\bibitem [{\citenamefont {{Even Tzur}}\ \emph {et~al.}(2023)\citenamefont {{Even Tzur}}, \citenamefont {Birk}, \citenamefont {Gorlach}, \citenamefont {Kr{\"{u}}ger}, \citenamefont {Kaminer},\ and\ \citenamefont {Cohen}}]{Tzur2022a}%
  \BibitemOpen
  \bibfield  {author} {\bibinfo {author} {\bibfnamefont {M.}~\bibnamefont {{Even Tzur}}}, \bibinfo {author} {\bibfnamefont {M.}~\bibnamefont {Birk}}, \bibinfo {author} {\bibfnamefont {A.}~\bibnamefont {Gorlach}}, \bibinfo {author} {\bibfnamefont {M.}~\bibnamefont {Kr{\"{u}}ger}}, \bibinfo {author} {\bibfnamefont {I.}~\bibnamefont {Kaminer}},\ and\ \bibinfo {author} {\bibfnamefont {O.}~\bibnamefont {Cohen}},\ }\bibfield  {title} {\bibinfo {title} {{Photon-statistics force in ultrafast electron dynamics}},\ }\href {https://doi.org/10.1038/s41566-023-01209-w} {\bibfield  {journal} {\bibinfo  {journal} {Nat. Photonics}\ }\textbf {\bibinfo {volume} {17}},\ \bibinfo {pages} {501} (\bibinfo {year} {2023})}\BibitemShut {NoStop}%
\bibitem [{\citenamefont {Tzur}\ \emph {et~al.}(2024)\citenamefont {Tzur}, \citenamefont {Birk}, \citenamefont {Gorlach}, \citenamefont {Kaminer}, \citenamefont {Kr{\"{u}}ger},\ and\ \citenamefont {Cohen}}]{Tzur2024}%
  \BibitemOpen
  \bibfield  {author} {\bibinfo {author} {\bibfnamefont {M.~E.}\ \bibnamefont {Tzur}}, \bibinfo {author} {\bibfnamefont {M.}~\bibnamefont {Birk}}, \bibinfo {author} {\bibfnamefont {A.}~\bibnamefont {Gorlach}}, \bibinfo {author} {\bibfnamefont {I.}~\bibnamefont {Kaminer}}, \bibinfo {author} {\bibfnamefont {M.}~\bibnamefont {Kr{\"{u}}ger}},\ and\ \bibinfo {author} {\bibfnamefont {O.}~\bibnamefont {Cohen}},\ }\bibfield  {title} {\bibinfo {title} {{Generation of squeezed high-order harmonics}},\ }\href {https://doi.org/10.1103/PhysRevResearch.6.033079} {\bibfield  {journal} {\bibinfo  {journal} {Phys. Rev. Res.}\ }\textbf {\bibinfo {volume} {6}},\ \bibinfo {pages} {033079} (\bibinfo {year} {2024})}\BibitemShut {NoStop}%
\bibitem [{\citenamefont {Rasputnyi}\ \emph {et~al.}(2024)\citenamefont {Rasputnyi}, \citenamefont {Chen}, \citenamefont {Birk}, \citenamefont {Cohen}, \citenamefont {Kaminer}, \citenamefont {Kr{\"{u}}ger}, \citenamefont {Seletskiy}, \citenamefont {Chekhova},\ and\ \citenamefont {Tani}}]{Rasputnyi2024}%
  \BibitemOpen
  \bibfield  {author} {\bibinfo {author} {\bibfnamefont {A.}~\bibnamefont {Rasputnyi}}, \bibinfo {author} {\bibfnamefont {Z.}~\bibnamefont {Chen}}, \bibinfo {author} {\bibfnamefont {M.}~\bibnamefont {Birk}}, \bibinfo {author} {\bibfnamefont {O.}~\bibnamefont {Cohen}}, \bibinfo {author} {\bibfnamefont {I.}~\bibnamefont {Kaminer}}, \bibinfo {author} {\bibfnamefont {M.}~\bibnamefont {Kr{\"{u}}ger}}, \bibinfo {author} {\bibfnamefont {D.}~\bibnamefont {Seletskiy}}, \bibinfo {author} {\bibfnamefont {M.}~\bibnamefont {Chekhova}},\ and\ \bibinfo {author} {\bibfnamefont {F.}~\bibnamefont {Tani}},\ }\bibfield  {title} {\bibinfo {title} {{High-harmonic generation by a bright squeezed vacuum}},\ }\href {https://doi.org/10.1038/s41567-024-02659-x} {\bibfield  {journal} {\bibinfo  {journal} {Nat. Phys.}\ }\textbf {\bibinfo {volume} {20}},\ \bibinfo {pages} {1960} (\bibinfo {year} {2024})}\BibitemShut {NoStop}%
\bibitem [{\citenamefont {{Even Tzur}}\ and\ \citenamefont {Cohen}(2024)}]{EvenTzur2024}%
  \BibitemOpen
  \bibfield  {author} {\bibinfo {author} {\bibfnamefont {M.}~\bibnamefont {{Even Tzur}}}\ and\ \bibinfo {author} {\bibfnamefont {O.}~\bibnamefont {Cohen}},\ }\bibfield  {title} {\bibinfo {title} {{Motion of charged particles in bright squeezed vacuum}},\ }\href {https://doi.org/10.1038/s41377-024-01381-w} {\bibfield  {journal} {\bibinfo  {journal} {Light Sci. Appl.}\ }\textbf {\bibinfo {volume} {13}},\ \bibinfo {pages} {41} (\bibinfo {year} {2024})}\BibitemShut {NoStop}%
\bibitem [{\citenamefont {Wang}\ \emph {et~al.}(2024)\citenamefont {Wang}, \citenamefont {Yu}, \citenamefont {Lai},\ and\ \citenamefont {Liu}}]{Wang2024h}%
  \BibitemOpen
  \bibfield  {author} {\bibinfo {author} {\bibfnamefont {S.~J.}\ \bibnamefont {Wang}}, \bibinfo {author} {\bibfnamefont {S.~G.}\ \bibnamefont {Yu}}, \bibinfo {author} {\bibfnamefont {X.~Y.}\ \bibnamefont {Lai}},\ and\ \bibinfo {author} {\bibfnamefont {X.~J.}\ \bibnamefont {Liu}},\ }\bibfield  {title} {\bibinfo {title} {{High harmonic generation from an atom in a squeezed-vacuum environment}},\ }\href {https://doi.org/10.1103/PhysRevResearch.6.033010} {\bibfield  {journal} {\bibinfo  {journal} {Phys. Rev. Res.}\ }\textbf {\bibinfo {volume} {6}},\ \bibinfo {pages} {033010} (\bibinfo {year} {2024})}\BibitemShut {NoStop}%
\bibitem [{\citenamefont {Lemieux}\ \emph {et~al.}(2025)\citenamefont {Lemieux}, \citenamefont {Jalil}, \citenamefont {Purschke}, \citenamefont {Boroumand}, \citenamefont {Hammond}, \citenamefont {Villeneuve}, \citenamefont {Naumov}, \citenamefont {Brabec},\ and\ \citenamefont {Vampa}}]{Lemieux2025}%
  \BibitemOpen
  \bibfield  {author} {\bibinfo {author} {\bibfnamefont {S.}~\bibnamefont {Lemieux}}, \bibinfo {author} {\bibfnamefont {S.~A.}\ \bibnamefont {Jalil}}, \bibinfo {author} {\bibfnamefont {D.~N.}\ \bibnamefont {Purschke}}, \bibinfo {author} {\bibfnamefont {N.}~\bibnamefont {Boroumand}}, \bibinfo {author} {\bibfnamefont {T.~J.}\ \bibnamefont {Hammond}}, \bibinfo {author} {\bibfnamefont {D.}~\bibnamefont {Villeneuve}}, \bibinfo {author} {\bibfnamefont {A.}~\bibnamefont {Naumov}}, \bibinfo {author} {\bibfnamefont {T.}~\bibnamefont {Brabec}},\ and\ \bibinfo {author} {\bibfnamefont {G.}~\bibnamefont {Vampa}},\ }\bibfield  {title} {\bibinfo {title} {{Photon bunching in high-harmonic emission controlled by quantum light}},\ }\href {https://doi.org/10.1038/s41566-025-01673-6} {\bibfield  {journal} {\bibinfo  {journal} {Nat. Photonics}\ }\textbf {\bibinfo {volume} {19}},\ \bibinfo {pages} {767} (\bibinfo {year} {2025})}\BibitemShut {NoStop}%
\bibitem [{\citenamefont {Gothelf}\ \emph {et~al.}(2025)\citenamefont {Gothelf}, \citenamefont {Lange},\ and\ \citenamefont {Madsen}}]{Gothelf2025}%
  \BibitemOpen
  \bibfield  {author} {\bibinfo {author} {\bibfnamefont {R.~V.}\ \bibnamefont {Gothelf}}, \bibinfo {author} {\bibfnamefont {C.~S.}\ \bibnamefont {Lange}},\ and\ \bibinfo {author} {\bibfnamefont {L.~B.}\ \bibnamefont {Madsen}},\ }\bibfield  {title} {\bibinfo {title} {{High-order harmonic generation in a crystal driven by quantum light}},\ }\href {https://doi.org/10.1103/PhysRevA.111.063105} {\bibfield  {journal} {\bibinfo  {journal} {Phys. Rev. A}\ }\textbf {\bibinfo {volume} {111}},\ \bibinfo {pages} {063105} (\bibinfo {year} {2025})}\BibitemShut {NoStop}%
\bibitem [{\citenamefont {Gerry}\ and\ \citenamefont {Mimih}(2010)}]{Gerry2010}%
  \BibitemOpen
  \bibfield  {author} {\bibinfo {author} {\bibfnamefont {C.~C.}\ \bibnamefont {Gerry}}\ and\ \bibinfo {author} {\bibfnamefont {J.}~\bibnamefont {Mimih}},\ }\bibfield  {title} {\bibinfo {title} {{The parity operator in quantum optical metrology}},\ }\href {https://doi.org/10.1080/00107514.2010.509995} {\bibfield  {journal} {\bibinfo  {journal} {Contemp. Phys.}\ }\textbf {\bibinfo {volume} {51}},\ \bibinfo {pages} {497} (\bibinfo {year} {2010})}\BibitemShut {NoStop}%
\bibitem [{\citenamefont {Birrittella}\ \emph {et~al.}(2021)\citenamefont {Birrittella}, \citenamefont {Alsing},\ and\ \citenamefont {Gerry}}]{Birrittella2021}%
  \BibitemOpen
  \bibfield  {author} {\bibinfo {author} {\bibfnamefont {R.~J.}\ \bibnamefont {Birrittella}}, \bibinfo {author} {\bibfnamefont {P.~M.}\ \bibnamefont {Alsing}},\ and\ \bibinfo {author} {\bibfnamefont {C.~C.}\ \bibnamefont {Gerry}},\ }\bibfield  {title} {\bibinfo {title} {{The parity operator: Applications in quantum metrology}},\ }\href {https://doi.org/10.1116/5.0026148} {\bibfield  {journal} {\bibinfo  {journal} {AVS Quantum Sci.}\ }\textbf {\bibinfo {volume} {3}},\ \bibinfo {pages} {014701} (\bibinfo {year} {2021})}\BibitemShut {NoStop}%
\bibitem [{\citenamefont {Xu}\ \emph {et~al.}(2024)\citenamefont {Xu}, \citenamefont {Sun}, \citenamefont {Chen}, \citenamefont {Rajteri}, \citenamefont {Garrone}, \citenamefont {Pepe}, \citenamefont {Li}, \citenamefont {Li}, \citenamefont {Zhang}, \citenamefont {Bu}, \citenamefont {Gao}, \citenamefont {Sun},\ and\ \citenamefont {Wang}}]{Xu2024a}%
  \BibitemOpen
  \bibfield  {author} {\bibinfo {author} {\bibfnamefont {X.}~\bibnamefont {Xu}}, \bibinfo {author} {\bibfnamefont {X.}~\bibnamefont {Sun}}, \bibinfo {author} {\bibfnamefont {J.}~\bibnamefont {Chen}}, \bibinfo {author} {\bibfnamefont {M.}~\bibnamefont {Rajteri}}, \bibinfo {author} {\bibfnamefont {H.}~\bibnamefont {Garrone}}, \bibinfo {author} {\bibfnamefont {C.}~\bibnamefont {Pepe}}, \bibinfo {author} {\bibfnamefont {W.}~\bibnamefont {Li}}, \bibinfo {author} {\bibfnamefont {J.}~\bibnamefont {Li}}, \bibinfo {author} {\bibfnamefont {M.}~\bibnamefont {Zhang}}, \bibinfo {author} {\bibfnamefont {T.}~\bibnamefont {Bu}}, \bibinfo {author} {\bibfnamefont {Y.}~\bibnamefont {Gao}}, \bibinfo {author} {\bibfnamefont {T.}~\bibnamefont {Sun}},\ and\ \bibinfo {author} {\bibfnamefont {X.}~\bibnamefont {Wang}},\ }\bibfield  {title} {\bibinfo {title} {{Development of Ti/Au Transition-Edge Sensors for Single-Photon Detection}},\ }\href {https://doi.org/10.1109/TASC.2024.3350569} {\bibfield  {journal} {\bibinfo  {journal} {IEEE
  Trans. Appl. Supercond.}\ }\textbf {\bibinfo {volume} {34}},\ \bibinfo {pages} {1} (\bibinfo {year} {2024})}\BibitemShut {NoStop}%
\bibitem [{\citenamefont {Song}\ \emph {et~al.}(2025)\citenamefont {Song}, \citenamefont {Barberena}, \citenamefont {Young}, \citenamefont {Chaparro}, \citenamefont {Chu}, \citenamefont {Agarwal}, \citenamefont {Niu}, \citenamefont {Young}, \citenamefont {Rey},\ and\ \citenamefont {Thompson}}]{Song2025}%
  \BibitemOpen
  \bibfield  {author} {\bibinfo {author} {\bibfnamefont {E.~Y.}\ \bibnamefont {Song}}, \bibinfo {author} {\bibfnamefont {D.}~\bibnamefont {Barberena}}, \bibinfo {author} {\bibfnamefont {D.~J.}\ \bibnamefont {Young}}, \bibinfo {author} {\bibfnamefont {E.}~\bibnamefont {Chaparro}}, \bibinfo {author} {\bibfnamefont {A.}~\bibnamefont {Chu}}, \bibinfo {author} {\bibfnamefont {S.}~\bibnamefont {Agarwal}}, \bibinfo {author} {\bibfnamefont {Z.}~\bibnamefont {Niu}}, \bibinfo {author} {\bibfnamefont {J.~T.}\ \bibnamefont {Young}}, \bibinfo {author} {\bibfnamefont {A.~M.}\ \bibnamefont {Rey}},\ and\ \bibinfo {author} {\bibfnamefont {J.~K.}\ \bibnamefont {Thompson}},\ }\bibfield  {title} {\bibinfo {title} {{A dissipation-induced superradiant transition in a strontium cavity-QED system}},\ }\href {https://doi.org/10.1126/sciadv.adu5799} {\bibfield  {journal} {\bibinfo  {journal} {Sci. Adv.}\ }\textbf {\bibinfo {volume} {11}},\ \bibinfo {pages} {eadu5799} (\bibinfo {year} {2025})}\BibitemShut {NoStop}%
\bibitem [{\citenamefont {Shen}\ \emph {et~al.}()\citenamefont {Shen}, \citenamefont {Ji}, \citenamefont {Guan}, \citenamefont {Qian}, \citenamefont {Chai}, \citenamefont {Duan}, \citenamefont {Wang},\ and\ \citenamefont {Xia}}]{Shen2025}%
  \BibitemOpen
  \bibfield  {author} {\bibinfo {author} {\bibfnamefont {Q.}~\bibnamefont {Shen}}, \bibinfo {author} {\bibfnamefont {W.}~\bibnamefont {Ji}}, \bibinfo {author} {\bibfnamefont {J.}~\bibnamefont {Guan}}, \bibinfo {author} {\bibfnamefont {L.}~\bibnamefont {Qian}}, \bibinfo {author} {\bibfnamefont {Z.}~\bibnamefont {Chai}}, \bibinfo {author} {\bibfnamefont {C.}~\bibnamefont {Duan}}, \bibinfo {author} {\bibfnamefont {Y.}~\bibnamefont {Wang}},\ and\ \bibinfo {author} {\bibfnamefont {K.}~\bibnamefont {Xia}},\ }\bibfield  {title} {\bibinfo {title} {{Investigation of Rare-Earth Ion-Photon Interaction and Strong Coupling in Optical Microcavities}},\ }\Eprint {https://arxiv.org/abs/2504.09863} {arXiv:2504.09863} \BibitemShut {NoStop}%
\bibitem [{\citenamefont {Askarani}\ \emph {et~al.}(2021)\citenamefont {Askarani}, \citenamefont {Das}, \citenamefont {Davidson}, \citenamefont {Amaral}, \citenamefont {Sinclair}, \citenamefont {Slater}, \citenamefont {Marzban}, \citenamefont {Thiel}, \citenamefont {Cone}, \citenamefont {Oblak},\ and\ \citenamefont {Tittel}}]{Askarani2021}%
  \BibitemOpen
  \bibfield  {author} {\bibinfo {author} {\bibfnamefont {M.~F.}\ \bibnamefont {Askarani}}, \bibinfo {author} {\bibfnamefont {A.}~\bibnamefont {Das}}, \bibinfo {author} {\bibfnamefont {J.~H.}\ \bibnamefont {Davidson}}, \bibinfo {author} {\bibfnamefont {G.~C.}\ \bibnamefont {Amaral}}, \bibinfo {author} {\bibfnamefont {N.}~\bibnamefont {Sinclair}}, \bibinfo {author} {\bibfnamefont {J.~A.}\ \bibnamefont {Slater}}, \bibinfo {author} {\bibfnamefont {S.}~\bibnamefont {Marzban}}, \bibinfo {author} {\bibfnamefont {C.~W.}\ \bibnamefont {Thiel}}, \bibinfo {author} {\bibfnamefont {R.~L.}\ \bibnamefont {Cone}}, \bibinfo {author} {\bibfnamefont {D.}~\bibnamefont {Oblak}},\ and\ \bibinfo {author} {\bibfnamefont {W.}~\bibnamefont {Tittel}},\ }\bibfield  {title} {\bibinfo {title} {{Long-Lived Solid-State Optical Memory for High-Rate Quantum Repeaters}},\ }\href {https://doi.org/10.1103/PhysRevLett.127.220502} {\bibfield  {journal} {\bibinfo  {journal} {Phys. Rev. Lett.}\ }\textbf {\bibinfo {volume} {127}},\ \bibinfo {pages}
  {220502} (\bibinfo {year} {2021})}\BibitemShut {NoStop}%
\bibitem [{\citenamefont {Gerry}\ \emph {et~al.}(2005)\citenamefont {Gerry}, \citenamefont {Benmoussa},\ and\ \citenamefont {Campos}}]{Gerry2005}%
  \BibitemOpen
  \bibfield  {author} {\bibinfo {author} {\bibfnamefont {C.~C.}\ \bibnamefont {Gerry}}, \bibinfo {author} {\bibfnamefont {A.}~\bibnamefont {Benmoussa}},\ and\ \bibinfo {author} {\bibfnamefont {R.~A.}\ \bibnamefont {Campos}},\ }\bibfield  {title} {\bibinfo {title} {{Quantum nondemolition measurement of parity and generation of parity eigenstates in optical fields}},\ }\href {https://doi.org/10.1103/PhysRevA.72.053818} {\bibfield  {journal} {\bibinfo  {journal} {Phys. Rev. A}\ }\textbf {\bibinfo {volume} {72}},\ \bibinfo {pages} {053818} (\bibinfo {year} {2005})}\BibitemShut {NoStop}%
\bibitem [{\citenamefont {Munro}\ \emph {et~al.}(2005)\citenamefont {Munro}, \citenamefont {Nemoto},\ and\ \citenamefont {Spiller}}]{Munro2005}%
  \BibitemOpen
  \bibfield  {author} {\bibinfo {author} {\bibfnamefont {W.~J.}\ \bibnamefont {Munro}}, \bibinfo {author} {\bibfnamefont {K.}~\bibnamefont {Nemoto}},\ and\ \bibinfo {author} {\bibfnamefont {T.~P.}\ \bibnamefont {Spiller}},\ }\bibfield  {title} {\bibinfo {title} {{Weak nonlinearities: a new route to optical quantum computation}},\ }\href {https://doi.org/10.1088/1367-2630/7/1/137} {\bibfield  {journal} {\bibinfo  {journal} {New J. Phys.}\ }\textbf {\bibinfo {volume} {7}},\ \bibinfo {pages} {137} (\bibinfo {year} {2005})}\BibitemShut {NoStop}%
\bibitem [{\citenamefont {Thekkadath}\ \emph {et~al.}(2020)\citenamefont {Thekkadath}, \citenamefont {Bell}, \citenamefont {Walmsley},\ and\ \citenamefont {Lvovsky}}]{Thekkadath2020}%
  \BibitemOpen
  \bibfield  {author} {\bibinfo {author} {\bibfnamefont {G.~S.}\ \bibnamefont {Thekkadath}}, \bibinfo {author} {\bibfnamefont {B.~A.}\ \bibnamefont {Bell}}, \bibinfo {author} {\bibfnamefont {I.~A.}\ \bibnamefont {Walmsley}},\ and\ \bibinfo {author} {\bibfnamefont {A.~I.}\ \bibnamefont {Lvovsky}},\ }\bibfield  {title} {\bibinfo {title} {{Engineering Schr{\"{o}}dinger cat states with a photonic even-parity detector}},\ }\href {https://doi.org/10.22331/q-2020-03-02-239} {\bibfield  {journal} {\bibinfo  {journal} {Quantum}\ }\textbf {\bibinfo {volume} {4}},\ \bibinfo {pages} {239} (\bibinfo {year} {2020})}\BibitemShut {NoStop}%
\bibitem [{\citenamefont {Bruynsteen}\ \emph {et~al.}(2023)\citenamefont {Bruynsteen}, \citenamefont {Gehring}, \citenamefont {Lupo}, \citenamefont {Bauwelinck},\ and\ \citenamefont {Yin}}]{Bruynsteen2023}%
  \BibitemOpen
  \bibfield  {author} {\bibinfo {author} {\bibfnamefont {C.}~\bibnamefont {Bruynsteen}}, \bibinfo {author} {\bibfnamefont {T.}~\bibnamefont {Gehring}}, \bibinfo {author} {\bibfnamefont {C.}~\bibnamefont {Lupo}}, \bibinfo {author} {\bibfnamefont {J.}~\bibnamefont {Bauwelinck}},\ and\ \bibinfo {author} {\bibfnamefont {X.}~\bibnamefont {Yin}},\ }\bibfield  {title} {\bibinfo {title} {{100-Gbit/s Integrated Quantum Random Number Generator Based on Vacuum Fluctuations}},\ }\href {https://doi.org/10.1103/PRXQuantum.4.010330} {\bibfield  {journal} {\bibinfo  {journal} {PRX Quantum}\ }\textbf {\bibinfo {volume} {4}},\ \bibinfo {pages} {010330} (\bibinfo {year} {2023})}\BibitemShut {NoStop}%
\bibitem [{\citenamefont {Bruynsteen}\ \emph {et~al.}(2021)\citenamefont {Bruynsteen}, \citenamefont {Vanhoecke}, \citenamefont {Bauwelinck},\ and\ \citenamefont {Yin}}]{Bruynsteen2021}%
  \BibitemOpen
  \bibfield  {author} {\bibinfo {author} {\bibfnamefont {C.}~\bibnamefont {Bruynsteen}}, \bibinfo {author} {\bibfnamefont {M.}~\bibnamefont {Vanhoecke}}, \bibinfo {author} {\bibfnamefont {J.}~\bibnamefont {Bauwelinck}},\ and\ \bibinfo {author} {\bibfnamefont {X.}~\bibnamefont {Yin}},\ }\bibfield  {title} {\bibinfo {title} {{Integrated balanced homodyne photonic^^e2^^80^^93electronic detector for beyond 20 GHz shot-noise-limited measurements}},\ }\href {https://doi.org/10.1364/OPTICA.420973} {\bibfield  {journal} {\bibinfo  {journal} {Optica}\ }\textbf {\bibinfo {volume} {8}},\ \bibinfo {pages} {1146} (\bibinfo {year} {2021})}\BibitemShut {NoStop}%
\bibitem [{\citenamefont {Lordi}\ \emph {et~al.}(2024)\citenamefont {Lordi}, \citenamefont {Tsao}, \citenamefont {Lind}, \citenamefont {Diddams},\ and\ \citenamefont {Combes}}]{Lordi2024}%
  \BibitemOpen
  \bibfield  {author} {\bibinfo {author} {\bibfnamefont {N.}~\bibnamefont {Lordi}}, \bibinfo {author} {\bibfnamefont {E.~J.}\ \bibnamefont {Tsao}}, \bibinfo {author} {\bibfnamefont {A.~J.}\ \bibnamefont {Lind}}, \bibinfo {author} {\bibfnamefont {S.~A.}\ \bibnamefont {Diddams}},\ and\ \bibinfo {author} {\bibfnamefont {J.}~\bibnamefont {Combes}},\ }\bibfield  {title} {\bibinfo {title} {{Quantum theory of temporally mismatched homodyne measurements with applications to optical-frequency-comb metrology}},\ }\href {https://doi.org/10.1103/PhysRevA.109.033722} {\bibfield  {journal} {\bibinfo  {journal} {Phys. Rev. A}\ }\textbf {\bibinfo {volume} {109}},\ \bibinfo {pages} {033722} (\bibinfo {year} {2024})}\BibitemShut {NoStop}%
\bibitem [{\citenamefont {Hubenschmid}\ \emph {et~al.}(2024)\citenamefont {Hubenschmid}, \citenamefont {Guedes},\ and\ \citenamefont {Burkard}}]{Hubenschmid2024}%
  \BibitemOpen
  \bibfield  {author} {\bibinfo {author} {\bibfnamefont {E.}~\bibnamefont {Hubenschmid}}, \bibinfo {author} {\bibfnamefont {T.~L.~M.}\ \bibnamefont {Guedes}},\ and\ \bibinfo {author} {\bibfnamefont {G.}~\bibnamefont {Burkard}},\ }\bibfield  {title} {\bibinfo {title} {{Optical Time-Domain Quantum State Tomography on a Subcycle Scale}},\ }\href {https://doi.org/10.1103/PhysRevX.14.041032} {\bibfield  {journal} {\bibinfo  {journal} {Phys. Rev. X}\ }\textbf {\bibinfo {volume} {14}},\ \bibinfo {pages} {041032} (\bibinfo {year} {2024})}\BibitemShut {NoStop}%
\bibitem [{\citenamefont {Yang}\ \emph {et~al.}()\citenamefont {Yang}, \citenamefont {Kizmann}, \citenamefont {Leitenstorfer},\ and\ \citenamefont {Moskalenko}}]{Yang2023b}%
  \BibitemOpen
  \bibfield  {author} {\bibinfo {author} {\bibfnamefont {G.}~\bibnamefont {Yang}}, \bibinfo {author} {\bibfnamefont {M.}~\bibnamefont {Kizmann}}, \bibinfo {author} {\bibfnamefont {A.}~\bibnamefont {Leitenstorfer}},\ and\ \bibinfo {author} {\bibfnamefont {A.~S.}\ \bibnamefont {Moskalenko}},\ }\bibfield  {title} {\bibinfo {title} {{Subcycle tomography of quantum light}},\ }\Eprint {https://arxiv.org/abs/2307.12812} {arXiv:2307.12812} \BibitemShut {NoStop}%
\bibitem [{\citenamefont {Kraus}(1983)}]{Kraus1983}%
  \BibitemOpen
  \bibfield  {author} {\bibinfo {author} {\bibfnamefont {K.}~\bibnamefont {Kraus}},\ }\href {https://doi.org/10.1007/3-540-12732-1} {\emph {\bibinfo {title} {{States, Effects, and Operations: Fundamental Notions of Quantum Theory}}}},\ edited by\ \bibinfo {editor} {\bibfnamefont {A.}~\bibnamefont {B{\"{o}}hm}}, \bibinfo {editor} {\bibfnamefont {J.~D.}\ \bibnamefont {Dollard}},\ and\ \bibinfo {editor} {\bibfnamefont {W.~H.}\ \bibnamefont {Wootters}},\ \bibinfo {series} {Lecture Notes in Physics}, Vol.\ \bibinfo {volume} {190}\ (\bibinfo  {publisher} {Springer Berlin Heidelberg},\ \bibinfo {address} {Berlin, Heidelberg},\ \bibinfo {year} {1983})\BibitemShut {NoStop}%
\bibitem [{\citenamefont {Wiseman}\ and\ \citenamefont {Milburn}(2009)}]{Wiseman2009}%
  \BibitemOpen
  \bibfield  {author} {\bibinfo {author} {\bibfnamefont {H.~M.}\ \bibnamefont {Wiseman}}\ and\ \bibinfo {author} {\bibfnamefont {G.~J.}\ \bibnamefont {Milburn}},\ }\href {https://doi.org/10.1017/CBO9780511813948} {\emph {\bibinfo {title} {{Quantum Measurement and Control}}}}\ (\bibinfo  {publisher} {Cambridge University Press},\ \bibinfo {address} {Cambridge, UK},\ \bibinfo {year} {2009})\BibitemShut {NoStop}%
\bibitem [{\citenamefont {Aharonov}\ \emph {et~al.}(1988)\citenamefont {Aharonov}, \citenamefont {Albert},\ and\ \citenamefont {Vaidman}}]{Aharonov1988}%
  \BibitemOpen
  \bibfield  {author} {\bibinfo {author} {\bibfnamefont {Y.}~\bibnamefont {Aharonov}}, \bibinfo {author} {\bibfnamefont {D.~Z.}\ \bibnamefont {Albert}},\ and\ \bibinfo {author} {\bibfnamefont {L.}~\bibnamefont {Vaidman}},\ }\bibfield  {title} {\bibinfo {title} {{How the result of a measurement of a component of the spin of a spin- 1/2 particle can turn out to be 100}},\ }\href {https://doi.org/10.1103/PhysRevLett.60.1351} {\bibfield  {journal} {\bibinfo  {journal} {Phys. Rev. Lett.}\ }\textbf {\bibinfo {volume} {60}},\ \bibinfo {pages} {1351} (\bibinfo {year} {1988})}\BibitemShut {NoStop}%
\bibitem [{\citenamefont {Caves}\ and\ \citenamefont {Milburn}(1987)}]{Caves1987}%
  \BibitemOpen
  \bibfield  {author} {\bibinfo {author} {\bibfnamefont {C.~M.}\ \bibnamefont {Caves}}\ and\ \bibinfo {author} {\bibfnamefont {G.~J.}\ \bibnamefont {Milburn}},\ }\bibfield  {title} {\bibinfo {title} {{Quantum-mechanical model for continuous position measurements}},\ }\href {https://doi.org/10.1103/PhysRevA.36.5543} {\bibfield  {journal} {\bibinfo  {journal} {Phys. Rev. A}\ }\textbf {\bibinfo {volume} {36}},\ \bibinfo {pages} {5543} (\bibinfo {year} {1987})}\BibitemShut {NoStop}%
\bibitem [{\citenamefont {Dalibard}\ \emph {et~al.}(1992)\citenamefont {Dalibard}, \citenamefont {Castin},\ and\ \citenamefont {M{\o}lmer}}]{Dalibard1992}%
  \BibitemOpen
  \bibfield  {author} {\bibinfo {author} {\bibfnamefont {J.}~\bibnamefont {Dalibard}}, \bibinfo {author} {\bibfnamefont {Y.}~\bibnamefont {Castin}},\ and\ \bibinfo {author} {\bibfnamefont {K.}~\bibnamefont {M{\o}lmer}},\ }\bibfield  {title} {\bibinfo {title} {{Wave-function approach to dissipative processes in quantum optics}},\ }\href {https://doi.org/10.1103/PhysRevLett.68.580} {\bibfield  {journal} {\bibinfo  {journal} {Phys. Rev. Lett.}\ }\textbf {\bibinfo {volume} {68}},\ \bibinfo {pages} {580} (\bibinfo {year} {1992})}\BibitemShut {NoStop}%
\bibitem [{\citenamefont {Wiseman}\ and\ \citenamefont {Milburn}(1993)}]{Wiseman1993}%
  \BibitemOpen
  \bibfield  {author} {\bibinfo {author} {\bibfnamefont {H.~M.}\ \bibnamefont {Wiseman}}\ and\ \bibinfo {author} {\bibfnamefont {G.~J.}\ \bibnamefont {Milburn}},\ }\bibfield  {title} {\bibinfo {title} {{Quantum theory of field-quadrature measurements}},\ }\href {https://doi.org/10.1103/PhysRevA.47.642} {\bibfield  {journal} {\bibinfo  {journal} {Phys. Rev. A}\ }\textbf {\bibinfo {volume} {47}},\ \bibinfo {pages} {642} (\bibinfo {year} {1993})}\BibitemShut {NoStop}%
\bibitem [{\citenamefont {Misra}\ and\ \citenamefont {Sudarshan}(1977)}]{Misra1977}%
  \BibitemOpen
  \bibfield  {author} {\bibinfo {author} {\bibfnamefont {B.}~\bibnamefont {Misra}}\ and\ \bibinfo {author} {\bibfnamefont {E.~C.~G.}\ \bibnamefont {Sudarshan}},\ }\bibfield  {title} {\bibinfo {title} {{The Zeno's paradox in quantum theory}},\ }\href {https://doi.org/10.1063/1.523304} {\bibfield  {journal} {\bibinfo  {journal} {J. Math. Phys.}\ }\textbf {\bibinfo {volume} {18}},\ \bibinfo {pages} {756} (\bibinfo {year} {1977})}\BibitemShut {NoStop}%
\bibitem [{\citenamefont {Fock}(1928)}]{Fock1928}%
  \BibitemOpen
  \bibfield  {author} {\bibinfo {author} {\bibfnamefont {V.}~\bibnamefont {Fock}},\ }\bibfield  {title} {\bibinfo {title} {{Verallgemeinerung und L{\"{o}}sung der Diracschen statistischen Gleichung}},\ }\href {https://doi.org/10.1007/BF01337923} {\bibfield  {journal} {\bibinfo  {journal} {Eur. Phys. J. A}\ }\textbf {\bibinfo {volume} {49}},\ \bibinfo {pages} {339} (\bibinfo {year} {1928})}\BibitemShut {NoStop}%
\bibitem [{\citenamefont {Bargmann}(1961)}]{Bargmann1961}%
  \BibitemOpen
  \bibfield  {author} {\bibinfo {author} {\bibfnamefont {V.}~\bibnamefont {Bargmann}},\ }\bibfield  {title} {\bibinfo {title} {{On a Hilbert space of analytic functions and an associated integral transform part I}},\ }\href {https://doi.org/10.1002/cpa.3160140303} {\bibfield  {journal} {\bibinfo  {journal} {Commun. Pure Appl. Math.}\ }\textbf {\bibinfo {volume} {14}},\ \bibinfo {pages} {187} (\bibinfo {year} {1961})}\BibitemShut {NoStop}%
\bibitem [{\citenamefont {Genes}\ \emph {et~al.}(2003)\citenamefont {Genes}, \citenamefont {Berman},\ and\ \citenamefont {Rojo}}]{Genes2003}%
  \BibitemOpen
  \bibfield  {author} {\bibinfo {author} {\bibfnamefont {C.}~\bibnamefont {Genes}}, \bibinfo {author} {\bibfnamefont {P.~R.}\ \bibnamefont {Berman}},\ and\ \bibinfo {author} {\bibfnamefont {A.~G.}\ \bibnamefont {Rojo}},\ }\bibfield  {title} {\bibinfo {title} {{Spin squeezing via atom-cavity field coupling}},\ }\href {https://doi.org/10.1103/PhysRevA.68.043809} {\bibfield  {journal} {\bibinfo  {journal} {Phys. Rev. A}\ }\textbf {\bibinfo {volume} {68}},\ \bibinfo {pages} {043809} (\bibinfo {year} {2003})}\BibitemShut {NoStop}%
\end{thebibliography}%
\end{document}